\documentclass[preprint,10pt,1p]{elsarticle}

\usepackage{amsfonts}
\usepackage{amssymb}
\usepackage{amsthm}
\usepackage{mathdots}
\usepackage{amsmath}
\usepackage{xcolor}
\usepackage[mathscr]{euscript}
\usepackage{mathrsfs}
\usepackage{lineno}
\usepackage{hyperref}
\usepackage{booktabs}
\usepackage{tabularx}
\usepackage{pifont}
\usepackage{appendix}

\usepackage{lipsum}
\makeatletter
\def\ps@pprintTitle{%
 \let\@oddhead\@empty
 \let\@evenhead\@empty
 \def\@oddfoot{}%
 \let\@evenfoot\@oddfoot}
\makeatother

\newcommand{\cmark}{\ding{51}}
\newcommand{\xmark}{\ding{53}}
\newcommand{\ie}{\emph{i.e.}}
\newcommand{\eg}{\emph{e.g.}}

\newcommand{\avg}[1]{\langle #1\rangle}
\newcommand{\ensG}{\mathcal{G}}
\newcommand{\mG}{\mathbf{G}}
\newcommand{\mW}{\mathbf{W}}
\newcommand{\mA}{\mathbf{A}}
\newcommand{\hmG}{\hat{\mG}}
\newcommand{\hmW}{\hat{\mW}}
\newcommand{\hmA}{\hat{\mA}}
\newcommand{\hw}{\hat w}
\newcommand{\ha}{\hat a}

\newcommand{\hL}{\hat L}
\newcommand{\hW}{\hat W}
\newcommand{\so}{s^{\mbox{\tiny out}}}
\newcommand{\si}{s^{\mbox{\tiny in}}}
\newcommand{\hso}{\hat{s}^{\mbox{\tiny out}}}
\newcommand{\hsi}{\hat{s}^{\mbox{\tiny in}}}
\newcommand{\ko}{k^{\mbox{\tiny out}}}
\newcommand{\ki}{k^{\mbox{\tiny in}}}
\newcommand{\hko}{\hat{k}^{\mbox{\tiny out}}}
\newcommand{\hki}{\hat{k}^{\mbox{\tiny in}}}

\newcommand{\yo}{y^{\mbox{\tiny out}}}
\newcommand{\yi}{y^{\mbox{\tiny in}}}
\newcommand{\xxo}{x^{\mbox{\tiny out}}}
\newcommand{\xxi}{x^{\mbox{\tiny in}}}

\journal{Physics Reports}

\begin{document}

\begin{frontmatter}

\title{Reconstruction methods for networks:\\the case of economic and financial 
systems}
\author[label1]{Tiziano Squartini}
\ead{tiziano.squartini@imtlucca.it}
\author[label1,label2,label3]{Guido Caldarelli\corref{cor1}}
\cortext[cor1]{Corresponding author: guido.caldarelli@imtlucca.it}
\author[label1,label2]{Giulio Cimini}
\ead{giulio.cimini@imtlucca.it}
\author[label2,label1]{Andrea Gabrielli}
\ead{andrea.gabrielli@roma1.infn.it}
\author[label1,label4]{Diego Garlaschelli}
\ead{garlaschelli@lorentz.leidenuniv.nl}

\address[label1]{IMT School for Advanced Studies Lucca, P.zza San Francesco 19, 
55100 Lucca (Italy)}
\address[label2]{Istituto dei Sistemi Complessi (ISC) - CNR, UoS Sapienza, 
Dipartimento di Fisica, \\Universit\`a ``Sapienza'', P.le Aldo Moro 5, 00185 
Rome (Italy)}
\address[label3]{European Centre for Living Technology (ECLT)
San Marco 2940, 30124 Venezia , ITALY}
\address[label4]{Lorentz Institute for Theoretical Physics, Leiden Institute of 
Physics, \\University of Leiden, Niels Bohrweg 2, 2333 CA Leiden (The 
Netherlands)}

\begin{abstract}

The study of social, economic and biological systems is often (when not always) limited by the partial information about the structure of the underlying networks. An example of paramount importance is provided by financial systems: information on the interconnections between financial institutions is privacy-protected, dramatically reducing the possibility of correctly estimating crucial systemic properties such as the resilience to the propagation of shocks. The need to compensate for the scarcity of data, while optimally employing the available information, has led to the birth of a research field known as {\em network reconstruction}. Since the latter has benefited from the contribution of researchers working in disciplines as different as mathematics, physics and economics, the results achieved so far are still scattered across heterogeneous publications. Most importantly, a systematic comparison of the network reconstruction methods proposed up to now is currently missing. This review aims at providing a unifying framework to present all these studies, mainly focusing on their application to economic and financial networks.
\end{abstract}

\begin{keyword}
entropy maximization \sep network reconstruction \sep statistical inference \sep 
economic and financial networks \sep systemic risk evaluation
\PACS 02.50.Tt \sep 89.65.Gh \sep 89.70.Cf \sep 89.75.Hc
\end{keyword}

\end{frontmatter}


\newpage

\vspace*{1cm}

\begin{flushright}
\small{The study of truth requires a considerable effort}\\
\small{which is why few are willing to undertake it out of love of 
knowledge}\\\medskip
\small{--- Thomas Aquinas, \emph{Summa Contra Gentiles}} \\
\end{flushright}

\newpage

\tableofcontents

\newpage

{\bf List of the symbols employed in the review}\\

\noindent Throughout this work we shall employ top-hatted letters (\eg, $\hso_i$, $\hsi_i$, $\hw_{ij}$, etc.) 
for the measured values of the empirical networks to be reconstructed, while we shall use the same letters (\eg, $\so_i$, $\si_i$, $w_{ij}$, etc.) 
without any addition to indicate the same quantities when considered as (deterministic or stochastic) variables. In particular:

\begin{itemize}
\item[$\bullet$] $\hmG$: the (observed) network to reconstruct;
\item[$\bullet$] $\ensG$: ensemble of network configurations used to reconstruct $\hmG$;
\item[$\bullet$] $N$: (fixed) number of nodes of $\hmG$ (and of all networks in the ensemble $\ensG$);
\item[$\bullet$] $\mG$: a generic network configuration belonging to $\ensG$;
\item[$\bullet$] $P(\mG)$: probability measure on the ensemble of configurations $\ensG$;
\item[$\bullet$] $I(\mG)$: information content of the configuration $\mG$;
\item[$\bullet$] $H(\mG)$: network Hamiltonian of the configuration $\mG$, i.e. linear combination of the constraints determining its probability $P(\mG)$;
\item[$\bullet$] $S$: Shannon entropy;
\item[$\bullet$] $\mathscr{L}$: Lagrangian functional, \ie, the constrained Shannon entropy;
\item[$\bullet$] $M$: number of constraints (excluding normalization) in $\mathscr{L}$;
\item[$\bullet$] $C_m$: $m$-th constraint;
\item[$\bullet$] $\lambda_m$: $m$-th Lagrange multiplier, controlling for the $m$-th constraint; 
\item[$\bullet$] $\hat{\lambda}_m$: estimation of the $m$-th Lagrange multiplier for $\hmG$;  
\item[$\bullet$] $\mathcal{L}(\mG|\vec{\lambda})$: log-likelihood of $\mG$ given the Lagrange multipliers defining $P(\mG)$;
\item[$\bullet$] $\avg{ X}$: expected value over the ensemble $\ensG$ of a generic quantity of interest $X$;
\item[$\bullet$] $\sigma_X$: ensemble standard deviation of a generic quantity of interest $X$;
\item[$\bullet$] $X^{\text{\tiny name}}$: estimate of quantity $X$ by the algorithm ``name'';
\item[$\bullet$] $L$: number of links in a network;
\item[$\bullet$] $\rho=L/[N(N-1)]$: density of links in a (directed) network, defined as the fraction of possible connections in the network that are actually realized;
\item[$\bullet$] $\mW$: weighted adjacency matrix representing a network, with generic element $w_{ij}$ denoting the weight of the link from node $i$ to node $j$;
\item[$\bullet$] $\mA$: adjacency matrix representing a binary version of $\mW$, with generic element $a_{ij}=1$ if $w_{ij}>0$ (and 0 otherwise);
\item[$\bullet$] $\so_i=\sum_{j=1}^N w_{ij}$ and $\si_i=\sum_{j=1}^N w_{ji}$: out-strength and in-strength of node $i$, 
or equivalently the {\em marginal} row and column sums of $\mW$;
\item[$\bullet$] $\ko_i=\sum_{j=1}^N a_{ij}$ and $\ki_i=\sum_{j=1}^N a_{ji}$: out-degree and in-degree of node $i$, or equivalently the {\em marginal} row and column sums of $\mA$;
\end{itemize}

\newpage

\section{Introduction}

\paragraph{Networks: the why and how of a theory} There is nowadays an 
overwhelming evidence that a large deal of complex systems around us can be 
successfully described by means of complex networks 
\cite{albert2002statistical,boccaletti2006complex,caldarelli2007scale}. Graph 
theory, from which complex networks theory originates, was firstly developed in 
the XVIII century as an application of discrete mathematics to the well-known 
``K\"onigsberg bridge problem'' \cite{carlson2014koenisberg} and for many years 
it remained relegated to merely solving puzzling topological problems 
\cite{biggs1998graph}.

A new burst of activity was registered in 1920, to provide mathematical support 
to the analysis of social networks 
\cite{moreno1934sociometry,moreno1941sociometry}. This laid the foundation 
for ``sociometry'', a discipline characterized by a mathematical description 
of social sciences\footnote{A similar approach coupled to modern technology 
constitutes the core of a newborn field, ``computational social 
science'' \cite{lazer2009social}, where mathematics is employed to analyze the 
huge amount of data generated by online social networks.}. Later on, the seminal papers 
of Erd\H{o}s and R\'enyi \cite{erdos1959random,erdos1960evolution}, lately extended to combinatorics by Bollob\'as \cite{bollobas1979graph}, opened the field of  ``random graph theory'' \cite{bollobas1985random}.

Only after the digital revolution, complex networks became part of everybody's life. 
Indeed current technology has made the presence of personal computers pervasive and also  
constantly reduced the cost of backup memories. These two features have produced immense 
databases, collecting information about a wide range of relationships. Just to name a few we have:
the Internet wiring \cite{faloutsos1999power,caldarelli2000fractal,huffaker2002distance}, 
the set of WWW connections \cite{page1999pagerank,kleinberg1999authoritative}, 
e-mail exchanges \cite{guimera2003self,newman2002email} \cite{caldarelli2004preferential}, 
mobile communication networks \cite{onnela2007structure,eagle2009inferring,miritello2011dynamical}, 
online social networks \cite{mislove2007measurements,borgatti2009network,kwak2010what,delvicario2016spreading}, 
protein-protein interactions \cite{mason2007network,STRING,koh2012analyzing}, 
food-webs and ecological networks \cite{williams2000simple,dunne2002food,proulx2005network},
production activities \cite{garlaschelli2005structure,hidalgo2007product,schweitzer2009economic,tacchella2012ec,cimini2014science,pugliese2017unfolding}, 
financial exchanges and stock investments \cite{iori2008network,glattfelder2009backbone,bordino2012web}.

The network representation clearly highlights qualitative universal behaviors, 
irrespectively from the specific case-studies \cite{barabasi2009scale}. In what 
follows we name some of the features shared by many real-world networks:

\begin{itemize}
\item a \emph{long-tail} (or even \emph{scale-free}) \emph{degree distribution} 
\cite{faloutsos1999power,barabasi1999emergence}; 
\item a \emph{small-world effect} \cite{watts1998collective}: distances are 
distributed around a characteristic average value, usually ``very small'', (i.e. $3-10$)
and scaling as the logarithm of the system size;
\item a \emph{large clustering} \cite{szabo2004clustering}: we spot the presence of densely-connected 
subgraphs  in many complex networks. The small-world and the 
large clustering effects co-exist thanks to presence of the so-called 
``weak-ties'' in the social networks literature, allowing for ``long-range'' 
interconnections without affecting the locally large density of links;
\item a distinct \emph{centrality} structure 
\cite{bonacich1987power} implying that some nodes appear 
to have a higher importance than others;
\item a well-defined \emph{assortativity structure} 
\cite{newman2002assortative}: the neighbors of each node have a degree that is 
either positively or negatively correlated to the degree of the node itself 
(more intuitively, ``my'' neighbors have a degree that is either very similar or 
very different from mine). In the case of bipartite networks, a well-defined 
{\em nestedness structure} has been observed, mainly in ecological and economic 
contexts \cite{almeida2008consistent,bascompte2003nestedness,johnson2013factors}.
\end{itemize}

\paragraph{The problem of missing information} After an initial activity aimed 
at determining the structure of real-world networks by measuring standard 
topological quantities, a more theoretical activity was started, aiming at both 
defining new quantities and devising proper models to explain observations \cite{watts1998collective,barabasi1999emergence,caldarelli2002scale,leskovec2008microscopic,medo2011temporal}.
Given the complexity that can arise even from a simple mathematical model based upon 
graphs, researchers have recently focused on the development of a {\em 
topological} theory: loosely speaking, topological quantities are 
employed to define statistical models, rather than reproduced from microscopic 
dynamical rules \cite{park2004statistical,bianconi2008entropy,garlaschelli2008maximum,squartini2011analytical,fronczak2014exponential}.

Unfortunately, when moving to the validation of such models a common problem 
arises: very often, the data available on the real network are either incomplete or 
imprecise (or both). This problem is particularly evident in the case of 
economic and financial networks: in this case, data collection suffers from the 
problem of partial accounting and the presence of disclosure requirements. In 
order to illustrate the importance of such an issue, let us think of a 
bipartite, financial network whose node sets represent investors and the 
investments they do. Although the knowledge of the whole network structure could 
help regulators to take immediate countermeasures to stop the propagation of 
financial distress, this information is seldom available (the knowledge of the 
whole network of investments would pose immense problems of privacy), thus 
hindering the possibility of providing a realistic estimate of the extent of the 
contagion. As confirmed by the analysis of the various papers reported in this 
review, the incompleteness of network instances seems to be unavoidable 
\cite{battiston2016price,anand2017missing}: since addressing the problem of estimating the 
resilience of financial networks cannot be addressed without knowing the 
structural details of national and cross-countries interbank networks, 
information theory seems indeed to provide the right framework to tackle this 
kind of problems. 

Finance is not the only domain affected by limitedness of information about 
nodes interdependencies: biological and ecological systems also exist (\eg, cell 
metabolic networks and ecological webs) whose interaction network is often only 
partially accessible due either to experimental limitations or observational 
constraints\footnote{In these cases, one is often more interested 
in the reconstruction of individual links from partial local information, 
a problem known as ``link prediction'' \cite{guimera2009missing,lu2011predicting}.}.

\paragraph{Approaching network reconstruction} In order to deal with the problem 
of missing information, many different approaches have been attempted so far. 
Some reconstruction procedures are based on a measured (or expected) statistical 
self-similarity of the network topology \cite{song2005self}. In this case, the 
observed behavior of a given topological property (\eg, the degree distribution) 
is supposed to be induced by some non-topological property assigned to nodes 
(\eg, a node ``fitness"\cite{caldarelli2002scale}) obeying the same behavior.

More often, the fundamental assumption grounding network reconstruction is {\em statistical 
homogeneity}. This means that the structure of the network observed are representative of 
statistical properties not depending on that specific portion. 
In the jargon of statistics, supposing that from similar observations we can infer 
similar regularities is equivalent to requiring that the  information available 
is representative of the whole network structure. This is particularly relevant 
when only limited information is available, and constitutes the physical reason 
to be confident that the efforts to define a statistically-grounded network 
reconstruction algorithm can indeed be successful. By using this ``symmetry'' it 
is then possible to provide some likely estimates of the missing quantities of 
the network under analysis. Clearly, while the homogeneity assumption minimizes 
the bias introduced by adopting arbitrary assumptions not supported by the 
available information (and, in principle, untestable), it also limits the 
accuracy of the reconstruction of real networks showing, instead, strong 
structural heterogeneity.

Given these premises, entropy maximization provides the unifying concept underlying 
all the reviewed methods. Entropy maximization is, in fact, an ubiquitous 
prescription for obtaining the least biased probability distribution consistent 
with some imposed constraints (\ie, a probability distribution not encoding other 
information than that represented by the constraints themselves 
\cite{cover2006elements,hanel2014multiplicity}). This principle has not only found hundreds of 
applications in statistical mechanics, information theory and statistics 
\cite{presse2013principles}, but it has been also argued to represent an evolutionary drive 
of out-of-equilibrium systems \cite{dewar2003information} (\eg, a relationship 
has been suggested between the dynamics of intelligent systems and 
entropy maximization \cite{wissner2013causal}).

\medskip

The outline of this report is the following. In section \ref{secinf}, 
\emph{Information theory as a basis for network reconstruction}, we present the 
tools that can be derived from classical approaches in statistical physics and 
information theory, mainly Gibbs' ensembles theory, Shannon's works on entropy 
and Jaynes' interpretation of statistical mechanics. Moving from these 
theoretical premises, in section \ref{secrec}, \emph{Reconstruction methods}, we 
present an overview of the different reconstruction methods, dividing them into 
{\bf a)} dense reconstruction methods, {\bf b)} density-tunable reconstruction 
methods, {\bf c)} exact-density methods, {\bf d)} alternative approaches. In 
section \ref{secfin}, \emph{Testing the network reconstruction}, we present in 
detail a number of indicators and metrics that can be used to test the accuracy of the 
achieved reconstruction; more specifically, we have distinguished three classes 
of indicators, of \emph{statistical}, \emph{topological} and \emph{dynamical} nature, 
respectively aiming at testing the accuracy in reconstructing the 
microscopic details, the macroscopic topological features and the dynamical properties of a 
given network. Concerning dynamical indicators, we put particular emphasis on the possibility 
of estimating the resilience of the system to processes of shocks propagation. 
In the financial context this is know as systemic risk, \ie, the likelihood that a consistent part of a given 
financial network may collapse (go bankrupt) as a consequence of a local failure. 
In section \ref{secrit}, \emph{Model selection criteria}, we 
describe some of the existing criteria to compare different models and the 
corresponding recipes for how to choose the most appropriate one. The report ends with 
section \ref{conc}, \emph{Conclusions and perspectives}, where we describe 
future possible applications of the reviewed algorithms.

\medskip

As a general remark, we would like to stress that almost every paper 
reviewed here has its own nomenclature for the (often similar) quantities of 
interest. Since we wanted to present the different contributions within a 
unified framework, our presentation might not reflect the original derivation of 
the results.

\section{Information theory as a basis for network reconstruction}\label{secinf}
Information theory provides the theoretical basis of our formalism \cite{anand2009entropy}. The concept 
of \emph{information} plays a fundamental role in network reconstruction, since 
\emph{reconstructing a network} ultimately means making optimal use of the 
available, partial, information. Otherwise stated, our task is that of inferring 
as much as possible about the system under analysis from the available data, 
while limiting the number of unsupported assumptions.

As stated before, real data are very often partial: thus, any data-driven 
inference procedure is bound to consider an enlarged set of plausible 
configurations, \ie, \emph{all configurations that are compatible with the 
available information}. In the language of statistical mechanics, this set is 
called \emph{ensemble}. Enlarging the set of allowable configurations means, in 
turn, increasing the degree of uncertainty about the actual one; the 
description, thus, becomes necessarily \emph{probabilistic}: a probability value 
must be assigned to each configuration compatible with the known information, 
that is, to all configurations belonging to the ensemble.

The degree of plausibility of a given configuration can be unambiguously 
quantified by recalling the concept of \emph{surprise}. Since a ``surprising'' 
event (deemed as highly improbable) is assumed to convey a large amount 
of information \cite{cover2006elements}. An operative definition of surprise 
should encode a (negative) correlation with the probability of realization of 
the event under consideration. The content of information of a given outcome 
$\mG$ out of the set of possible outcomes $\ensG$ can be thus 
quantified as

\begin{equation}
I(\mG)=-\ln P(\mG),
\end{equation}
a definition pointing out that the occurrence of an event that is certain (\ie, 
characterized by $P(\mG)=1$) brings no information and comes with 
no surprise, whereas, the occurrence of an (almost) impossible event (\ie, 
characterized by $P(\mG)\simeq0$) conveys an (almost) infinite amount of 
information and causes an (almost) ``infinite'' amount of surprise 
\cite{cover2006elements}. The average degree of surprise which 
accompanies the events belonging to the set $\ensG$ can be then quantified by 
averaging over the ensemble itself. Such an operation leads to the basic concept of the \emph{Shannon entropy}:

\begin{equation}
S=\avg{ I}=\sum_{\mG\in 
\ensG}P(\mG)I(\mG)=\sum_{\mG\in 
\ensG}-P(\mG)\ln P(\mG).
\label{sh-ent}
\end{equation}

Another interpretation of eq. \eqref{sh-ent} comes from information theory 
\cite{shannon1948mathematical1,shannon1948mathematical2}. Given an alphabet of 
symbols (as a language) to be transmitted across a channel, shorter codes should 
be assigned to symbols met with larger frequency, while longer codes should be 
employed for symbols met with smaller frequency (see Appendix A). Looking for 
the average code length needed to transmit a given message leads to $S$ as well.

From an axiomatic point of view, Shannon entropy is the {\em only} functional 
that satisfies a number of properties known as the Shannon-Khinchin axioms 
\cite{hanel2014multiplicity}:

\begin{enumerate}
\item Shannon entropy is a continuous functional of all its arguments: this 
ensures that small deformations $\delta P(\mG)$ of the probability 
distribution $P(\mG)$ induce small changes in $S$;
\item Shannon entropy attains its maximum in correspondence of the uniform 
distribution over the set of possible configurations;
\item Shannon entropy is invariant under the addition of events with zero 
probability;
\item For a system composed by two independent subsystems $A$ and $B$, whose 
ensembles of possible configurations $\ensG_A$ and $\ensG_B$ have 
probability measures $W_A$ and $W_B$, the entropy is additive: 
$S(W_{A+B})=S(W_{A}W_{B})=S(W_{A})+S(W_{B})$, \ie the entropy of the whole 
system is the sum of the entropies of the two subsystems\footnote{In 
\cite{jaynes1957information} this axiom is replaced by the so-called 
``composition law''.}. On the other hand, if the two subsystems are not 
independent, $S(W_{A+B})=S(W_{A})+S(W_{B|A})$, with $W_{B|A}$ indicating the 
probability measure for the configurations of the subsystem $B$, conditioned to 
the realization of subsystem $A$.
\end{enumerate}

In other words, Shannon entropy is a functional of the probability distribution 
of an arbitrary set of random variables (in our case, the configurations 
$\mG$ within the aforementioned ensemble $\ensG$) and quantifies 
the (un)evenness of the distribution itself \cite{lesne2014shannon}. As an 
example, if no information on the system is available, uncertainty about it is 
maximal and Shannon entropy prescribes to assign a uniform 
distribution over $\ensG$. By converse, any statistical information gained on the system 
reduces the uniform character of the probability distribution, which becomes 
progressively more peaked in correspondence of the configurations conveying the given information.

\subsection{Setting the problem: constraining Shannon entropy}

E. T. Jaynes first pointed out the possibility of using Shannon entropy to 
define a novel inference procedure \cite{jaynes1957information}, by extending 
the recipe proposed by Gibbs in the context of statistical mechanics. Jaynes 
proposed to carry out a constrained maximization of Shannon entropy, \ie, to 
maximize the functional

\begin{equation}\label{eqp}
\mathscr{L}[P]=S-\lambda_0\left[\sum_{\mG\in\ensG}P(\mG
)-1\right]-\sum_{m=1}^M\lambda_m\left[\sum_{\mG\in\ensG}P(\mathbf{G
})C_m(\mG)-\avg{C_m}\right],
\end{equation}
with the $M$ quantities $\{C_m(\mG)\}_{m=1}^M$ that sum up the available 
knowledge on the system acting as constraints to be satisfied by the probability 
distribution $P(\mG)$ itself. Maximizing Shannon entropy ensures that 
\emph{our ignorance about the system to be reconstructed is maximized, except 
for what is known} or, equivalently, that the number of unjustified assumptions 
about the system itself is minimized. Indeed, it can be proven that the 
probability distribution that maximizes Shannon entropy is maximally 
non-committal with respect to the unknown information 
\cite{jaynes1957information} (see Appendix B).

Since {\em constrained} Shannon entropy maximization represents a sort of 
``guessing'' process about the unknown information, characterized by the least 
amount of statistical bias, it can be viewed as an updated version of the 
Laplace principle of insufficient reason \cite{jaynes1957information}. The 
latter states that in absence of any information about the system under analysis, 
there is no reason to prefer any particular configuration which, thus, is  
equiprobable to any other. This principle is nothing else than a particular case of eq. 
\eqref{eqp} with no constraints but the normalization condition---which actually leads 
to the uniform distribution over the ensemble 
$P(\mG)=\frac{1}{|\ensG|},\:\forall\:\mG\in\ensG$. As 
already mentioned, Shannon entropy attains its maximum in this situation.

The framework sketched above is general enough to allow physical systems as well 
as networks to be analyzed. While in the first case the $M$ constraints 
represent physical quantities (\eg, the mean energy of the system), in the second case they 
represent purely topological quantities as nodes degrees, the network's 
reciprocity, and so on. We would also like to stress that the Gibbs-Jaynes approach 
has been originally defined within the realm of equilibrium statistical 
mechanics, and this is also the spirit that has guided its application to networks. 
Generally speaking, however, networks do not satisfy the equilibrium conditions 
valid for thermodynamic systems. Consequently, in the networks realm the 
adoption of an entropy-maximization approach for the reconstruction of 
higher-order statistical properties from lower-order constraints is rather 
justified by the minimization of arbitrary statistical assumptions on 
the network structure not supported by the available information.

\medskip

Let us now explicitly consider $\ensG$ as the set of all possible 
network configurations with $N$ vertices. In most cases we consider networks that 
are both weighted and directed, which implies that the interlinkages 
characterizing them can be represented by an asymmetric $N\times N$ matrix with real entries:

\begin{equation}
\mW=\left( 	
\begin{array}{cccccc}
	w_{11} & \dots & w_{1i} & \dots & w_{1N} \\
	\vdots & \ddots & \vdots & \ddots & \vdots \\
	w_{i1} & \dots & w_{ii} & \dots & w_{iN} \\
	\vdots & \ddots & \vdots & \ddots & \vdots \\
	w_{N1} & \dots & w_{Ni} & \dots & w_{NN} \\
\end{array} 
\right)
\end{equation}
where $w_{ij}\ge 0$ represents the weight of the link from node $i$ to node $j$. 
Matrix $\mW$ induces a second matrix $\mA$, known as the 
{\em adjacency matrix} of the network. Formally, the generic element $a_{ij}$ 
of $\mA$ is equal to 1 if $w_{ij}>0$, and 0 otherwise. In other words, 
the matrix $\mA$ simply indicates the presence/absence of connections 
between node pairs.

The problem of network reconstruction arises whenever 
the weights $\hw_{ij}$ of an empirical network $\hmW$ are not directly observable, 
and instead only aggregate ({\em marginal}) information on the network is accessible. More precisely, 
only the sum of the rows and/or the columns are typically known:
\begin{equation}\label{margins}
\left\{
\begin{array}{ll}
\hso_i=\sum_{j=1}^N \hw_{ij} & \mbox{(out-strength)}\\
&\\
\hsi_i=\sum_{j=1}^N \hw_{ji} & \mbox{(in-strength)}
\end{array}
\right.\:\forall\:i.
\end{equation}
Note that rows and column sums of $\hmA$, namely the out-degree 
$\hko_i=\sum_{j=1}^N \ha_{ij}$ and in-degree 
$\hki_i=\sum_{j=1}^N \ha_{ji}$ of each node $i$ are typically unknown.

In the financial context, $\hmW$ typically represents 
a matrix of interbank exposures, also named {\em liability matrix}. 
The entries $\hw_{ij}$ of this matrix are the loans and borrowings between banks, 
protected by privacy issues, while marginals are publicly released in balance sheets. 
$\hso_i$ then quantifies the total interbank {\em assets} of node $i$, 
and $\hsi_i$ its total interbank {\em liabilities} (see section \ref{balshee}). 
Another classical example is the World Trade Network (WTN), 
where these quantities represent the total export and import of countries.

Generally speaking, any algorithm aimed at reconstructing a weighted directed 
network outputs two matrices, $\mathbf{P}=\{p_{ij}\}_{i,j=1}^N$ and $\mW=\{w_{ij}\}_{i,j=1}^N$:
while the generic entry $p_{ij}$ of the first matrix describes the probability that any two nodes $i$ and $j$ are 
connected, the generic entry of the second matrix provides an estimate of the 
weight $w_{ij}$ of the corresponding link. We can say that, in certain conditions, 
the probabilities and weights estimates of the methods considered in 
the present review are functions of the accessible information and can, in general, be written as 
$p_{ij}(\hso_i,\hsi_j)$ and $w_{ij}(\hso_i,\hsi_j)$. As a consequence, 
the entries $a_{ij}$ can be interpreted as Bernoulli variables that are $1$ with 
a certain probability $p_{{ij}}$ and $0$ otherwise.

Since the number of available data in the cases we consider here is $O(N)$ ($2N$ 
if only out- and in-strengths are known), the problem of reconstructing an adjacency matrix 
of $N^2$ real numbers is under-determined. In what follows, we 
shall review methods that adopt a probabilistic approach to tackle these kinds of 
problems, making use of tools and concepts developed within the 
information-theoretic framework introduced in the previous subsection.

\subsection{Exponential Random Graphs}\label{ERG-sect}

Exponential Random Graphs (ERG) \cite{park2004statistical,fronczak2014exponential} 
occupy a central role in most of network reconstruction algorithms. Indeed, 
ERG are defined as the ensemble of graphs whose 
probability $P(\mG)$ is obtained by maximizing of the constrained entropy 
functional of eq. \eqref{eqp}. More specifically, solving the functional 
differential equation $\frac{\delta \mathscr{L}[P]}{\delta P(\mG)}=0$ 
with respect to $P(\mG)$ leads to the following formula:

\begin{equation}\label{eqerg}
P(\mG|\vec{\lambda})=e^{-1-\lambda_0-\sum_{m=1}^M\lambda_mC_m(\mG)}
\end{equation}
that describes an exponential distribution (whence the name of the formalism) 
over the set $\ensG$ of all possible network configurations. Notice that the coefficients 
$\{\lambda_m\}_{m=0}^M$ are nothing else than the Lagrange multipliers in eq. 
\eqref{eqp}, whose values are fixed by the set of equations

\begin{equation}\label{eqlagrange}
\sum_{\mG\in\ensG}P(\mG|\vec{\lambda})C_m(\mG)=\avg{C_m}
\end{equation}
$\forall m=0\dots M$, where $C_0({\bf G})=1$ sets the normalization condition

\begin{equation}\label{eqergnorm}
e^{1+\lambda_0}\equiv Z(\vec{\lambda})=\sum_{\mG\in\ensG} e^{-\sum_{m=1}^M\lambda_mC_m(\mG)}.
\end{equation}
Using the above relation, we can eliminate $\lambda_0$ and obtain the 
standard expression of the ERG probability distribution that is analogous to 
that of the canonical ensemble in statistical physics:

\begin{equation}\label{eqerg2}
P(\mG|\vec{\lambda})=\frac{e^{-\sum_{m=1}^M\lambda_mC_m(\mG)}}{Z(\vec{\lambda})}.
\end{equation}

The quantity $H(\mG|\vec{\lambda})=\sum_{m=1}^M\lambda_mC_m(\mG)$ at the 
exponent is called \emph{graph Hamiltonian} and 
$Z(\vec{\lambda})=\sum_{\mG\in\ensG} 
e^{-\sum_{m=1}^M\lambda_mC_m(\mG)}$ is the \emph{partition function}, 
which properly normalizes the probability distribution. It can be easily shown 
that eq. \eqref{eqerg} not only makes the first functional derivative of 
$\mathscr{L}[P]$ vanish, but also makes the second derivative $\frac{\delta^2 
\mathscr{L}[P]}{\delta P(\mG)\delta P(\mathbf{G'})}$ negative definite, 
so that $P(\mathbf G)$ is indeed a maximum of $\mathscr{L}[P]$ (see Appendix C).

The ERG formalism can be fruitfully used to analyze real-world networks by 
supposing that a given observed configuration $\hmG$ has been 
drawn from $\ensG$ and, as such, can be consistently assigned the 
probability coefficient $P(\hmG|\vec{\lambda})$. However, we still 
need a recipe to set the mean values $\{\avg{C_m}\}_{m=1}^M$ or, 
equivalently, the Lagrange multipliers $\{\lambda_m\}_{m=1}^M$, in an optimal way. To 
this aim, we can invoke the {\em maximum-likelihood principle}, prescribing to maximize 
$P(\hmG|\vec{\lambda})$ as a function of $\vec{\lambda}$, 
or equivalently the {\em log-likelihood function}

\begin{equation}
\mathcal{L}(\hmG|\vec{\lambda})=\ln P(\hmG|\vec{\lambda})=-\sum_{m=1}^M\lambda_mC_m(\hmG)-\ln Z(\vec{\lambda}).
\end{equation}

Using eqs. \eqref{eqlagrange} and \eqref{eqerg2}, it is possible to show that the 
set of values $\{\hat{\lambda}_m\}_{m=1}^{M}$ satisfying the set of equations 
$\frac{\partial \mathcal{L}(\hmG|\vec{\lambda})}{\partial 
\lambda_m}=0,\:m=1\dots M$ are those ensuring that

\begin{equation}\label{eqmaxl}
\avg{C_m}=\hat{C}_m
\end{equation}
$\forall m=1\dots M$, namely that the ensemble averages of constraints match their values observed in $\hmG$. 
\cite{squartini2011analytical}. Remarkably, likelihood maximization consistently 
prescribes that the only information usable to make inference is the one we have 
access to. Moreover, it is simple to show that the same choice makes the Hessian 
of $\mathcal{L}(\hmG|\vec{\lambda})$ negative definite, implying that 
the position in eq. \eqref{eqmaxl} corresponds to a maximum of the likelihood.

The whole ERG recipe can be thus summarized as joining two optimization 
principles: 

\begin{itemize}
\item entropy maximization, which guarantees that the derived probability 
distribution encodes information only from the chosen constraints;
\item likelihood maximization, which guarantees that the value of the imposed 
constraints matches the observed one, without any statistical bias.
\end{itemize}

In what follows, we shall dwell into the details of various network reconstruction 
methods. Note however that, generally speaking, the ERG formalism can be also fruitfully used for a 
second purpose, \ie, analyzing known real-world networks in order to detect the 
level of randomness affecting their topological structure. 
Indeed, suppose to have an empirical network $\hmG$ whose structure is completely 
known. We can then ask whether this network is a ``typical'' configuration of 
a particular ERG derived from imposing an arbitrary set of network observables 
$\{C_m(\hmG)\}_{m=1}^M$. In other words, we can check if the 
network $\hmG$ is ``maximally random''---the level of randomness being determined 
by the chosen constraints, so that 
the information brought by any other network observable can be reduced to the 
information encoded in the constraints. We shall briefly discuss this approach in subsection \ref{wmb}.
\footnote{Entropy-maximization can be also used to 
assign probabilities to (dynamical) pathways. In such a context, the principle 
is known as maximum-caliber \cite{presse2013principles}. An application of this principle 
concerns the so-called \emph{origin-destination} networks, \ie, graphs defined by 
an origin node, a destination node and a set of connections linking them, 
representing the pathways along which information flows \cite{squartini2015information}.}

\section{Reconstruction methods}\label{secrec}

The reconstruction methods reviewed in the present work can be classified 
according to the link density of the output configurations. 
Indeed, the reconstructed networks can be fully connected (or, at least, very dense), 
with a tunable density, or exactly reproducing the observed number of links.

\subsection{Dense reconstruction methods}\label{dense}

\begin{figure*}
\centerline{
\includegraphics[width=0.8\textwidth]{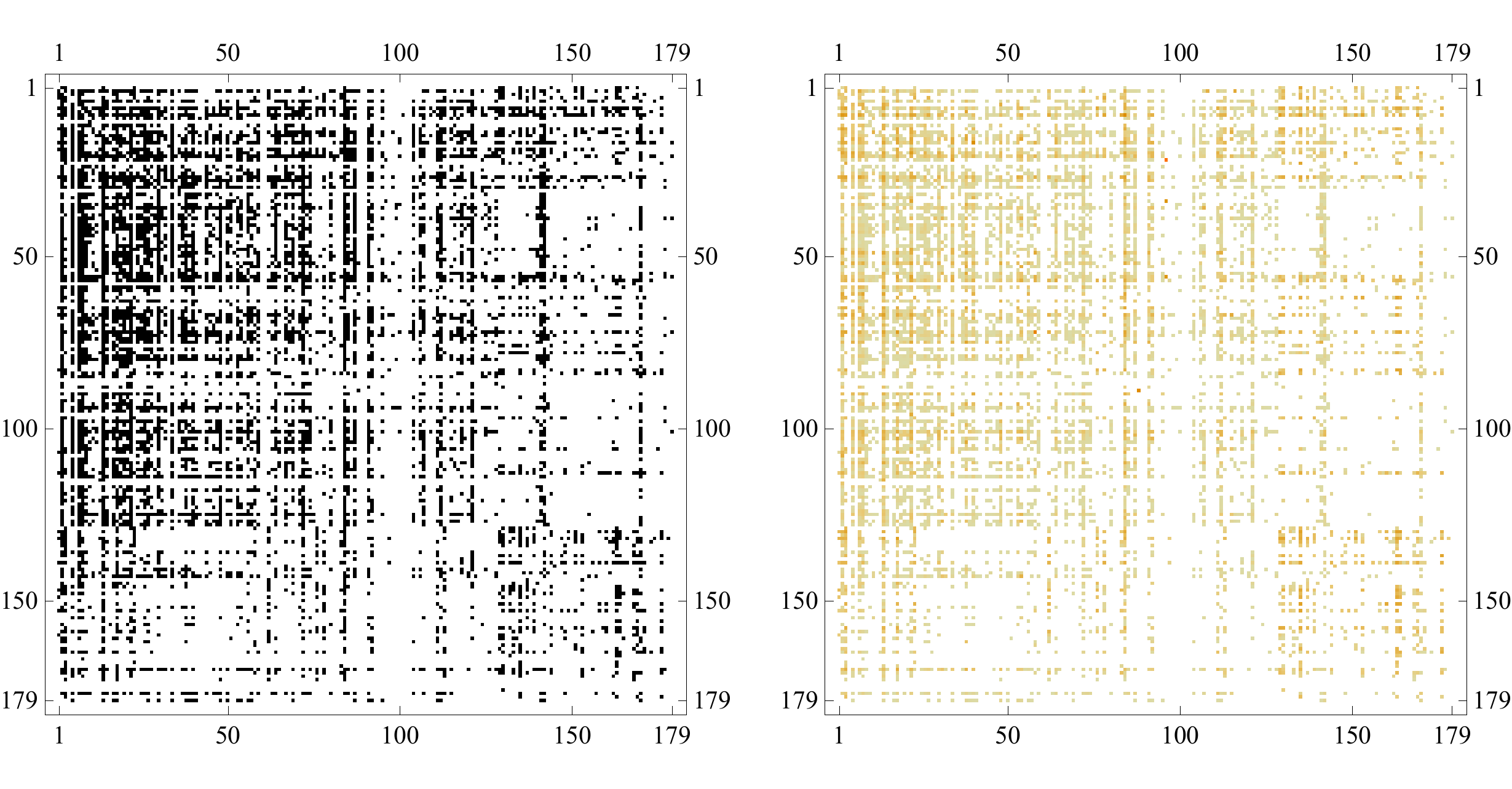}
}
\centerline{
\includegraphics[width=0.8\textwidth]{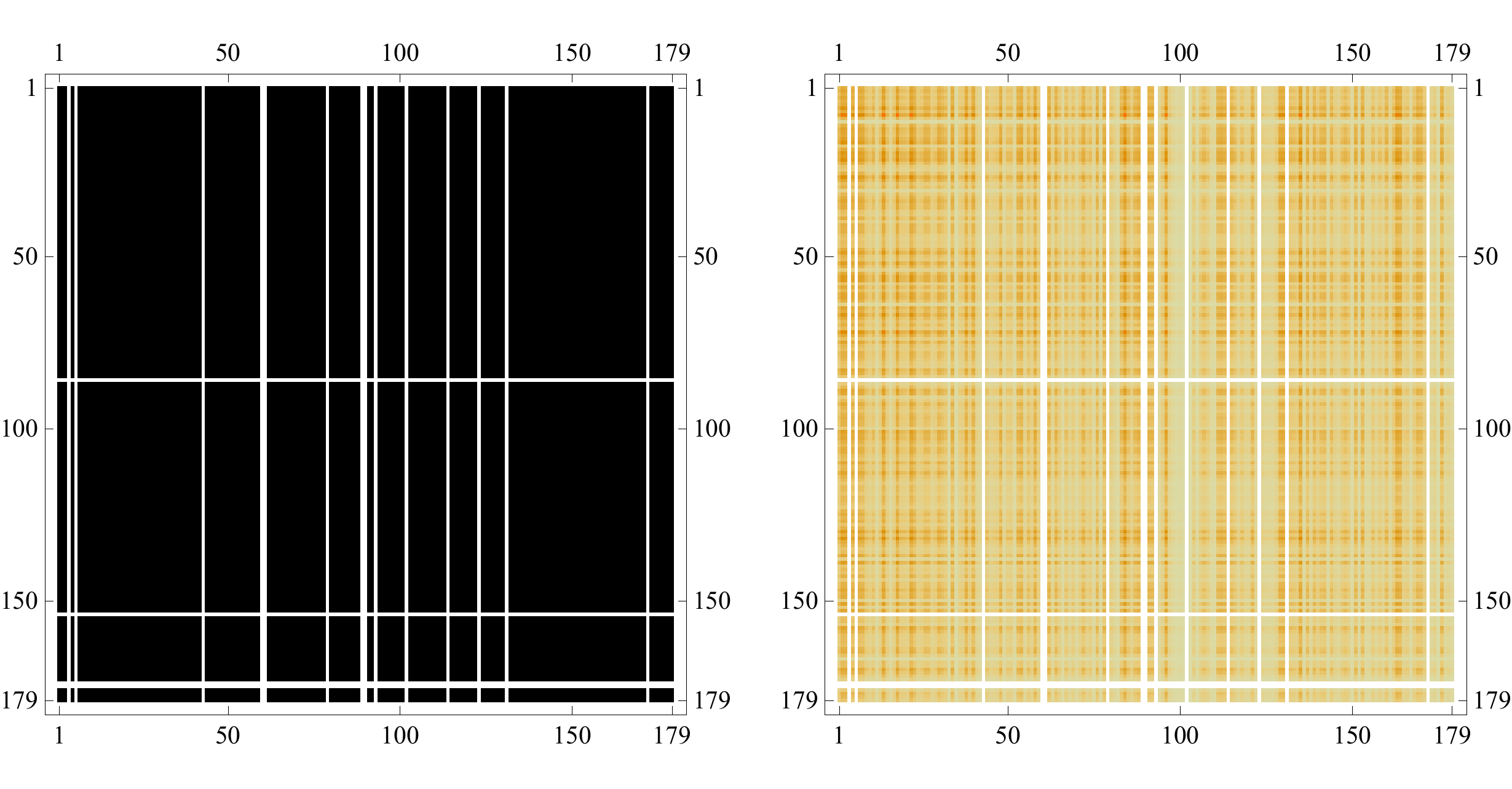}
}
\caption{Comparison between the observed adjacency matrix of the eMID network in 
2003 (top panels) and its reconstructed version according to the MaxEnt method 
described in section \ref{maxent} (bottom panels). 
Left panels represent binary adjacency matrices with black/white denoting the presence/absence of connections, 
whereas, right panels represent weighted adjacency matrices with color intensity denoting the weight of connections}
\label{fig1}
\end{figure*}

\subsubsection{The MaxEnt algorithm}\label{maxent}

The MaxEnt algorithm \cite{wells2004financial,upper2011simulation} represents 
the simplest and, probably, the best known method for reconstructing networks. 
It prescribes to maximize the functional

\begin{equation}\label{MC}
S=-\sum_{i=1}^N\sum_{j=1}^Nw_{ij}\ln w_{ij}
\end{equation}
under the constraints represented by eqs. (\ref{margins}). Equation (\ref{MC}) 
defines a particular kind of entropy in which the random variables are the 
matrix entries themselves. However, the absence of a proper normalization 
condition prevents eq. (\ref{MC}) from returning a genuine probability 
distribution. The solution to the aforementioned constrained maximization 
problem is, in fact,

\begin{equation}\label{maxe}
w_{ij}^{\text{\tiny ME}}=\frac{\hso_i\hsi_j}{\hW}\qquad\forall i,j,
\end{equation}
where $\hW=\sum_{i=1}^N \hso_i\equiv\sum_{j=1}^N \hsi_j$ is the total weight of the 
observed network $\hmG$. It is easy to verify that the constraints 
are satisfied, since $\hso_i=\sum_{j=1}^N w_{ij}^{\text{\tiny ME}}$ and $\hsi_i=\sum_{j=1}^N 
w_{ji}^{\text{\tiny ME}}$ $\forall i$. The summation index, however, has to run over all 
values $j=1\dots N$, including the ones corresponding to the diagonal entries. 
Note that eq. \eqref{maxe} implies that, unless either $\hso_i=0$ or $\hsi_i=0$ 
for some nodes, no entries can be zero and the resulting matrix is fully 
connected (see fig. \ref{fig1} where this algorithm has been applied to a 
snapshot of the Italian electronic market for interbank deposits eMID \cite{iori2008network}).
This feature represents the main limitation of the method, for a twofold reason. 
The first one is that real-world networks are often very sparse, and thus MaxEnt cannot possibly reproduce their topology.
The second one is that systemic risk is underestimated in dense networks \cite{mistrulli2011assessing,mastromatteo2012reconstruction}. 
Yet, the MaxEnt prescription provides quite accurate estimates whenever only the magnitude of weights are considered 
\cite{squartini2017network,mazzarisi2017methods}. The latter is the reason why MaxEnt is 
widely used in economics---the simple {\em gravity model} (without distance) has the same 
functional form of the MaxEnt estimate \cite{squartini2013jan}, and in finance---where 
it takes the same form of the {\em capital asset pricing model} (CAPM) 
\cite{sharpe1964capm,digiangi2016assessing}. As a final remark, we stress that the MaxEnt 
algorithm generates a unique reconstructed configuration, thus being 
classifiable as a deterministic algorithm.

\subsubsection{Overcoming the problem of missing connections: the IPF algorithm}\label{rasalgo}

A first step in the description of networks which are not fully connected (\ie, 
characterized by some null entries of the corresponding adjacency matrix) is given by the \emph{iterative 
proportional fitting} (IPF) procedure. It is a simple recipe to obtain a matrix 
$\mW$ that

\begin{itemize}
\item lies at the ``minimum distance" from the MaxEnt matrix 
$\mW^{\text{\tiny ME}}$ defined by eq. (\ref{maxe});
\item satisfies eqs. (\ref{margins});
\item admits the presence of a set of zero entries---in the typical case, the diagonal ones.
\end{itemize}

Let us call ${\bf W}^{\text{\tiny IPF}}$ the matrix satisfying these conditions. 
In the case of null diagonal entries, the method formally defines $\mW^{\text{\tiny IPF}}$ as:

\begin{equation}\label{IPF}
\min_{\mW}\left\{\sum_{i=1}^N\sum_{j(\neq 
i)=1}^Nw_{ij}\ln\left(\frac{w_{ij}}{w_{ij}^{\text{\tiny 
ME}}}\right)\right\}=\sum_{i=1}^N\sum_{j(\neq i)=1}^Nw_{ij}^{\text{\tiny 
IPF}}\ln\left(\frac{w_{ij}^{\text{\tiny IPF}}}{w_{ij}^{\text{\tiny ME}}}\right)
\end{equation}
\ie, the matrix that minimizes the Kullback-Leibler 
(KL) divergence \cite{kullback1951information} between a generic non-negative 
${\mW}$ with null diagonal entries and ${\mW}^{\text{\tiny 
ME}}$. The KL divergence is an asymmetric measure of ``distance'' between 
any two probability distributions and quantifies the amount of information lost 
when ${\mW}$ is approximated by ${\mW}^{\text{\tiny ME}}$.

A numerical recipe guaranteeing that the three requests above are met is 
provided by the iterative process whose basic steps at the $n$-th and $(n+1)$-th 
iterations are

\begin{equation}\label{ras}
w_{ij}^{(n+1)}=\hsi_j\left(\frac{w_{ij}^{(n)}}{\sum_{k(\neq j)} 
w_{kj}^{(n)}}\right),\qquad w_{ij}^{(n)}=\hso_i\left(\frac{w_{ij}^{(n-1)}}{\sum_{
k(\neq i)} w_{ik}^{(n-1)}}\right),
\end{equation}
so that $w_{ij}^{\text{\tiny IPF}}=\lim_{n\to\infty}w_{ij}^{(n)}$, 
and $w_{ij}^{(0)}$ represents the matrix used to initialize the algorithm. 
In a nutshell, the IPF algorithm iteratively distributes the known matrix marginals 
across the non-zero entries of the initial matrix. As long as this initial matrix 
is irreducible (meaning that it cannot be permuted into a block upper triangular matrix, 
or equivalently that the network is represents is strongly connected), eqs. (\ref{ras}) always yield 
a unique matrix that satisfies the marginals \cite{bacharach1965estimating}. 
As a first consistency check, let us consider the case in which the initial matrix is 
defined by $w_{ij}^{(0)}=w_{ij}^{\text{\tiny ME}}$ $\forall i,j$. 
Without restricting the sum to the non-diagonal terms, we would obtain 
$w_{ij}^{(1)}=w_{ij}^{(2)}=\dots=w_{ij}^{\text{\tiny ME}}$ $\forall i,j$. As a 
second check, let us consider the case in which the initial matrix is taken 
to be $w_{ij}^{(0)}=1$ $\forall i,j$, a position that is equivalent to 
immediately maximizing the functional in eq. (\ref{maxe}). We obtain 
$w_{ij}^{(1)}=\frac{\hso_i}{N}$ and $w_{ij}^{(2)}=\frac{\hso_i\hsi_j}{\hW}$ $\forall i,j$, 
hence the MaxEnt estimation is correctly recovered after just two iterations.

\subsubsection{The Directed Weighted Configuration Model}\label{secdwcm}

Like MaxEnt, the IPF has the major drawback of generating a single deterministic configuration, 
so that it is difficult to statistically evaluate the accuracy of 
the provided reconstruction. A more rigorous statistical method to 
evaluate the probability that nodes $i$ and $j$ are connected by a link is 
the ERG-based approach known as Directed Weighted Configuration 
Model (DWCM) \cite{squartini2011analytical}. The method constrains 
the out-strength $\so_i$ and in-strength $\si_i$ (defined as in eqs. (\ref{margins})) of each node $i$ 
of the network, and the Hamiltonian takes the form

\begin{equation}\label{H-DWCM}
H(\mW|\vec{\gamma},\vec{\delta})=\sum_{i=1}
^N\left(\gamma_i\so_i+\delta_i\si_i\right).
\end{equation}
Substituting the definitions of $\so_i$ and $\si_i$ in eq. 
(\ref{H-DWCM}) leads to a probability distribution 
$P(\mW|\vec{\gamma},\vec{\delta})$ which factorizes into the product of $N(N-1)$ 
pair-specific distributions

\begin{equation}
P(\mW|\vec{\gamma},\vec{\delta})=\prod_{i=1}^N\prod_{j(\neq 
i)=1}^Nq_{ij}^{\text{\tiny DWCM}}(w).
\label{DWCM-P}
\end{equation}

In the simple case of weights $w_{ij}$ taking only non-negative integer values, 
the probability distribution governing the behavior of the random variable $w_{ij}$ 
is geometric \cite{squartini2011analytical}:

\begin{equation}\label{DWCM-w}
q_{ij}^{\text{\tiny DWCM}}(w)=(\yo_i\yi_j)^w(1-\yo_i\yi_j)\mbox{ for }w\in\mathbb{Z}_+
\end{equation}
where $\yo_i=e^{-\gamma_i}$ and $\yi_i=e^{-\delta_i}$. From eq. (\ref{DWCM-w}), we immediately find that 

\begin{equation}\label{geo}
\avg{w_{ij}}^{\text{\tiny DWCM}}=\frac{\yo_i\yi_j}{1-\yo_i\yi_j}
\end{equation}
and, by definition, the probability $p_{ij}$ that a directed link from node $i$ 
to $j$ is present is $p_{ij}\equiv\sum_{w=1}^\infty q_{ij}(w)$, which in view of 
eq. (\ref{DWCM-w}) becomes:

\begin{equation}\label{DWCM-pij}
p_{ij}^{\text{\tiny DWCM}}=\yo_i\yi_j.
\end{equation}
Finally, the Lagrange multipliers are found by solving the corresponding $2N$ 
equations deriving from the likelihood-maximization principle: $\forall i$, 
$\hso_i=\avg{\so_i}\equiv\sum_{j(\neq i)}\avg{w_{ij}}^{\text{\tiny DWCM}}$ 
and $\hsi_i=\avg{\si_i}\equiv\sum_{j(\neq i)}\avg{w_{ji}}^{\text{\tiny DWCM}}$.

The DWCM falls into the category of dense reconstruction methods because the 
observed marginals are usually so large that the induced link probability 
between any two nodes $i$ and $j$ becomes very close to 1. However, differently 
from the MaxEnt and the IPF, the DWCM algorithm produces a whole ensemble of 
networks, by treating link as independent variables and drawing the 
corresponding weights from the geometric distributions described by eq. 
(\ref{geo}).

\subsubsection{Combining MaxEnt and ERG frameworks}\label{digiangi}

An approach combining the MaxEnt and the ERG frameworks has been recently developed, 
under the name of Maximum Entropy CAPM (MECAPM) \cite{digiangi2016assessing}. 
The idea is to maximize the Shannon entropy constraining 
not the expected values of the matrix marginals, but rather 
the expected value of each link weight. Similarly to the case of the DWCM, this leads to 
\begin{equation}\label{geo2}
q_{ij}^{\text{\tiny MECAPM}}(w)=(y_{ij})^{w}(1-y_{ij})
\end{equation}
where $y_{ij}$ is the Lagrange multiplier controlling for the weight of the link from $i$ to $j$. 
This framework can be used for network reconstruction provided that the imposed expected weights 
depend only on the matrix marginals, which is the only information available on the system. 
This is naturally achieved by the MaxEnt recipe, hence:
\begin{equation}
\avg{w_{ij}}^{\text{\tiny MECAPM}}=\frac{y_{ij}}{1-y_{ij}}\equiv w_{ij}^{\text{\tiny ME}},
\end{equation}
a position allowing for the Lagrange multipliers to be readily estimated as the link probabilities:
\begin{equation}\label{p_mecapm}
p_{ij}^{\text{\tiny MECAPM}}=y_{ij}=\frac{w_{ij}^{\text{\tiny ME}}}{1+w_{ij}^{\text{\tiny ME}}}.
\end{equation}

This algorithm falls into the category of dense reconstruction methods, 
since the MaxEnt weights are usually sufficiently large to induce $p_{ij}\simeq 1$ $\forall i,j$. 
And as the DWCM, the MECAPM algorithm produces a whole ensemble of networks.

\subsection{Density-tunable reconstruction methods}\label{density-tune}

The MaxEnt, the DWCM and the MECAPM methods suffer from the same limitation: the 
predicted configurations are often too dense to faithfully describe real-world 
networks. Therefore other reconstruction methods have been proposed. The 
rationale driving the algorithms described below is to produce configurations 
that are sparser than the ones obtained through the aforementioned algorithms.

\subsubsection{The IPF algorithm: generic formulation}

As we have seen, the IPF algorithm basically acts by distributing the known marginals 
across the positive entries of the matrix. Hence, it requires that the position of the null entries 
is known in advance. This limitation is the reason why the method is 
often used in combination with other algorithms that estimate the positions of the zeros. 
Once these positions are known, the IPF algorithm adjusts the 
positive entries (typically initialized as MaxEnt estimates) to match the constraints.

Indeed, the freedom to choose the topological details turns out to be 
fundamental whenever an algorithm able to {\em generate} realistic 
configurations is needed. The general formulation of the IPF 
algorithm give us this freedom. Indeed, in order to account for either known or guessed subsets of entries 
(which do not necessarily need to be zero), it is enough to 
i) subtract them from the known marginals $\{\hso_i\}_{i=1}^N$ and $\{\hsi_i\}_{i=1}^N$, and 
ii) modify eqs. (\ref{IPF}) and (\ref{ras}) by using these rescaled marginals and 
explicitly excluding known entries from the sums at the denominator \cite{bacharach1965estimating}. 
In the most general case, the IPF estimation can be written as the infinite product

\begin{equation}
w_{ij}^{\text{\tiny IPF}}=\prod_{n=0}^{\infty}\frac{\hso_i}{[\so_i]^{(2n)}}w_{ij}^{(0)}\frac{\hsi_j}{[\si_j]^{(2n+1)}}
\end{equation}
where $[\so_i]^{(2n)}=\sum_{j(\neq i)} w_{ij}^{(2n)}$ and $[\si_i]^{(2n+1)}=\sum_{j(\neq i)} w_{ji}^{(2n+1)}$.\footnote{The IPF 
algorithm is also known with the name of RAS algorithm, because the form of the solution devised in \cite{bishop2007discrete} 
is written as the product of three matrices whose symbols are $\mathbf{R}$, $\mA$, $\mathbf{S}$. 
More specifically, $\mA^{(2n+1)}=\mathbf{R}^{(n+1)}\mA^{(2n)}$ and 
$\mA^{(2n+2)}=\mA^{(2n+1)}\mathbf{S}^{(n+1)}$, with 
$\mA^{(0)}$ being the initial adjacency matrix and 
$\mathbf{R}$, $\mathbf{S}$ being the two diagonal matrices 
$\mathbf{R}$, $\mathbf{S}$ being the two diagonal matrices
$\mathbf{R}^{(n+1)}=\mbox{diag}\left(\hso_i/[\so_i]^{(2n)}\right)$ and 
$\mathbf{S}^{(n+1)}=\mbox{diag}\left(\hsi_i/[\si_i]^{(2n+1)}\right)$.} 

\begin{figure*}
\centerline{
\includegraphics[width=0.8\textwidth]{fig1.pdf}
}
\centerline{
\includegraphics[width=0.8\textwidth]{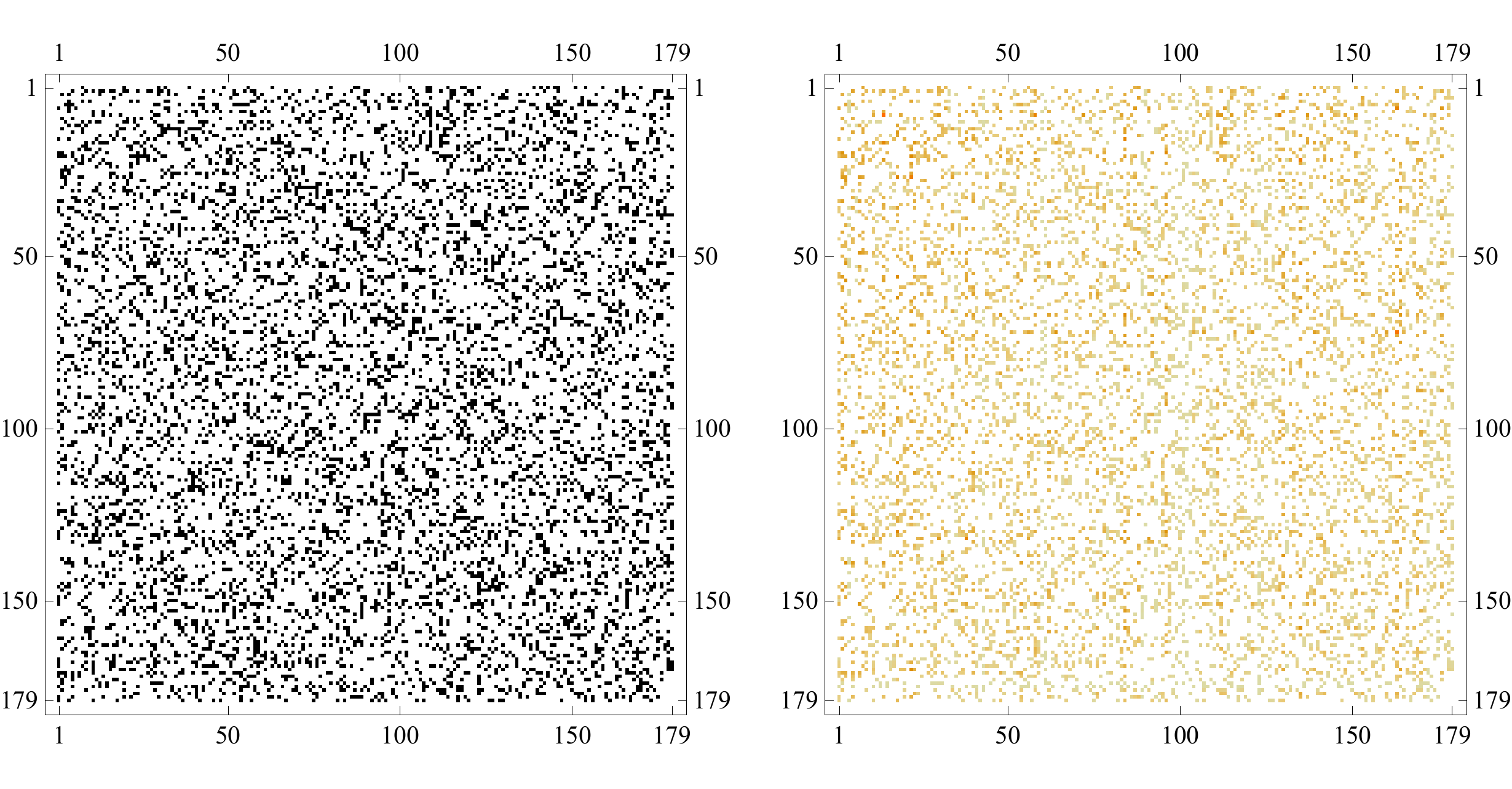}
}
\caption{Comparison between the observed adjacency matrix of the eMID network in 
2003 (top panels) and its reconstructed version according to the method by Drehmann \& Tarashev
described in section \ref{drehmann} (bottom panels).
Left panels represent binary adjacency matrices with black/white denoting the presence/absence of connections, 
whereas, right panels represent weighted adjacency matrices with color intensity denoting the weight of connections.}
\label{fig2}
\end{figure*}

\subsubsection{The Drehmann \& Tarashev approach}\label{drehmann}

In the absence of a clear recipe to estimate the network density, a number 
of algorithms exploring the whole range $[0,1]$ of possible density values 
have been proposed. Drehmann \& Tarashev devise a simple approach \cite{drehmann2013measuring} 
to perturb the MaxEnt matrix and obtain sparse network, following three steps: 

\begin{itemize}
\item  choosing a random set of off-diagonal entries to be zero, 
thus manually setting a desired link density; 
\item  treating the remaining non-zero entries as random variables, uniformly 
distributed between zero and twice their MaxEnt estimated value:
\begin{equation}
w_{ij}^{\text{\tiny D-T}}\sim U(0,2w_{ij}^{\text{\tiny ME}})
\end{equation}
(so that the expected value of weights under this distribution coincides 
with the MaxEnt estimates $w_{ij}^{\text{\tiny ME}}=\hso_i\hsi_j/\hW$);
\item running the IPF algorithm to correctly restore the value of the 
marginals.
\end{itemize}

The value of the network density can be tuned to generate arbitrarily sparse 
configurations. Yet a drawback of the method lies in the completely random 
nature of the obtained topological structure(s) (see fig. \ref{fig2}).

\subsubsection{The Mastromatteo, Zarinelli \& Marsili approach}\label{mastro}

Another approach to generate reconstructed networks with an arbitrary 
density of links has been formulated in \cite{mastromatteo2012reconstruction}. 
This method consists in sampling uniformly the set of network configurations that are compatible 
with the constraints defined by eqs. (\ref{margins}) and have a given value of the density. 
Two additional assumptions are made on the matrix $\hmW$: 
i) the entries larger than a certain threshold $\theta$ (supposed to be at most 
of order $N$) are considered to be known---an hypothesis that in the case of financial 
networks meets the recent disclosure policies adopted for some markets \cite{ESMA}); 
ii) the unknown entries are rescaled by the threshold $\theta$ itself, 
and thus become bounded in the range $[0,1]$. 
Known entries can be completely omitted in the description of the method, 
and the focus can be kept only on the ensemble generated by the variability 
of the unknown entries.

The fundamental question the authors tackle is the following: given an 
arbitrary value of $\rho$, how many matrices exist that satisfy eqs. 
(\ref{margins}) and whose density matches the chosen one? Clearly, while the 
maximum density value allowing for (at least) one configuration to exist is 
$\rho_{max}=1$ (the MaxEnt algorithm always satisfies  eqs. (\ref{margins})), 
finding $\rho_{min}$ for a given value of the constraints is non-trivial. 
The measure introduced in \cite{mastromatteo2012reconstruction} 
to fairly sample the space of binary adjacency matrices 
that are {\em compatible} with a given $\rho$ is $P_0({\bf 
A}|z)=\frac{1}{Z}z^{L(\mA)}$, where $Z$ is a normalization constant, 
$L(\mA)$ is the number of links in $\mA$ and 
the parameter $z$ sets the average density

\begin{equation}
\avg{\rho}=\sum_{\mA}P_0({\bf A}|z)\rho(\mA)
\end{equation}
to a desired value (notice that the sum runs over the compatible configurations only). 

However, when $\avg{\rho}<1$ it is hard to analytically 
evaluate the sampling probability distribution $P_0({\bf A}|z)$.
Consequently, an approximated probability distribution is introduced:

\begin{equation}\label{q}
P(\mA|z)=\frac{e^{-\beta H(\mA)}z^{L(\mA)}}{Z(\beta, z)}
\end{equation}
where $H(\mA)=\sum_i(\Theta[\so_i-\ko_i]+\Theta[\si_i-\ki_i])$. 
The parameter $z$ plays the role of the fugacity in statistical 
physics and fixes the mean value of the density over the {\em whole} set of 
adjacency matrices. Since the unknown entries of $\bf{W}$ range between 0 and 1, 
the number of out-going (in-coming) links of each node in any compatible 
configuration with the chosen density value is always larger than its 
out-strength (in-strength): this brings to the condition $H(\mA)=0$. 
Conversely, the probability of any infeasible configuration (\ie, with 
$H(\mA)=1$) to appear is suppressed by a coefficient $e^{-\beta}$ which 
vanishes as $\beta\rightarrow\infty$. Since an analytical treatment of the 
distribution in eq. (\ref{q}) is also infeasible, the authors implement a {\em 
message-passing algorithm} \cite{mastromatteo2012reconstruction} to calculate 
the marginal link probabilities, which are then independently sampled to build 
a candidate binary network. And once the binary topology is obtained, 
weights are inferred via the IPF algorithm.

As the authors explicitly notice \cite{mastromatteo2012reconstruction}, being 
able to tune the network density implies being able to consider a whole range of 
structures characterized by (potentially) different degrees of robustness to, 
\eg, financial shocks. However, as a result of the authors' ``compatibility 
analysis'', no allowable configurations can be found below a certain value of 
the network density. And while the MaxEnt method produces configurations believed 
to maximize the network robustness, sparser configurations may, on the other hand, 
provide lower bounds to it.

\subsubsection{The Moussa \& Cont approach}\label{mussa_cont}

Another algorithm that combines the MaxEnt and the ERG frameworks is the one 
presented in \cite{moussa2011contagion}, where the authors propose two different 
versions of their method---both similar to the entropy-based approaches described 
in the previous subsections. The major difference lies in the amount of 
available information required, which is now substantially larger: namely, 
the out- and in-degree distributions of the network, as well as the out- and in-strength distributions, 
all of them supposedly well-described by power-laws (whose 
parameters are tuned to reproduce stylized facts of financial networks). 
In a nutshell, the method generates a whole ensemble of $R$ different 
binary network configurations (\ie, the \emph{prior} configurations), 
whose topological structure is then ``adjusted'' \emph{a posteriori} 
to match the constraints represented by eqs. (\ref{margins}).

\paragraph*{The ``exact'' approach} The first version of the method is designed 
to meet the constraints \emph{exactly} in each configuration of the 
ensemble. Prior configurations are characterized by the same degree 
distributions but different topological structures (the algorithm used to 
generate the ensemble is the generalization of the preferential 
attachment algorithm to directed networks \cite{bollobas2003directed}). Once a 
binary configuration is generated, the IPF algorithm is used to 
assign weights to the realized links. 
The problem can thus be formally stated as that of determining, 
for each binary configuration in the ensemble, the set of weights 
$\{w^{\text{\tiny M-C}}_{ij}\}$ such that

\begin{equation}
\min_{{\mW}}\left\{\sum_{\{a_{ij}=1\}}w_{ij}\ln\left(\frac{w_{ij}}{w_{ij}
^{\text{\tiny ME}}}\right)\right\}=\sum_{\{a_{ij}=1\}}w^{\text{\tiny 
M-C}}_{ij}\ln\left(\frac{w^{\text{\tiny M-C}}_{ij}}{w_{ij}^{\text{\tiny 
ME}}}\right),
\end{equation}
where the sum runs over the set of non-zero entries of any configuration in the 
ensemble (for all null entries we trivially have $w_{ij}^{\text{\tiny M-C}}=0$). 
Using a normalization condition that the entries to be estimated satisfy $\sum_{\{a_{ij}=1\}} 
w_{ij}=1$, the solution to the optimization problem above can be formally written as

\begin{equation}\label{mc1}
w_{ij}^{\text{\tiny M-C}}=\hso_i\hsi_j\left(\frac{e^{\gamma_i+\delta_j}}{\sum_{\{a_{kl}=1\}}\hso_k\hsi_ke^{\gamma_k+\delta_l}}\right)\,,
\end{equation}
where $\{\gamma_i\}_{i=1}^N$ and $\{\delta_i\}_{i=1}^N$ are, respectively, the 
Lagrange multipliers related to the constraints on nodes out- and in-strengths, found by solving 
$\avg{\so_i}=\hso_i$ and $\avg{\si_i}=\hsi_i$ $\forall i$.

\paragraph*{The ``average'' approach} However, as the authors explicitly notice, 
dealing with too sparse matrices may prevent the IPF algorithm to converge, 
because a solution of the IPF algorithm exists and is unique 
if and only if the matrix is irreducible (\ie, the network is strongly connected) 
\cite{bacharach1965estimating}. Moreover, there is no guarantee 
that the numerical values of weights assigned by IPF are distributed according to 
some empirical observations (\eg, following a power law) \cite{moussa2011contagion}.

For this reason, a second version of the algorithm is proposed. The ensemble of 
configurations is now composed by weighted networks for which both degrees and 
weights distributions are specified. The probability distribution on this set of 
configurations, however, is not uniform: each configuration $r=1,\dots,R$ is, in fact, 
characterized by a statistical weight $\nu_r\in(0,1)$ such that the constraints 
represented by eqs. (\ref{margins}) are satisfied on average. The ensemble 
probability distribution $\{\nu_r\}_{r=1}^R$ is derived by minimizing its KL 
divergence from the uniform distribution $1/R$

\begin{equation}\label{me3}
\sum_{r=1}^R\nu_r\ln\left(\frac{\nu_r}{1/R}\right)
\end{equation}
under the constraints $\hso_i=\sum_r\nu_r[\so_i]^{(r)}$ and 
$\hsi_i=\sum_r\nu_r[\si_i]^{(r)}$ $\forall i$ (where 
$\{[\so_i]^{(r)}\}_{i=1}^N$ and $\{[\si_i]^{(r)}\}_{i=1}^N$ are the out- and 
in-strengths for the $r^{th}$ configuration in the ensemble). Similar calculations to 
those used for eq. (\ref{mc1}) lead to the analytical expression of the 
probability coefficients

\begin{equation}
\nu_r=\frac{e^{\sum_{i=1}^N\gamma_i[\so_i]^{(r)}+\sum_{j=1}^N\delta_j[\si_j]^{(r)}}}{Z(\vec{\gamma},\vec{\delta})}
\end{equation}
with $Z(\vec{\gamma},\vec{\delta})=\sum_r e^{\sum_{i=1}^N\gamma_i[\so_i]^{(r)}+\sum_{j=1}^N\delta_j[\si_j]^{(r)}}$. 
As usual, the numerical value of the Lagrange multipliers can be found by solving the 
constraints equations $\avg{\so_i}=\hso_i$ and $\avg{\si_i}=\hsi_i$ $\forall i$. 
The estimation of any quantity of interest is 
then carried out by computing the ensemble average of the quantity itself.

\subsection{Exact density methods}\label{ex}

The reconstruction algorithms described in the previous sections are 
attempts to circumvent the lack of information on the actual network density 
and, most importantly, to avoid predicting too dense configurations. The methods 
described below, instead, explicitly require the knowledge of the observed 
network density or, at least, the link density for a subset of nodes. This is because, as 
shown in \cite{cimini2015estimating,mazzarisi2017methods,gandy2017adjustable}, 
adding this piece of information can dramatically increases the performance of a 
reconstruction algorithm.
We now introduce a series of algorithms of this kind, 
all strictly following the ERG formalism introduced in Section \ref{ERG-sect}.

\subsubsection{The density-corrected DWCM}\label{secdwcm2}

The first example of ERG-based reconstruction method we have met in subsection 
\ref{secdwcm} is the DWCM, obtained by constraining the out-strength and in-strength sequences. 
As we have seen, the DWCM-induced ensemble is basically characterized by fully-connected configurations, 
with a density of links which cannot be tuned independently from the distribution of weights 
(see eq. (\ref{DWCM-pij}). So the outcome of the DWCM is very close to that of the MaxEnt method, 
and indeed the DWCM can be seen a sort of stochastic generalization of MaxEnt. 
In order to overcome this limitation, the authors of \cite{mazzarisi2017methods} propose a {\em 
density-corrected} version of the DWCM, defined by constraining the total number 
of links $L$ beside the out-strength and in-strength sequences:

\begin{equation}
H(\mW|\vec{\gamma},\vec{\delta},\zeta)=\sum_{i=1}
^N\left(\gamma_i\so_i+\delta_i\si_i\right)+\zeta L.
\label{dwcm-c}
\end{equation}
In analogy with the DWCM, also in this case links turn out to be statistically 
independent. When weights can assume only positive integer values, the weight 
probability distribution can be written as

\begin{equation}\label{DWCM-m-w}
q_{ij}^{\mbox{\tiny dcDWCM}}(w)=p_{ij}^{\mbox{\tiny dcDWCM}}(\yo_i\yi_j)^{w-1}(1-\yo_i\yi_j)
\end{equation}
where
\begin{equation}\label{dwcm2-pij}
p_{ij}^{\mbox{\tiny dcDWCM}}=\frac{z\yo_i\yi_j}{1+z\yo_i\yi_j-\yo_i\yi_j}
\end{equation}
in which $\yo_i$ and $\yi_j$ are defined as in 
the DWCM and $z=e^{-\zeta}$ (thus, for $\zeta=0$ we recover the standard DWCM). Notice that 
eq. (\ref{DWCM-m-w}) defines a composition of a single Bernoulli trial, 
controlling for the existence of a link between any two nodes $i$ and $j$, and a 
geometric distribution for its weight, whose mean value reads

\begin{equation}
\avg{w_{ij}}^{\mbox{\tiny dcDWCM}}=\frac{p_{ij}^{\mbox{\tiny dcDWCM}}}{1-\yo_i\yi_j}.
\end{equation}
Finally the Lagrange multipliers are fixed as in the DWCM, while $\zeta$ is obtained by $\hL=\avg{L}\equiv \sum_{i\neq j}p_{ij}^{\mbox{\tiny dcDWCM}}$.

This method ideally refines the DWCM by explicitly adding to the recipe a piece of 
topological information. By doing so, the occurrence probability of a network in the ensemble 
still depends on the marginals, but also on the imposed number of links, 
hence very dense configurations become highly improbable. 

\subsubsection{The Directed Enhanced Configuration Model}\label{DECM-sec}

Beyond the total number of links, also the degree heterogeneity can be explicitly taken into account. 
The Directed Enhanced Configuration Model (DECM) \cite{mastrandrea2014enhanced} is defined by:

\begin{equation}
\label{H-DECM}
H(\mW|\vec{\alpha},\vec{\beta},\vec{\gamma},\vec{\delta})=\sum_{i=1}
^N\left(\alpha_i\ko_i+\beta_i\ki_i+\gamma_i\so_i+\delta_i\si_i\right)
\end{equation}
and encompasses many ERG-based models as special cases (for instance, 
the degree-corrected DWCM is obtained when $\alpha_i=\beta_i=\zeta/2$, $\forall i$). 
The DECM probability distribution can be written as:

\begin{equation}
P(\mW|\vec{\alpha},\vec{\beta},\vec{\gamma},\vec{\delta})=\prod_{i=1}^N\prod_{j(\neq i)=1}^Nq_{ij}^{\mbox{\tiny DECM}}(w)
\label{DECM-P}
\end{equation}
with

\begin{equation}\label{qq}
q_{ij}^{\mbox{\tiny DECM}}(w) = \left\{ \begin{array}{ll} 
1-p_{ij}^{\mbox{\tiny DECM}} & \mbox{if } w=0,\\
p_{ij}^{\mbox{\tiny DECM}}(\yo_i\yi_j)^{w-1}(1-\yo_i\yi_j) & \mbox{if } w>0
\end{array} \right.
\end{equation}
and

\begin{equation}\label{decm}
p_{ij}^{\mbox{\tiny DECM}}=\frac{\xxo_i\xxi_j\yo_i\yi_j}{1+\xxo_i\xxi_j\yo_i\yi_j-\yo_i\yi_j}
\end{equation}
(where $\xxo_i=e^{-\alpha_i}$, $\xxi_i=e^{-\beta_i}$, $\yo_i=e^{-\gamma_i}$ and $\yi_i=e^{-\delta_i}$). 
Notice that from eq. (\ref{qq}) it is simple to evaluate 
the average value of the generic link weight as 

\begin{equation}\label{qqq}
 \avg{w_{ij}}^{\mbox{\tiny DECM}}=\frac{p_{ij}^{\mbox{\tiny DECM}}}{1-\yo_i\yi_j}. 
\end{equation}
Lagrange multipliers are as usual found by solving the corresponding $4N$ 
equations derived from the likelihood-maximization principle: $\forall i$, 
$\hko_i=\avg{\ko_i}\equiv\sum_{j(\neq i)}p_{ij}^{\mbox{\tiny DECM}}$, 
$\hki_i=\avg{\ki_i}\equiv\sum_{j(\neq i)}p_{ji}^{\mbox{\tiny DECM}}$, 
$\hso_i=\avg{\so_i}\equiv\sum_{j(\neq i)}\avg{w_{ij}}^{\mbox{\tiny DECM}}$, 
$\hsi_i=\avg{\si_i}\equiv\sum_{j(\neq i)}\avg{w_{ji}}^{\mbox{\tiny DECM}}$.

As for the density-corrected DWCM, eq. (\ref{qq}) can be interpreted as a 
combination of a Bernoulli trial, with probability $p_{ij}^{\mbox{\tiny DECM}}$, and a drawing from 
a geometric distribution, with parameter $\yo_i\yi_j$; in this case, however, the link probability depends also 
on the degrees of nodes $i$ and $j$.

The DECM method has a simple interpretation when a specific economic system is 
considered, namely the World Trade Network (WTN). In economic terms, the two 
aforementioned processes respectively describe the tendency of a generic country 
$i$ either to establish a new export towards country $j$ (with probability 
$p_{ij}$) or to reinforce an existing one (with probability $\yo_i\yi_j$, 
by rising the exchanged amount of goods of, so to say, ``one unit'' of trade). 
In order to understand which process is more probable, 
we can study the behavior of the ratio $p_{ij}^{\mbox{\tiny DECM}}/(\yo_i\yi_j)$.
Whenever this quantity is greater than 1, country $i$ 
would probably establish a new export relation towards $j$, and 
at the same time experience a certain resistance to reinforce it; 
otherwise, country $i$ would experience a certain resistance to start exporting to $j$, 
but once such relation were established it would be characterized by a 
relatively low ``friction'', inducing the involved partners to strengthen it 
\cite{almog2015double}. 
Note that the case $p_{ij}^{\mbox{\tiny DECM}}/(\yo_i\yi_j)=1$ 
implies reducing eq. (\ref{qq}) to eq. (\ref{DWCM-w}) of the DWCM. 
In other words, the DWCM does not discriminate between the first link and the subsequent ones, 
reducing $q_{ij}(w)$ to a simple geometric distribution. As shown in 
\cite{mastrandrea2014enhanced}, the DWCM fails in reproducing the observed 
properties of the WTN precisely because it cannot give the right 
importance to the very first link, which is treated as a simple unit of weight. 
This observation hints at the importance of the information encoded into nodes degrees, 
to be considered as a fundamental ingredient (together with nodes strengths) 
for a faithful reconstruction of real-world networks.

\subsubsection{Simplifying the DECM: a two-step model}\label{DECMts-sec}

A simplified version of the DECM can be derived by noticing that the 
estimation of the topological structure of a network can be, in some circumstances, 
disentangled from the estimation of its weighted structure. This 
observation rests upon the evidence that the link probabilities of the 
DECM show a large (positive) correlation with the analogous probabilities of a 
much simpler ERG model, namely the Directed Binary Configuration Model (DBCM) 
obtained by constraining only the out- and in-degree sequences \cite{almog2015double}. 
The DBCM is thus defined by the Hamiltonian

\begin{equation}\label{DBGM-H}
H(\mA|\vec{\alpha},\vec{\beta})=\sum_{i=1}^N\left(\alpha_i\ko_i+\beta_i\ki_i\right),
\end{equation}
which leads to the connection probability
\begin{equation}
p_{ij}^{\text{\tiny DBCM}}=\frac{\xxo_i\xxi_j}{1+\xxo_i\xxi_j}
\end{equation}
with $\xxo_i=e^{-\alpha_i}$ and $\xxi_i=e^{-\beta_i}$, determined via the usual $2N$ equations 
$\hko_i=\avg{\ko_i}\equiv\sum_{j(\neq i)=1}^Np_{ij}^{\text{\tiny DBCM}}$ and 
$\hki_i=\avg{\ki_i}\equiv\sum_{j(\neq i)=1}^Np_{ji}^{\text{\tiny DBCM}}$, $\forall i$.

Putting this expression into eqs. (\ref{qq}) and (\ref{qqq}) leads to 

\begin{equation}
q_{ij}^{\mbox{\tiny 2s-DECM}}(w)=p_{ij}^{\text{\tiny DBCM}}(\yo_i\yi_j)^{w-1}(1-\yo_i\yi_j),\qquad
\avg{w_{ij}^{\mbox{\tiny 2s-DECM}}}=\frac{p_{ij}^{\mbox{\tiny DBCM}}}{1-\yo_i\yi_j},
\end{equation}
which defines a ``two-step'' version of the DECM, and whose Lagrange multipliers are found by imposing, $\forall i$, 
$\hko_i=\avg{\ko_i}^{\mbox{\tiny DBCM}}$ and $\hki_i=\avg{\ki_i}^{\mbox{\tiny DBCM}}$ first and 
then $\hso_i=\avg{\so_i}^{\mbox{\tiny 2s-DECM}}$ and $\hsi_i=\avg{\si_i}^{\mbox{\tiny 2s-DECM}}$.

\subsubsection{Fitness-induced Exponential Random Graphs}\label{FiERG}

Despite the previous findings, we note that it is impossible to use either the DECM or its two-step version when the degrees of nodes are not known, which is unfortunately a rather common situation. Nevertheless, these cases can be treated by resorting to the \emph{fitness} ansatz, which states that the link probability between any two nodes 
depends on non-topological features of the involved nodes, typical of the system under analysis. 
More precisely, it is assumed that the ``activity'' of each node $i$ in the network is summed up by an 
``intrinsic'' quantity, called \emph{fitness}\cite{caldarelli2002scale}, which is directly related 
to the Lagrange multiplier $x_i$ controlling the degree of that node through a monotone functional relation. 
Note that such relation between fitness values and degrees lies at the basis of the the so-called 
\emph{good-gets-richer} mechanism, according to which ``better'' nodes (those characterized by 
a higher fitness value) have more chances to ``attract'' connections \cite{caldarelli2002scale}.

For instance, in the case of the (undirected) WTN, where nodes represent countries 
and links represent trade relationships between them, a strong linear correlation can be observed 
between the Lagrange multipliers of nodes' degrees and the Gross Domestic Product (GDP) 
values of the respective countries: 
$x_i\simeq\sqrt{z}\:\textrm{GDP}_i$ $\forall i$ \cite{garlaschelli2004fitness,garlaschelli2008maximum}. 
Consequently, the link probability between nodes $i$ and $j$ can be rewritten as
\begin{equation}\label{gv1}
p_{ij}^{\text{\tiny UBCM}}\simeq\frac{z\,\text{GDP}_i\,\text{GDP}_j}{1+z\,\text{GDP}_i\,\text{GDP}_j},
\end{equation}
where UBCM stands for Undirected Binary Configuration Model. 
Similar fitness ansatzs have been successfully tested for financial networks, such as interbank markets 
\cite{demasi2006fitness,cimini2015systemic,cimini2015estimating} and shareholding networks \cite{garlaschelli2005scale,squartini2017stock},

The validity of the fitness ansatz has profound implications on the kind of information that is necessary 
to have in order to accurately reconstruct a network, but in general leads to face the problem of finding 
node observables that are correlated with degrees. Remarkably, strengths often work well as fitnesses \cite{barrat2004architecture},
a ``stylized fact" implying that the DBCM Lagrange multipliers can be expressed as 
$\xxo_i=f(\hso_i)$ and $\xxi_i=f(\hsi_i)$. 
As the many empirical analyses of economic and financial systems mentioned above have pointed out, 
the functional form $\xxo_i=\sqrt{z}(\hso_i)^b$ 
and $\xxi_i=\sqrt{z}(\hsi_i)^b$ with exponent $b=1$ 
is often accurate enough for all practical purposes.

\paragraph*{Estimating the degrees} The above assumption of linear proportionality 
allows to estimate the unknown degrees in a straightforward way. Indeed, 
connection probabilities or the fitness-induced DBCM assume the form
\begin{equation}\label{pLp}
p_{ij}^{\text{\tiny f-DBCM}}=\frac{z\hso_i\hsi_j}{1+z\hso_i\hsi_j},
\end{equation}
so that the only variable left is the proportionality constant $z$.\footnote{Note that 
the MECAPM connection probabilities defined in eq. (\ref{p_mecapm}) are recovered here by the particular choice $z=\hW^{-1}$}. 
The latter can be simply estimated provided that 
the total number of links $\hL$ of the empirical network is known. 
Imposing $\hL=\avg{ L}$ in fact means solving only one equation 
\begin{equation}\label{L}
\hL=\sum_{i=1}^N\sum_{j(\neq i)=1}^N\frac{z\hso_i\hsi_j}{1+z\hso_i \hsi_j}
\end{equation}
which has a single solution $z>0$ \cite{cimini2015systemic,cimini2015estimating}.
Once $z$ is found, the degrees of any node $i$ in the network can be estimated as 
\begin{equation}\label{estdeg}
\avg{\ko_i}^{\mbox{\tiny f-DBCM}}=\sum_{j(\neq i)=1}^N p_{ij}^{\text{\tiny f-DBCM}}, \qquad 
\avg{\ki_i}^{\mbox{\tiny f-DBCM}}=\sum_{j(\neq i)=1}^N p_{ij}^{\text{\tiny f-DBCM}}.
\end{equation}

An estimate of $z$ can be obtained also using the information on the 
connectivity of a subset $I$ of nodes, for instance the total number $\hL_I$ of links 
internal to $I$, or the degrees of all nodes belonging to $I$ \cite{musmeci2013bootstrapping,squartini2017network}. 
In both cases, in fact, the likelihood-maximization principle leads to an equation 
similar-in-spirit to eq. (\ref{L}). In the first case, we have
\begin{equation}\label{oneest}
\hL_I=\sum_{i\in I}\sum_{j(\neq i)\in I}\frac{z\hso_i \hsi_j}{1+z\hso_i\hsi_j},
\end{equation}
while in the second case it is
\begin{equation}\label{twoest}
\sum_{i\in I}(\hko_i+\hki_i)=
\sum_{i\in I}\sum_{j=1\atop{(j\neq i)}}^N\left(\frac{z\hso_i\hsi_j}{1+z\hso_i\hsi_j}+\frac{z\hso_j\hsi_i}{1+z\hso_j\hsi_i}\right).
\end{equation}
As shown in \cite{blagus2015empirical}, the way a specific subset of nodes is 
selected does matter. Whenever a faithful estimation of the network density 
is needed, nodes must be sampled according to a random selection scheme \cite{squartini2017network}, 
any other procedure being biased towards larger or smaller density values.

\subsubsection{Combining fitness-induced DBCM and IPF}

An reconstruction method combining the fitness-induced ERG formalism and the IPF recipe is 
proposed in \cite{battiston2016leveraging}. Here, the authors assume to 
know only the out- and in-strengths $\{\hso_i, \hsi_i\}_{i=1}^N$ and the total 
number of links $\hL$. The algorithm then consists of two steps: 
\begin{itemize}
\item the presence of a link between any two nodes $i$ and $j$ is estimated as 
in the fitness-induced DWCM, \ie, via eq. (\ref{pLp}) using as $z$ the solution of eq. (\ref{L});
\item the weights are placed on each generated binary configuration according to the IPF algorithm, 
and hence the constraints of eqs. (\ref{margins}) are always met exactly.
\end{itemize}

\subsubsection{Combining fitness-induced DBCM and DECM} 
A more rigorous way to assign weights to the fitness-induced ERG formalism 
consists in solving ``bootstrapped'' version of the DECM \cite{cimini2015estimating}. More precisely, the system of equations 
to be solved to find the Lagrange multipliers $\{\xxo_i,\xxi_i,\yo_i,\yi_i\}_{i=1}^N$ of the DECM becomes
\begin{equation}\label{becm}
\left\{ \begin{array}{cl}
\avg{\ko_i}^{\mbox{\tiny f-DBCM}}&=\:\:\:\sum_{j(\neq i)=1}^Np_{ij}^{\mbox{\tiny DECM}}\\
\avg{\ki_i}^{\mbox{\tiny f-DBCM}}&=\:\:\:\sum_{j(\neq i)=1}^Np_{ji}^{\mbox{\tiny DECM}}\\
\hso_i&=\:\:\:\sum_{j(\neq i)=1}^N p_{ij}^{\mbox{\tiny DECM}}(1-\yo_i\yi_j)^{-1}\\
\hsi_i&=\:\:\:\sum_{j(\neq i)=1}^N p_{ji}^{\mbox{\tiny DECM}}(1-\yo_j\yi_i)^{-1}
\end{array} \right. \qquad\forall i
\end{equation}
where $p_{ij}^{\mbox{\tiny DECM}}$ is defined in eq. (\ref{decm}) and $\avg{\ko_i}^{\mbox{\tiny f-DBCM}}$ and 
$\avg{\ki_i}^{\mbox{\tiny f-DBCM}}$ are the fitness-induced DBCM estimates defined by eqs. (\ref{estdeg}). 
The name {\em bootstrapped DECM} comes from the double role played by node out- and in-strengths, 
which are first used to estimate the degrees, and then imposed as complementary constraints.

\begin{figure*}
\centerline{
\includegraphics[width=0.8\textwidth]{fig1.pdf}
}
\centerline{
\includegraphics[width=0.8\textwidth]{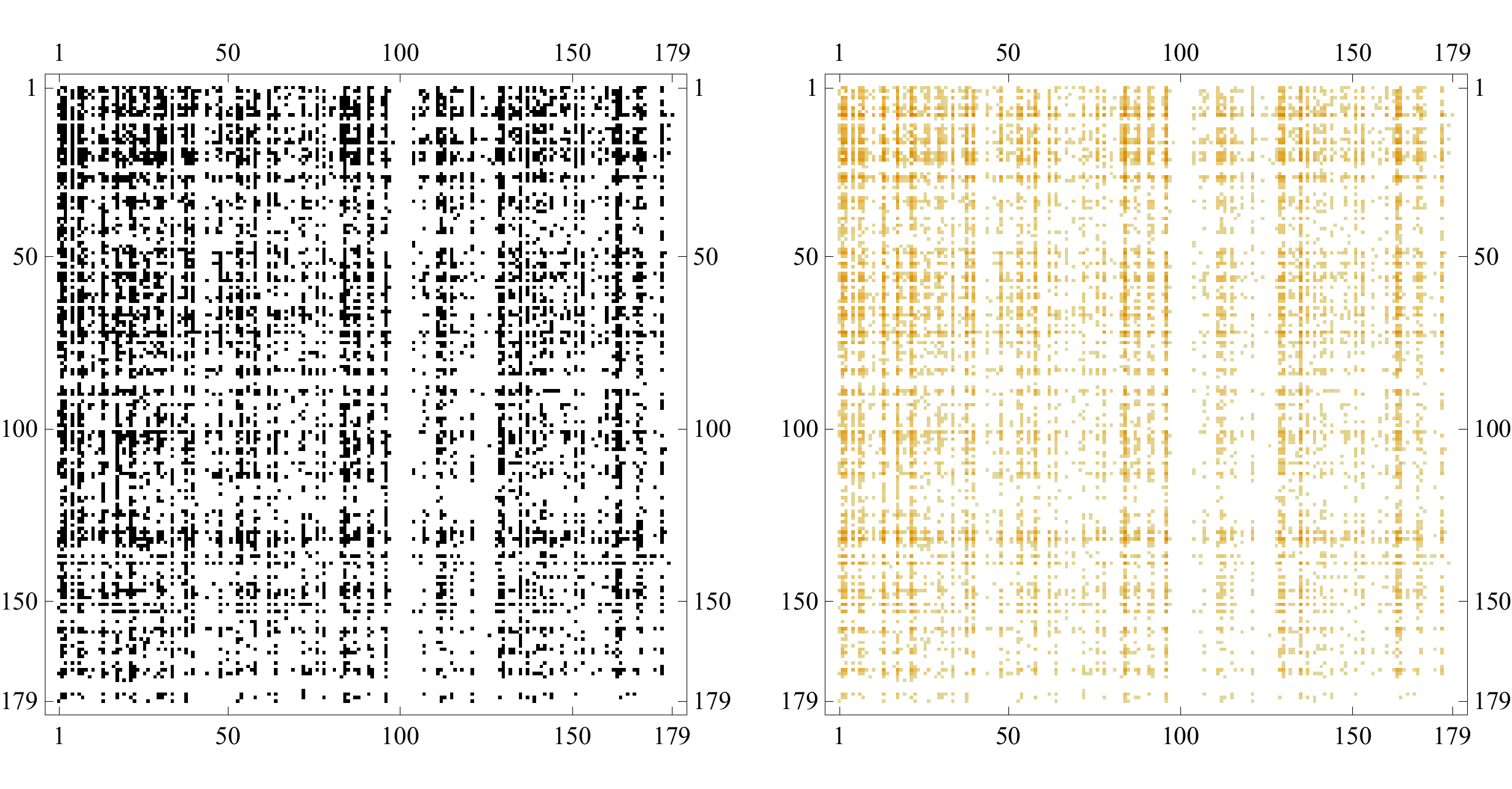}
}
\caption{Comparison between the observed adjacency matrix of the eMID network in 
2003 (top panels) and its reconstructed version according to the dcGM method described in section \protect{\ref{dcGM}} (bottom panels). 
Left panels represent binary adjacency matrices with black/white denoting the presence/absence of connections, 
whereas, right panels represent weighted adjacency matrices with color intensity denoting the weight of connections.}
\label{fig3}
\end{figure*}

\subsubsection{The degree-corrected gravity model}\label{dcGM}

Although the DECM (both in its original and ``bootstrapped'' versions) 
represents a very accurate reconstruction method \cite{mastrandrea2014enhanced,cimini2015estimating}, 
its numerical resolution can represent a computationally-demanding 
task\footnote{It should, however, be noticed that the estimation 
procedure leading to eqs. (\ref{estdeg}) has a regularizing effect on the 
values of degrees, which become smooth monotonic functions of the strengths 
\cite{cimini2015systemic}. This, in turn, may lead to a 
smaller computational effort to solve eqs. (\ref{becm}).}. Building upon the 
MaxEnt recipe, economics provides the main inspiration for a simpler 
alternative. Indeed, although the MaxEnt method performs 
poorly in reproducing the topological structure of many real-world networks, the 
observed weights are nicely reproduced by eq. (\ref{maxe}) \cite{squartini2017network,mazzarisi2017methods}. 
A straightforward way to both retain the explanatory power of the gravity model 
and avoid ending up with a complete network is provided by the heuristic recipe 
of the ``degree-corrected gravity model'' (dcGM) \cite{cimini2015systemic}:

\begin{equation}\label{dcgm}
w_{ij}^{\mbox{\tiny dcGM}} = \left\{ \begin{array}{ll}
\qquad 0 & \mbox{with probability }1-p_{ij}^{\text{\tiny f-DBCM}}\\
w_{ij}^{\mbox{\tiny ME}} (p_{ij}^{\text{\tiny f-DBCM}})^{-1} & \mbox{with probability } p_{ij}^{\text{\tiny f-DBCM}}
\end{array} \right.\quad \forall i\neq j.
\end{equation}
This equation ``corrects'' the MaxEnt recipe by placing the weight $w_{ij}^{\mbox{\tiny ME}}$ only with probability 
$p_{ij}^{\mbox{\tiny f-DBCM}}$ (\ie, conditional to the existence of the link), 
and rescaled in order to have $\avg{w_{ij}^{\mbox{\tiny dcGM}}} = w_{ij}^{\mbox{\tiny ME}}$. 
In this way, both the network marginals and the link density are correctly reproduced on average (see fig. \ref{fig3}). 
However, as for the original MaxEnt, this happens only if non-zero diagonal entries are considered as well. 
Otherwise, by restricting the sums over $i\neq j$, the expected values of the strengths obtained from the dcGM 
would require an extra-term to reproduce the observed marginals: $\forall i$, we would get 
$\avg{\so_i}^{\mbox{\tiny dcGM}}=\hso_i-\hso_i\hsi_i/\hW$ and $\avg{\si_i}^{\mbox{\tiny dcGM}}=\hsi_i-\hso_i\hsi_i/\hW$, 
and the missing term would be precisely the expected diagonal weight $\avg{w_{ii}^{\mbox{\tiny dcGM}}}=\hso_i\hsi_i/\hW\equiv w_{ii}^{\mbox{\tiny ME}}$.

The authors in \cite{squartini2017network} devise a procedure, inspired by the IPF algorithm, 
to redistribute the diagonal terms across the non-diagonal entries of the network. 
Precisely, the correction term $\delta w_{ij}^{(n)}$ to be added to the second line of eq. \eqref{dcgm} at the 
$n$-th iteration of the IPF-like algorithm is defined as

\begin{equation}
\delta w_{ij}^{(n+1)}=\frac{\hso_j\hsi_j}{\hW}\left(\frac{\delta 
w_{ij}^{(n)}}{\sum_{k(\neq j)}\delta w_{kj}^{(n)}}\right),\qquad\delta 
w_{ij}^{(n)}=\frac{\hso_i\hsi_i}{\hW}\left(\frac{\delta 
w_{ij}^{(n-1)}}{\sum_{k(\neq i)} \delta w_{ik}^{(n-1)}}\right),
\end{equation}
and where the algorithm is initialized at $w_{ij}^{(0)}=1-\delta_{ij}$. 
Once the asymptotic corrections $\delta w_{ij}^{(\infty)}$ 
are determined, the heuristic recipe of eq. (\ref{dcgm}) is replaced by

\begin{equation}
w_{ij}^{\mbox{\tiny dcGM}} = \left\{ \begin{array}{ll}
\qquad\qquad 0 & \mbox{with probability } 1-p_{ij}^{\text{\tiny f-DBCM}}\\
(w_{ij}^{\mbox{\tiny ME}}+\delta w_{ij}^{(\infty)}) (p_{ij}^{\text{\tiny f-DBCM}})^{-1} & \mbox{with probability } p_{ij}^{\text{\tiny f-DBCM}}
\end{array} \right.\quad \forall i\neq j.
\end{equation}

For all practical purposes, a small number of iterations is often enough to 
achieve a satisfactory degree of accuracy. Here we explicitly report the 
functional form of the first three iterations:

\begin{eqnarray}\label{corr}
w_{ij}^{(1)}&=&\frac{\hso_i\hsi_i}{\hW}\left(\frac{1}{N-1}\right);\nonumber\\
w_{ij}^{(2)}&=&\frac{\hso_i\hsi_i}{\hW}\left(\frac{\hso_j\hsi_j}{\sum_{l(\neq 
j)}\hso_l\hsi_l}\right);\\
w_{ij}^{(3)}&=&\frac{\hso_i\hsi_i}{\hW}\left(\frac{\hso_j\hsi_j}{\sum_{l(\neq 
j)}\hso_l\hsi_l}\right)\left(\frac{1}{\sum_{k(\neq 
i)}\frac{\hso_k\hsi_k}{\sum_{m(\neq k)}\hso_m\hsi_m}}\right).\nonumber
\end{eqnarray}

\paragraph*{A bipartite degree-corrected gravity model} The degree-corrected 
gravity model can be straightforwardly extended to the case of bipartite (undirected) networks \cite{squartini2017stock}. 
It is in fact enough to adapt eq. (\ref{L}) and eq. (\ref{dcgm}) to the new problem setup. In 
particular, the equation to determine the unknown coefficient $z$ relating the 
known and the expected total number of links is 

\begin{equation}
\hL=\sum_{i=1}^{N_1}\sum_{\alpha=1}^{N_2}\frac{z\hat{s}^{\mbox{\tiny [1]}}_i\hat{s}^{\mbox{\tiny [2]}}_\alpha}{1+z\hat{s}^{\mbox{\tiny [1]}}_i\hat{s}^{\mbox{\tiny [2]}}_\alpha}
\end{equation}
with $N_1$ and $N_2$ denoting the cardinality of the two layers of the network, and 
$\{\hat{s}^{\mbox{\tiny [1]}}_i\}_{i=1}^{N_1}$ and $\{\hat{s}^{\mbox{\tiny [2]}}_\alpha\}_{\alpha=1}^{N_2}$ indicating the {\em 
known} strength sequences of nodes belonging to the first and second layer, 
respectively. Notably, correction terms as the ones defined by eqs. (\ref{corr}) 
are no longer needed, since diagonal terms are now absent by definition.

\subsubsection{Reconciling ERG and gravity models}

Remarkably, the ERG framework allows the ``economic'' information defining 
gravity models to be translated into opportunely-defined fitnesses. Equation 
(\ref{gv1}) provides a clear example. Another example is provided by the 
following definition

\begin{equation}\label{gv2}
p_{ij}^{\text{\tiny GM}}=\frac{z\,\text{GDP}_i\,\text{GDP}_j\,e^{-\phi 
f(d_{ij})}}{1+z\,\text{GDP}_i\,\text{GDP}_j\,e^{-\phi f(d_{ij})}}
\end{equation}
where $f(d_{ij})$ is an increasing function of the geographic distance $d_{ij}$ 
between countries $i$ and $j$. The simplest functional form $f(d_{ij})=d_{ij}$ 
comes from considering the Hamiltonian  

\begin{equation}
\label{gravity-ham}
H(\mW|z,\phi)=-\sum_{i=1}^N\left(k_i\ln\text{GDP}_i\right)-L\ln z- 
F\phi
\end{equation}
with $F(\mA)=\sum_i\sum_{j(\neq i)}a_{ij}d_{ij}$. The latter term quantifies 
to what extent the topological structure of the network fills the geometric 
space in which the network itself is embedded \cite{ruzzenenti2012spatial}. As usual, the 
unknown parameters must be estimated by solving the equations $\hL=\avg{L}$ and $\hat F=\avg{F}$. 
In a sense, eq. (\ref{gv2}) defines the 
closest network-based model to traditional gravity models.

The main difference between the approach proposed here and the traditional 
economic one becomes evident upon sketching the derivation of the so-called 
\emph{zero-inflated gravity models} (ZIGM) \cite{duenas2013modeling}. In order 
to prevent this model from predicting a fully-connected network (the same limitation 
of the MaxEnt recipe), a probability coefficient reading

\begin{equation}\label{zigv}
p_{ij}^{\text{\tiny ZIGM}}=\frac{1}{1+e^{-\vec{\phi}\cdot\vec{C}_{ij}}}
\end{equation}
is assumed to control for the presence of a link between any two nodes $i$ and 
$j$. The vector $\vec{\phi}$ of unknowns is estimated by considering 
the elements $a_{ij}$ of the adjacency matrix to be the dependent variables, and 
the quantities usually employed to fit a gravity model (as the countries GDP, 
their geographic distances, etc.---\ie, a whole vector $\vec{C}$ of quantities 
for each pair of nodes) to play the role of explanatory variables. Both are 
then used to define the likelihood function for the actual network 
configuration $\hmG$, which is as usual maximized with respect to 
$\vec{\phi}$. Once a matrix of probability coefficients is 
obtained, only the nodes pairs satisfying the condition $p_{ij}^{\text{\tiny 
ZIGM}}\geq \hat{\rho}$ (where $\hat{\rho}$ is the known density of links) 
are actually linked \cite{duenas2013modeling}.

Zero-inflated gravity models and network-based gravity models differ in the 
amount of information needed to be fully specified. While the former require the 
whole adjacency matrix of a given (economic) network to be fully specified, the 
latter only require the knowledge of global (marginal) information. Remarkably, 
despite the much smaller amount of information needed, 
network-based gravity models perform much better than zero-inflated gravity 
models \cite{duenas2013modeling}.

\subsubsection{A remark on the ensemble methods}\label{comm}

One of the reasons of the attractiveness of the ``ensemble methods'' lies in the 
possibility they offer to generate \emph{different} topological structures that 
satisfy the \emph{same} (weighted) constraints. This feature can be used to 
disentangle the impact of marginals such as the {\em balance sheets} of a financial system 
and of network structural details on the outcome of a dynamical process like 
the spreading of financial distress \cite{gandy2017adjustable}.

In order to generate realistic scenarios, however, some kind of topological 
information must be accessible. In the optimal case, the whole degree sequence 
of a real network would be available (and, thus, used as additional constrain 
beside the weighted marginals, via the DECM or its two-step version). 
Otherwise, if a more aggregate knowledge on the system is available (like the total number of 
links, or the degrees of a particular subset of nodes), an additional assumption 
is needed to make the best use out of such information. From the many 
attempts done so far, it seems that a preliminary estimation of the degree 
sequence (as in the bootstrapped DECM scheme) enhances the performance of a 
given reconstruction method, an evidence that explains the superiority of the 
algorithm in \cite{cimini2015systemic} with respect to, \eg, the algorithm in 
\cite{digiangi2016assessing}---although both are defined by exactly the same 
information \cite{mazzarisi2017methods}. From a quantitative point of 
view, providing a realistic estimate of the degrees of nodes from aggregate 
information implies having a good fitness ansatz and a realistic estimate of the whole 
network density. The latter requirements ultimately means having an estimate for the 
parameter $z$ to be used in eq. (\ref{pLp}).

\subsection{Shannon-like approaches to reconstruction}

The reconstruction algorithms presented in the previous subsections 
build on the constrained maximization of Shannon entropy, or closely-related 
functionals like the KL divergence. Shannon entropy is however only 
one out of many possible functionals that can be taken to extremes 
under the constraints representing the accessible information.

\subsubsection{Spectral entropy}\label{spectre}

Among Shannon-like functionals, entropic measures exist that are inspired by quantum 
physics. \emph{Spectral entropy}, also known as \emph{Von Neumann entropy}, 
deserves a special mention \cite{anand2011shannon}. It is defined as

\begin{equation}
S^{\text{VN}}=\text{Tr}[\mathbf{\Xi}\ln\mathbf{\Xi}]=\sum_{m=1}
^N\xi_m(\mathbf{\Xi})\ln \xi_m(\mathbf{\Xi})
\end{equation}
\ie, as the Shannon entropy of a probability distribution induced by the eigenvalues $\{\xi_m\}_{m=1}^N$ of the matrix $\mathbf{\Xi}$. In quantum physics, the density matrix $\mathbf{\Xi}$ describes a system that can be found in one of a set of pure states with different probabilities (precisely defined by the eigenvalues of $\mathbf{\Xi}$): in order to employ this concept in network theory, the density matrix has to be re-expressed in terms of network quantities. A network-based version of the density matrix satisfying the properties of positive semi-definiteness and trace unitarity has been defined as \cite{dedomenico2016spectral}\footnote{Other proposals like $\mathbf{\Xi}=\frac{\mathbf{L}}{\text{Tr}[\mathbf{L}]}$ do not satisfy the (sub)additivity property \cite{dedomenico2016spectral}.}:

\begin{equation}
\mathbf{\Xi}=\frac{e^{-\beta\mathbf{L}}}{Z}.
\end{equation}
where $\mathbf{L}=\mathbf{D}-\mA$ is the Laplacian matrix 
($\mathbf{D}$ is a diagonal matrix of nodes degrees) with elements 
$L_{ij}=k_{i}\delta_{ij}-a_{ij}$ $\forall\:i,j$ and $Z=\text{Tr}[e^{-\beta\mathbf{L}}]$. 

This approach ultimately boils down to the calculation of the divergence between the spectral density of an operator associated with the empirical graph and that of the corresponding operator associated with a graph model \cite{dedomenico2016spectral}. In principle this allows to optimize the parameters of the model using a sort of quantum analogue of the method described in section \ref{ERG-sect}.

\subsubsection{The Cressie-Read family of power divergences}\label{creed}

A whole family of functionals to be extremized to reconstruct partially known networks, generalizing the usual Shannon entropy or Kullback-Leibler divergence (see subsection \ref{rasalgo}), is represented by the so-called {\it Cressie-Read power divergences}. The latter can be compactly written as

\begin{equation}\label{cr}
I(\mathbf{P},\mathbf{Q},\gamma)=\frac{1}{\gamma(\gamma+1)}\sum_{\mG
\in\ensG}P(\mG)\left[\left(\frac{P(\mG)}{Q(\mG)}
\right)^\gamma-1\right]
\end{equation}
with the real parameter $\gamma$ indexing the members of the family. Equation 
(\ref{cr}) generalizes the KL divergence and provides an alternative measure of 
``distance'' between any two probability distributions $\mathbf{P}$ and $\mathbf{Q}$. Notice that even if 
$I(\mathbf{P},\mathbf{Q},\gamma)$ is not a proper metric distance for all values 
of $\gamma$, the properties it satisfies are nonetheless useful for quantifying 
the extent to which any two distributions differ \cite{judge2011information}. 
More specifically, 

\begin{itemize}
\item $I(\mathbf{P},\mathbf{Q},\gamma)$ is a continuous function of all its 
arguments $\{P(\mG)\}_{\mG\in\ensG}$, 
$\{Q(\mG)\}_{\mG\in\ensG}$;
\item $I(\mathbf{P},\mathbf{Q},\gamma)\geq0$, with equality if and only if 
$P(\mG)=Q(\mG),\:\forall\:\mG$; 
\item $I(\mathbf{P},\mathbf{Q},\gamma)$ is invariant under the addition of 
events with zero probability;
\item $I(\mathbf{P},\mathbf{Q},\gamma)$ is log-additive\footnote{The 
log-additivity property reads 
$\ln[1+\theta(\theta+1)I(\mathbf{P},\mathbf{Q},\gamma)]+\ln[
1+\theta(\theta+1)I(\mathbf{R},\mathbf{S},\gamma)]=\ln[
1+\theta(\theta+1)I(\mathbf{P\times R},\mathbf{Q\times S},\gamma)]$ with 
$\mathbf{P}\times\mathbf{R}$ and $\mathbf{Q}\times\mathbf{S}$ indicating the 
tensor product of the two involved probability distributions 
\cite{cressie1984multinomial}.};
\item the functionals indexed by values of the parameter $\gamma\in(-1,0)$ 
satisfy the triangle inequality;
\item the only functional representing a proper metric distance (related to the Matusita distance) is the one characterized by $\gamma=-1/2$.
\end{itemize}

Since $\mathbf{Q}$ is often intended as summarizing the prior information about 
the system, the prescription to search for the probability distribution $\mathbf{P}$ which is 
maximally non-committal with respect to the missing information can be 
translated into the request of \emph{minimizing} the divergence from $\mathbf{Q}$ to $\mathbf{P}$. In the case constraints are present, this (first) optimization step leads to recover an expression for $\mathbf{P}$ which depends on a vector of unknown Lagrange multipliers:

\begin{equation}\label{cro}
\min_{\mathbf{P}}\left\{I(\mathbf{P},\mathbf{Q},
\gamma)-\lambda_0\left[\sum_{\mG\in\ensG}P(\mG)-1\right]
-\sum_{m=1}^M\lambda_m\left[\sum_{\mG\in\ensG}P(\mG
)C_m(\mG)-\avg{C_m}\right]\right\};
\end{equation}
the second step of the whole procedure prescribes to substitute the recovered expression of $\mathbf{P}$ into $I$ itself and optimize $I(\vec{\lambda})$ with respect to the unknown parameters $\vec{\lambda}$ \cite{judge2011information}.

Equation (\ref{cro}) generalizes the two principles lying at the basis of the ERG formalism introduced in the previous sections. As $\gamma$ varies, the functional describing the divergence between $\mathbf{P}$ and $\mathbf{Q}$ varies as well. Two noteworthy examples are retrieved by solving the following limits

\begin{eqnarray}
\lim_{\gamma\rightarrow0} 
I(\mathbf{P},\mathbf{Q},\gamma)&=&D_{\text{KL}}(\mathbf{P}||\mathbf{Q})=\sum_{
\mG\in\ensG}P(\mG)\ln\left(\frac{P(\mG)}{Q(\mG
)}\right),\\
\lim_{\gamma\rightarrow-1} 
I(\mathbf{P},\mathbf{Q},\gamma)&=&D_{\text{KL}}(\mathbf{Q}||\mathbf{P})=\sum_{
\mG\in\ensG}Q(\mG)\ln\left(\frac{Q(\mG)}{P(\mG
)}\right)
\label{dklqp}
\end{eqnarray}
\ie, the KL divergence between $\mathbf{P}$ and $\mathbf{Q}$ and between 
$\mathbf{Q}$ and $\mathbf{P}$. Whenever the maximally uninformative prior is 
adopted, 
$Q(\mG)=\frac{1}{|\ensG|}$ $\forall$ $\mG\in\ensG$, the 
well-known expressions

\begin{eqnarray}
D_{\text{KL}}(\mathbf{P}||\mathbf{Q})&=&\sum_{\mG\in\ensG}P(\mathbf
{G})\ln P(\mG)+\ln|\ensG|,\\
D_{\text{KL}}(\mathbf{Q}||\mathbf{P})&=&-\sum_{\mG\in\ensG}\frac{
\ln P(\mG)}{|\ensG|}+\ln|\ensG|,
\end{eqnarray}
are recovered, respectively defining (up to a sign) the Shannon entropy functional and the likelihood functional \cite{judge2011information}. Notice that, for $\gamma\rightarrow0$, minimizing $I$ consistently translates into maximizing Shannon entropy, thus retrieving the procedure described previously.

In order to use the framework described above for reconstruction purposes, the most general problem of inferring the entries $\{n_{ij}\}_{i=1\dots I,\:j=1\dots J}$ of a rectangular matrix by using the information provided by marginals $n_{i\cdot}=\sum_jn_{ij}$ $\forall i$ and $n_{\cdot 
j}=\sum_in_{ij}$ $\forall j$ must be restated in probabilistic terms. As illustrated in table \ref{table1}, upon introducing the variables 
$p_{ij}=n_{ij}/n_{i\cdot}$ $\forall i,j$, and further dividing all entries by $n$ (thus inducing the definitions $x_{i}=n_{i\cdot}/n$ $\forall i$ and 
$y_{i}=n_{\cdot i}/n$ $\forall i$), providing a numerical estimate of the table entries translates into estimating the entries of the matrix $\mathbf{P}$ appearing within the set of linear equations

\begin{equation}
\mathbf{y}=\mathbf{x}\mathbf{P}.
\end{equation}
Indeed the entries of $\mathbf{P}$ can be \emph{formally} interpreted as probability coefficients, defined as fractions of marginals. This position, in turn, allows a problem formally analogous to the one stated in eq. \ref{cro} to be defined as

\begin{equation}
\min_{{p}_{jk}}\left\{I(\{
{p}_{jk}\},\{q_{jk}\},\gamma)-\sum_j\beta_j\!\!\left(\sum_k{p}_{jk}
-1\right)-\sum_k\alpha_k\!\!\left(\sum_j{p}_{jk}x_j-y_k\right)\right\}
\end{equation}
and a solution of the form $I(\{{p}_{jk}\},\{q_{jk}\},\gamma)=\frac{1}{\gamma(\gamma+1)}
\sum_j\sum_k{p}_{jk}\left[\left(\frac{{p}_{jk}}{q_{jk}}
\right)^\gamma-1\right]$ to be found. Notice that choosing $\gamma\rightarrow0$ and a 
maximally uninformative prior reduces to the usual exponential 
form coming from the minimization of the KL divergence\footnote{This solution 
is formally equivalent to the MaxEnt one. However, 
since this approach is typically used to infer election percentages or estimate 
the purchases of a basket of commodities (\ie, to reconstruct tables where zero 
entries are practically never observed), the evidence that non-zero marginals 
cannot induce zero entries does not constitute a problem 
\cite{squartini2015information}.}

\begin{equation}\label{croe}
p_{jk}^{(0)}=\frac{e^{\alpha_kx_j}}{\sum_{k'}e^{\alpha_{k'}x_j}}.
\end{equation}
Other choices, instead, lead to coefficients described by different functional 
forms. As an example, adopting the functional induced by the choice 
$\gamma\rightarrow-1$ leads to the expression

\begin{equation}
p_{jk}^{(-1)}=-\frac{1}{\alpha_kx_j+\beta_j}.
\end{equation}
In general, for an arbitrary choice of the exponent $\gamma$, the functional form of the entries of $\mathbf{P}$ induced by the functional $I(\{p_{jk}\}, \{q_{jk}\}, \gamma)$ differs substantially from the ``usual'' exponential one shown in eq. (\ref{croe}). This, in turn, induces a reconstruction procedure whose performance is potentially very different from that of the KL-based approach.

\begin{table}[t!]
\centering
\begin{tabular}{|cc|c}
\noalign{\smallskip}
\hline
$n_{11}$ & $n_{12}$ & $n_{1\cdot}$\\
$n_{21}$ & $n_{22}$ & $n_{2\cdot}$\\
$n_{31}$ & $n_{32}$ & $n_{3\cdot}$\\
\hline
$n_{\cdot 1}$ & $n_{\cdot 2}$ & $n$
\end{tabular}$\:\:\:\longrightarrow\:\:\:$
\begin{tabular}{|cc|c}
\noalign{\smallskip}
\hline
$p_{11}n_{1\cdot}$ & $p_{12}n_{1\cdot}$ & $n_{1\cdot}$\\
$p_{21}n_{2\cdot}$ & $p_{22}n_{2\cdot}$ & $n_{2\cdot}$\\
$p_{31}n_{2\cdot}$ & $p_{32}n_{2\cdot}$ & $n_{3\cdot}$\\
\hline
$n_{\cdot 1}$ & $n_{\cdot 2}$ & $n$
\end{tabular}$\:\:\:\longrightarrow\:\:\:$
\begin{tabular}{|cc|c}
\noalign{\smallskip}
\hline
$p_{11}x_{1}$ & $p_{12}x_{1}$ & $x_{1}$\\
$p_{21}x_{2}$ & $p_{22}x_{2}$ & $x_{2}$\\
$p_{31}x_{3}$ & $p_{32}x_{3}$ & $x_{3}$\\
\hline
$y_{1}$ & $y_{2}$ & $1$
\end{tabular}
\caption{Pictorial representation of the ill-posed inverse problem concerning 
the inference of the entries of a table, using only the information provided by 
marginals. Upon introducing the variables $p_{ij}=n_{ij}/n_{i\cdot}$ and 
dividing the entries by $n$ (whence the definitions $x_{i}=n_{i\cdot}/n$, 
$y_{i}=n_{\cdot i}/n$) a constrained-optimization problem naturally emerges \cite{cho2015information}.\label{table1}}
\end{table}

\subsubsection{Other entropic families}\label{tsrenyi}

Just like the Cressie-Read functionals provide a generalization of the Kullback-Leibler divergence, generalizations of the Shannon entropy are provided by \emph{Renyi entropies} \cite{renyi1961entropy} and \emph{Tsallis entropies} \cite{tsallis1988possible}. These entropies depend on a free parameter and include Shannon entropy as a limiting case. More specifically, Renyi entropies are defined as

\begin{equation}
S_\alpha=\frac{\ln\left[\sum_{\mG\in\ensG}P(\mG
)^\alpha\right]}{1-\alpha}
\end{equation}
(with $\alpha$ being a non-negative parameter, different from 1) and satisfy the 
additivity property\footnote{The additivity property reads 
$S_\alpha(\mathbf{P}\times\mathbf{Q})=S_\alpha(\mathbf{P})+S_\alpha(\mathbf{Q})$ 
with $\mathbf{P}\times\mathbf{Q}$ indicating the tensor product of the two 
involved probability distributions.}.

Tsallis entropies can be axiomatically defined upon generalizing the fourth 
Shannon-Khinchin axiom (see section \ref{secinf}). While this axiom unambiguously 
identifies Shannon entropy, substituting it with the requirement that 
$S_q(W_{A+B})=S_q(W_{A})+S_q(W_{B|A})+(1-q)S_q(W_{A})S_q(W_{B|A})$ leads to the \emph{only} functional that satisfies such a new set of axioms\footnote{The parameter $q$ quantifies the degree of non-extensivity of such a functional.}:

\begin{equation}
S_q=\frac{1-\sum_{\mG\in\ensG}P(\mG)^q}{q-1}.
\end{equation}

Remarkably, $S_q$ can be employed to define a non-extensive version of the ERG 
formalism, whose derivation proceeds along similar lines. For example, 
imposing only the normalization condition leads to the functional 
$\mathscr{L}_q[P]=S_q-\lambda_0\left[\sum_{\mG\in\ensG}P(\mG
)-1\right]$ which is maximized by the uniform distribution 
$P(\mG)=\frac{1}{|\ensG|}$ $\forall \mG$. 
Imposing less trivial constraints, however, has not been attempted yet: as a 
consequence, a thorough comparison between the goodness of the reconstruction 
performances induced by extensive and non-extensive entropies is still missing.

\subsection{Beyond Shannon entropy: alternative approaches to reconstruction}

After having revised the existing Shannon-based and Shannon-like approaches to 
reconstruction, we now review algorithms that are not based on the maximization 
of Shannon-inspired functionals.

\subsubsection{The ``copula'' approach}\label{copulaapp}

The first ``alternative'' approach to entropy-based reconstruction is, actually, the closest 
one to traditional MaxEnt. The ``copula'' method, in fact, adopts the same 
philosophy and uses the entries of a given matrix to define the support of a 
probability distribution to be estimated; 
at the same time, however, it provides a more general solution to the problem.

The MaxEnt prescription represents the simplest method for estimating a bivariate probability 
distribution $P_{xy}(X,Y)$, given the two marginal distributions $P_x(X)$ and 
$P_y(Y)$. Indeed, maximizing the Shannon entropy

\begin{equation}
S=-\sum_i\sum_jP_{xy}(X_i,Y_j)\ln P_{xy}(X_i,Y_j)
\end{equation}
under the constraints represented by normalization 
$\sum_i\sum_jP_{xy}(X_i,Y_j)=1$ and the two marginal distributions 
$P_x(X_i)=\sum_jP_{xy}(X_i,Y_j) $ $\forall i$ and 
$P_y(Y_j)=\sum_iP_{xy}(X_i,Y_j)$ $\forall j$ leads precisely to the MaxEnt-like estimation

\begin{equation}
P^{\text{\tiny MC}}_{xy}(X_i,Y_j)=P_x(X_i)P_y(Y_j).
\end{equation}

The recipe above, however, can be generalized by introducing the so-called {\em 
copula functions}. The rationale for employing copulas is provided by Sklar's 
theorem, which states that every multivariate cumulative distribution function (CDF) can be expressed in terms of its 
marginal CDFs\footnote{Sklar's theorem requires the marginals to be continuous. 
Whenever discrete datasets are considered, the results described here can be 
thought as being applied to the kernel density estimations of the corresponding 
(discrete) CDFs.} (say, $F_x(X)$, $F_y(Y)$, etc.) and a copula function $\mathcal{C}$ 
which, as the name suggests, ``couples'' them:

\begin{equation}\label{copula}
F_{xy\dots}(X,Y\dots)=\mathcal{C}[F_x(X),F_y(Y)\dots].
\end{equation}

In our case, the marginal CDFs are those of the constraints (defined by eqs. (\ref{margins})) to be 
estimated from data. The choice of the particular copula function, on 
the other hand, is completely arbitrary\footnote{Notably, a 
maximum-entropy recipe for estimating copulas has been recently proposed 
\cite{piantadosi2012copulas}.}. The authors in 
\cite{baral2012estimating} use the Gumbel copula, defined as

\begin{equation}
\mathcal{C}^{\text{Gumbel}}_{ij}(\theta)=e^{-\left[\left(-\ln 
F_x(\so_i)\right)^\theta+\left(-\ln F_y(\si_j) 
\right)^\theta\right]^\frac{1}{\theta}}
\end{equation}
where the only parameter $\theta$ quantifies the dependence between the 
marginals. Remarkably, the parameter estimation can be carried out by 
maximizing the likelihood-like function

\begin{equation}
\mathcal{L}(\hmG|\theta)=\sum_{i=1}^N\sum_{j=1}^N\ln 
\mathcal{C}[F_x(\hso_i),F_y(\hsi_j)\dots]
\end{equation}
with respect to $\theta$. Once the model parameter has been estimated, a matrix whose 
entries are (interpreted as) probability coefficients is obtained. Finally, the IPF method 
is employed to readjust the sums along rows and columns and recover the observed marginals. 

Note that if the so-called ``independent'' copula function, defined by $\mathcal{C}[F_x(X),F_y(Y)\dots]=F_x(X)F_y(Y)\dots$,  
is used, then the MaxEnt estimation is recovered. And as for MaxEnt, the copula approach 
cannot reproduce the topological structure of sparse networks \cite{baral2012estimating}.

\subsubsection{A Bayesian approach to network reconstruction}\label{veraart}

The major difference between likelihood-based methods (as those described in the 
previous sections) and Bayesian methods lies in the role played 
by model parameters. Very broadly speaking, while 
likelihood-based methods provide a recipe for estimating the unknown parameters 
on the basis of the observations, Bayesian approaches treat the unknown 
parameters as (additional) random variables, described by properly-defined {\em 
prior} probability distributions, whose parameters (called {\em 
meta-parameters}) are chosen \emph{a priori}.

The first example of this second kind of approach to network modeling is provided by the 
\emph{fitness model} \cite{caldarelli2002scale}, resting upon the 
same ideas lying at the basis of the ERG approach:

\begin{itemize}
\item each node $i$ is described by a hidden variables $x_i$, representing 
its ``fitness''; generally speaking, this is a real numbers supposedly 
quantifying the importance of that node in the network, and is drawn from a given 
probability distribution $\nu(x)$;
\item any two nodes $i$ and $j$ establish a connection according to a coupling function 
$f(x_i,x_j)$ that, for undirected networks, is symmetric in the hidden variables assigned to nodes $i$ and $j$.
\end{itemize}

The main difference with respect to the ERG approach lies in the \emph{a priori} 
choice of both the functional form of the coupling function $f$ and the 
distribution $\nu$ from which fitnesses are drawn. The 
fitness model can be straightforwardly implemented by adapting the discrete formulas 
derived within the ERG framework to the continuous case. For instance, 
in the undirected case, the Bayesian derivation of the nodes degrees and of the total number of links reads

\begin{equation}
k_i=\sum_{j(\neq i)=1}^N f(x_i,x_j) \longrightarrow k_i(x)=(N-1)\int 
f(x,y)\nu(y)dy,
\end{equation}
\begin{equation}\label{bayeL}
L=\sum_{i=1}^N\sum_{j(<i)=1}^N f(x_i,x_j) \longrightarrow 
L=\frac{N(N-1)}{2}\int\int f(x,y)\nu(x)\nu(y)dxdy,
\end{equation}
where the integrations over the support of the distribution $\nu$ are necessary to account for the fitness variability.

Remarkably, several combinations of $f$ and $\nu$ lead to recover power-law 
degree distributions. In particular, both the intuitive combination

\begin{equation}\label{b1}
f(x,y)=zxy,\qquad \nu(x)\propto x^{-2}
\end{equation}
and the highly non-trivial combination

\begin{equation}\label{fitexp}
f(x,y)=\Theta(x+y-z),\qquad \nu(x)=e^{-x}
\end{equation}
lead to $P(k)\propto k^{-2}$. This result points out that a number of topological properties believed to arise only 
as a consequence of microscopic dynamic processes (as the one described by the preferential 
attachment mechanism and its variants) can, instead, be replicated also via a static model 
\cite{caldarelli2002scale}. In other words, whenever preferential attachment 
does not represent a plausible mechanism, it is reasonable to imagine that any 
two vertices establish a connection when the link creates a mutual benefit, depending on some intrinsic node property.

\medskip

The aforementioned approach has been recently extended to account also 
for weights, through an algorithm which is not dissimilar in spirit from the DECM. 
More specifically, the model introduced in \cite{gandy2016bayesian} is described by the following probability distribution for link weights:

\begin{equation}\label{bayes}
q_{ij}^{\mbox{\tiny Bayes}}(w) = \left\{ \begin{array}{ll}
1-p^{\mbox{\tiny Bayes}}_{ij} & \mbox{if }w=0,\\
p^{\mbox{\tiny Bayes}}_{ij}\theta_{ij}e^{-\theta_{ij}w} & \mbox{if }w>0.
\end{array} \right.
\end{equation}
(with $\theta_{ij}>0$). Hence, as in the DECM, distinct links are independent, 
and while the presence of the link is described by a Bernoulli trial, the value of its weight 
is set according to an exponential distribution. Note that the latter is the continuous version of the geometric 
distribution, which is obtained by ERG-based models upon assuming discrete weights. 
The philosophy of the fitness model is then encoded into the choice of functional forms reading

\begin{equation}
p_{ij}^{\mbox{\tiny Bayes}}=f(x_i+x_j),\qquad \theta_{ij}=G^{-1}_{\zeta,\eta}(e^{-x_i})+G^{-1}_{\zeta,\eta}(e^{-x_j}).
\end{equation}
In these expressions, $G^{-1}_{\zeta,\eta}$ is the quantile function of the Gamma distribution 
with positive shape and scale parameters $\zeta$ and $\eta$, drawn from an a prior distribution $\pi(\zeta,\eta)$, 
while fitnesses are drawn from the distribution $\nu(x)=e^{-x}$ and $f$ is defined as in 
\cite{gandy2016bayesian} such that the degree distribution exhibits a power law. 
An homogeneous version of the model has been also 
introduced, defined by the choices $p_{ij}=p\sim\text{Beta}(\alpha,\beta)$ and 
$\theta_{ij}=\theta\sim\text{Gamma}(\gamma,\delta)$.

Since this model induces an entire ensemble of networks, the expected value of 
the quantities of interest can be computed only after introducing a sampling 
procedure on the ensemble. The authors of \cite{gandy2016bayesian} adopt 
a {\em Gibbs sampler} working as follows.

\begin{itemize}
\item The sampler is initialized with a matrix $\mW^{(0)}$. When 
considering the homogeneous version, the initial matrix is generated via the 
Erd\H{o}s-R\'enyi model whose only parameter is required to match the observed 
average degree. Since the initial configuration is required to satisfy the 
conditions $\so_i(\mW^{(0)})<\hso_i$ and 
$\si_i(\mW^{(0)})<\hsi_i$ $\forall i$, the maximum-flow algorithm 
\cite{schrijver2002history} is employed to obtain a matrix 
$\mW^{(1)}$ matching the constraints exactly.
\item The whole ensemble of configurations is obtained by ``perturbing'' 
$\mW^{(1)}$. Such perturbations generalize the {\em local rewiring 
algorithm} according to the following rules: 1) the dimension $k$ of the 
sub-matrices to be updated is chosen and a set $\kappa$ of $k$ pairs of indices 
is selected; 2) the entries of $\mW^{(1)}$ are updated according to the 
rule

\begin{equation}
\mW^{(n+1)}_{\kappa_i}=\mW^{(n)}_{\kappa_i}+(-1)^{i+1}\Delta
\end{equation}
with $\kappa_i$ indicating the $i$-th pair of indices (\eg, the $r$-th row and 
the $c$-th column) and 
$\Delta\in[-\min_{i,\:\text{odd}}\mW^{(n)}_{\kappa_i}, 
\min_{i,\:\text{even}}\mW^{(n)}_{\kappa_i}]$.
\end{itemize}

Such a sampling process does not leave unexplored regions of the space of configurations: 
the existence of a sequence of Gibbs moves allowing for a transition from any matrix compatible 
with the given constraints to any other is, in fact, guaranteed \cite{gandy2016bayesian}). 
And although the algorithm allows one to generate networks characterized by different topological structures, 
every configuration satisfies the constraints defined by eqs. (\ref{margins}) exactly.

\paragraph*{An ``empirical'' Bayesian approach to network reconstruction} The 
same authors of \cite{gandy2016bayesian} have recently developed an ``adjustable'' 
version of their Bayesian approach \cite{gandy2017adjustable}. In this model, the linking probability 
between any two nodes $i$ and $j$ is assumed to be 
$p_{ij}^{\mbox{\tiny E-Bayes}}=\frac{z x_i x_j}{1+z x_i x_j}$, with 
$x_i=\hso_i+\hsi_i$ (naturally, this position better models undirected networks). 

\subsubsection{A comment on the Bayesian approaches to reconstruction}

Although the three aforementioned algorithms have been labeled as Bayesian, they 
share features of both likelihood-based and genuinely Bayesian methods. 
All of them are, in fact, characterized by the 
presence of one, or more, free parameters not to be drawn from \emph{a priori} 
distributions, but to be estimated via properly-defined likelihood conditions. 

For what concerns the model in \cite{caldarelli2002scale}, the only free 
parameter $z$ is fixed by imposing the condition $\hL=\avg{ L}$, \ie, by 
substituting either eq. (\ref{b1}) or eq. (\ref{fitexp}) into eq. (\ref{bayeL}) 
and solving the resulting equation for $z$. A main difference with the fitness-induced ERG formalism 
\cite{cimini2015systemic} remains in the way fitnesses are dealt with. In a sense, 
the fitness-induced ERG formalism represents the likelihood-based analogue of 
the Bayesian approach discussed here: there, the information on the 
fitness distribution is completely ignored and just a point-estimation 
is carried out; here, the fitness variability is completely accounted for. 

The model in \cite{gandy2016bayesian}, on the other hand, needs to be calibrated 
whenever used to reconstruct real-world networks. After 
assuming that many of the free parameters coincide, the authors are able to 
solve the equation $\hW=\avg{W}$, which results in  
$\theta_{ij}\equiv\theta=\sum_{ij}p_{ij}^{\mbox{\tiny Bayes}}/\hW$. 
Similarly, in \cite{gandy2017adjustable} the free parameter $z$ is ``adjusted'' in 
order to ensure that the expected density matches the observed one.

This discussion highlights the main limitation of using Bayesian approaches as reconstruction methods. 
Indeed, the freedom coming from treating model parameters as random variables does not necessarily help in 
reproducing the features of \emph{specific} real-world configurations. As the 
authors in \cite{gandy2016bayesian} explicitly recognize, Bayesian models need 
to be (at least partially) tuned in order to be used as reconstruction models. 
Whenever this step is missing, the arbitrariness in choosing the {\em a priori} 
distributions can be better employed to generate scenarios---useful, for instance, 
to obtain confidence intervals for stress tests outcomes.

\subsubsection{The Montagna \& Lux approach}\label{montagna}

Link probability coefficients can be also defined {\em ad hoc}, 
without any explicit derivation from a given optimization principle. 
For instance, the authors in \cite{montagna2017contagion} consider the following forms:
\begin{eqnarray}
p_{ij}^{\mbox{\tiny M-L-1}}&=&d_1(\hso_i)^\alpha(\hso_j)^\beta,\\
p_{ij}^{\mbox{\tiny M-L-2}}&=&d_2(\hso_i+\hso_j),\\
p_{ij}^{\mbox{\tiny M-L-3}}&=&d_3\Theta(\hso_i+\hso_j-z),
\end{eqnarray}
where the parameters $d_1$, $d_2$, $d_3$ are used to adjust the density of the network. 
This model thus follows the philosophy of the fitness model, 
by assuming that any two nodes $i$ and $j$ are connected with a probability $p_{ij}(\hso_i, \hso_j)$.
An ensemble of network configurations is then generated according to 

\begin{equation}
a_{ij} = \left\{ \begin{array}{ll}
0 & \mbox{with probability } 1-p_{ij}^{\mbox{\tiny M-L-h}}\\
1 & \mbox{with probability } p_{ij}^{\mbox{\tiny M-L-h}}
\end{array} \right.
\end{equation}
(with $h=1,2,3$ and with the additional rule of eliminating loops generated when $a_{ij}=a_{ji}=1$). 
Once a topological structure has been generated, the weights of the realized connections 
are set proportionally to the ``size'' of involved nodes as

\begin{equation}
w_{ij}^{\mbox{\tiny M-L-h}}=\hso_i\left(\frac{p_{ij}^{\mbox{\tiny M-L-h}}}{\sum_{\{a_{ij}=1\}
}p_{ij}^{\mbox{\tiny M-L-h}}}\right).
\end{equation}

In the original paper \cite{montagna2017contagion}, this approach is not used for network reconstruction, 
but rather to generate a financial interbank network assuming banks' total interbank assets to be power-law distributed---a choice that 
also leads to power-law distributed degrees. Additionally, in the original paper total assets $\hat{a}_i$ are used instead of total interbank assets $\hso_i$ 
as fitness values for each node $i$ (see section \ref{balshee}). The formulation we present here is equivalent though, 
since a linear proportionality relation $\hat{a}_i(1-\theta)=\hso_i$ is assumed \cite{nier2007network,montagna2017contagion}.

\begin{figure*}
\centerline{
\includegraphics[width=0.8\textwidth]{fig1.pdf}
}
\centerline{
\includegraphics[width=0.8\textwidth]{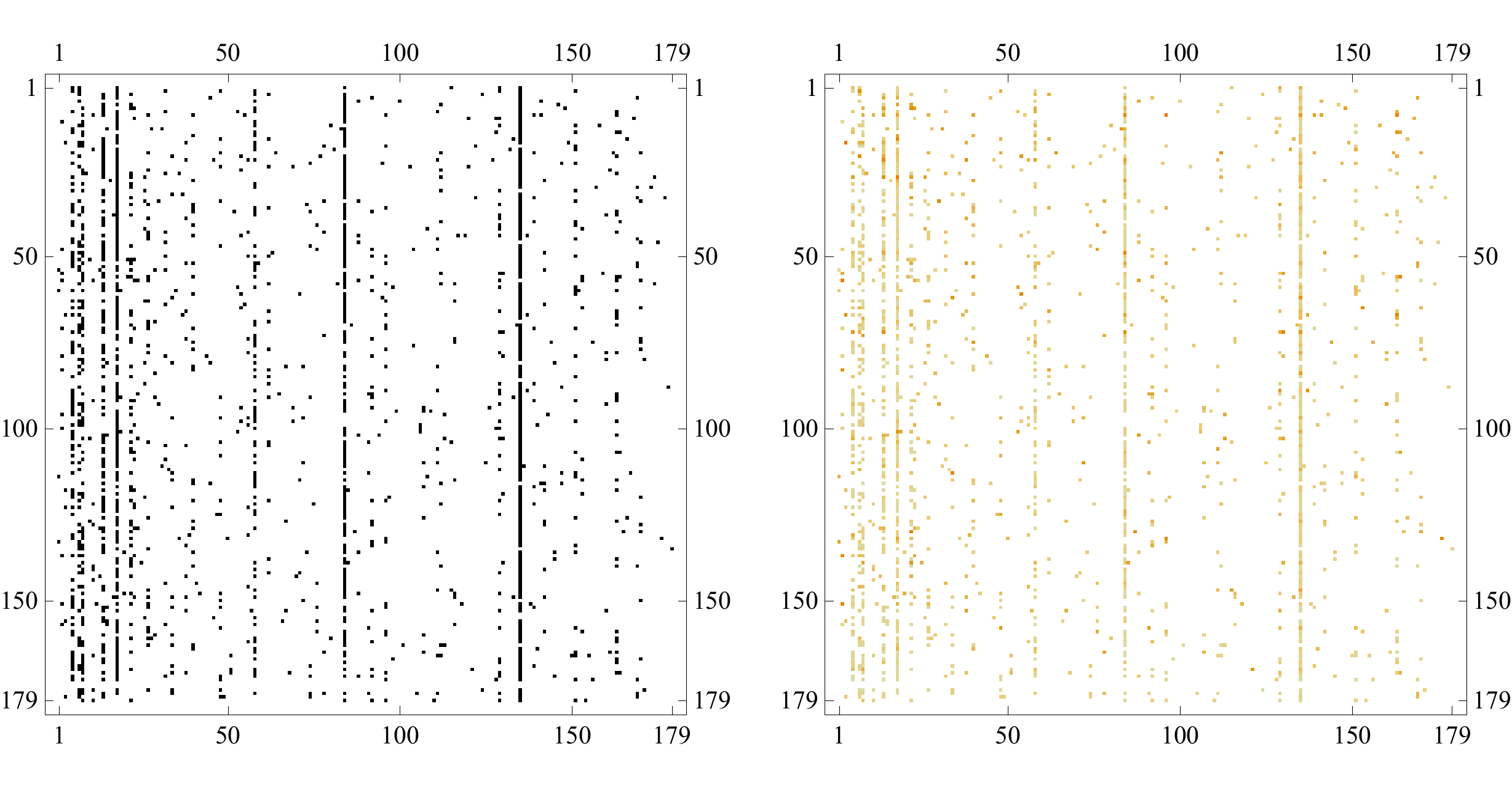}
}
\caption{Comparison between the observed adjacency matrix of the eMID network in 
2003 (top panels) and its reconstructed version according to the method by Ha\l{}aj \& Kok 
described in section \protect{\ref{halaj}} (bottom panels).
Left panels represent binary adjacency matrices with black/white denoting the presence/absence of connections, 
whereas, right panels represent weighted adjacency matrices with color intensity denoting the weight of connections.}
\label{fig4}
\end{figure*}

\subsubsection{The probability map of Ha\l{}aj \& Kok}\label{halaj}

Ha\l{}aj \& Kok further define link probabilities assuming the additional information on nodes' \emph{membership} to groups \cite{halaj2013assessing}. 
In the specific model implementation for interbank networks, each node (bank) belongs to a geographic area (\ie, a country), 
and link probability are expressed as fraction of node out-strengths aggregated within countries (which are assumed to be known): 
\begin{equation}
p_{ij}^{\mbox{\tiny H-K}}=\frac{\hw_{g_i,g_j}}{\hw_{g_i,\cdot}},
\end{equation}
where $\hw_{g_i,g_j}=\sum_{i\in g_i}\sum_{j\in g_j}\hw_{ij}$ is the total observed weight from area $g_i$ to area $g_j$ and 
$\hw_{g_i,\cdot}=\sum_{i\in g_i}\sum_{j}\hw_{ij}$ is the total weight going out from area $g_i$. 

Although this algorithm is similar in spirit to fitness models (and specifically to the \emph{stochastic block-model}, see section \ref{wmb}), 
its formalism differs from that of ERG. Indeed, the network structure is determined according to the following steps:

\begin{itemize}
\item a pair of nodes is randomly drawn out of the set of possible ones; 
\item the link is realized according to the corresponding probability $p_{ij}^{\mbox{\tiny H-K}}$; 
\item if the link is retained, a random number $r_{ij}\in[0,1]$ is generated, in 
order to determine the percentage of out-strength value of node $i$, and in-strength 
value of node $j$, assigned to the weight $w_{ij}$;
\item the residual magnitude of the out-strength of $i$ and the in-strength of $j$ is 
updated accordingly, \ie, $[\so_i]^{(n)}=(1-r_{ij}^{(n)})[\so_i]^{(n-1)}$ and 
$[\si_j]^{(n)}=(1-r_{ij}^{(n)})[\si_j]^{(n-1)}$ (with $n$ indicating the $n$-th 
iteration of the algorithm); 
\item the steps above are repeated until $\so_i\simeq \so_i$ and 
$\hsi_i\simeq \si_i$ $\forall i$ (constraints may be satisfied only 
approximately, because of the purely numerical nature of the last step of the algorithm).
\end{itemize}

Besides requiring a substantial amount of additional information with respect to other reconstruction method, 
this approach treats all nodes belonging to the same geographic area as equivalent 
(the only variability being provided by the (random) allocation of fraction of weights): 
The structure of the sub-network linking any two geographic areas is random, and may not reflect its observed counterpart (see fig. \ref{fig4}).

\subsubsection{The Minimum Density algorithm}\label{min_dens}

As we have already stressed, the main reason for defining reconstruction algorithms alternative 
to MaxEnt is the densely-connected nature of the configurations predicted by eq. (\ref{maxe}), which misrepresent the actual network 
structures (and may lead to underestimate the systemic risk). In order to 
overcome the intrinsic limitations of a complete network structure, the opposite 
approach of \emph{minimizing} the link density while satisfying the observed constraints has been 
recently devised \cite{anand2014filling}.

\begin{figure*}
\centerline{
\includegraphics[width=0.8\textwidth]{fig1.pdf}
}
\centerline{
\includegraphics[width=0.8\textwidth]{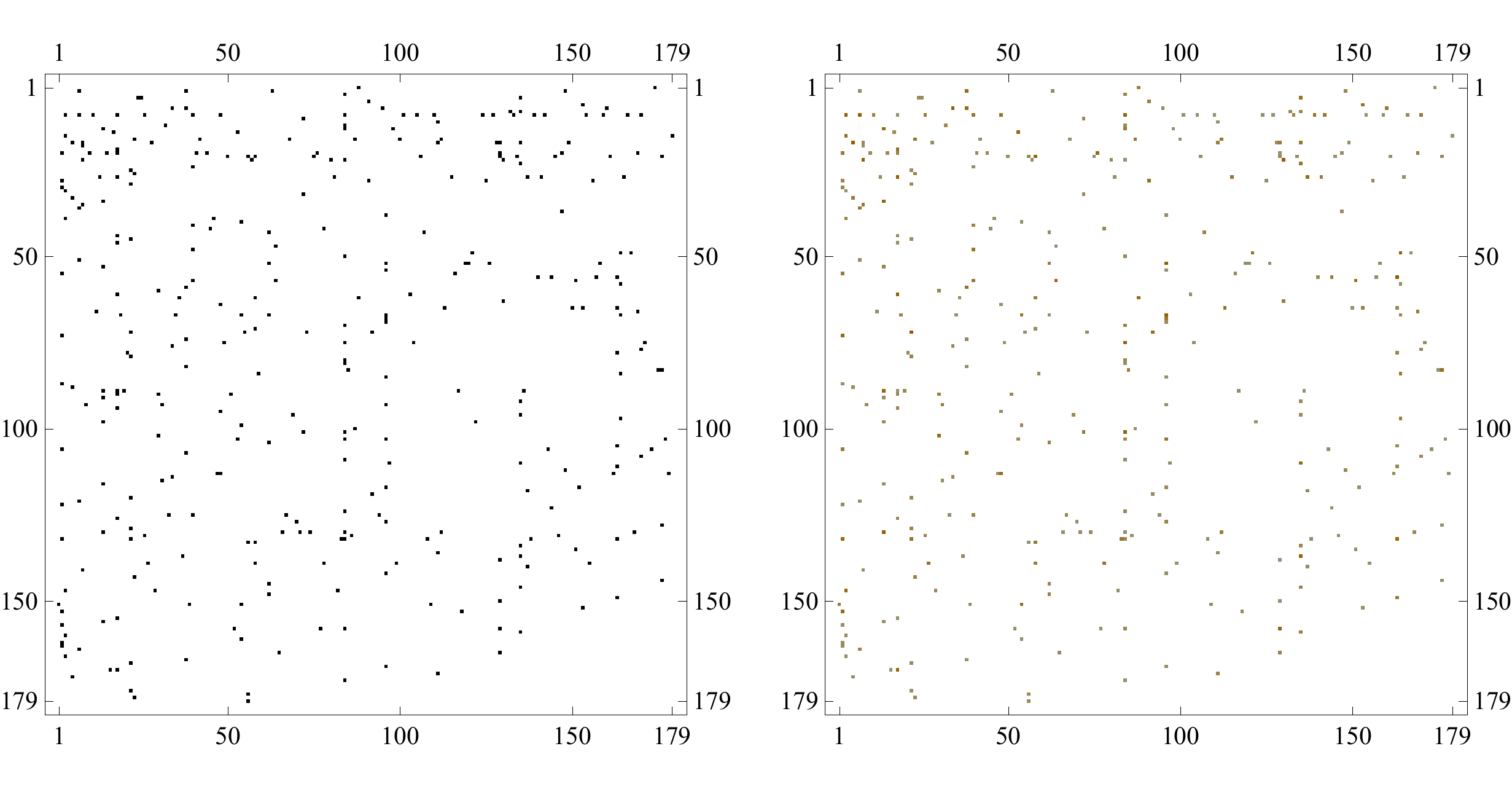}
}
\caption{Comparison between the observed adjacency matrix of the eMID network in 
2003 (top panels) and its reconstructed version according to the Minimum Density method 
described in section \protect{ \ref{min_dens}} (bottom panels).
Left panels represent binary adjacency matrices with black/white denoting the presence/absence of connections, 
whereas, right panels represent weighted adjacency matrices with color intensity denoting the weight of connections.}
\label{fig5}
\end{figure*}

Contrarily to MaxEnt, which evenly shares the marginals across all connections, 
the Minimum Density (MD) algorithm allocates the marginals 
over the minimum possible number of links (see fig. \ref{fig5}). MD does not rest upon 
the maximization of Shannon entropy, but on an optimization principle based on minimizing 
the cost of maintaining connections. The algorithm, in fact, works as follows.

\begin{itemize}
\item Define deviations from the marginals to be satisfied

\begin{equation}
\Delta_{\so_i}^{(n)}=\hso_i-[\so_i]^{(n)}=\hso_i-\sum_{j(\neq 
i)=1}^Nw_{ij}^{(n)},
\end{equation}
\begin{equation}
\Delta_{\si_i}^{(n)}=\hsi_i-[\si_i]^{(n)}=\hsi_i-\sum_{j(\neq 
i)=1}^Nw_{ji}^{(n)}
\end{equation}
where $w^{(n)}$ is the matrix obtained at the $n$-th iteration of the MD algorithm.

\item Choose a pair of nodes according to the probability coefficients

\begin{equation}
q_{ij}^{(n)}\propto\max\left\{\frac{\Delta_{\so_i}^{(n)}}{\Delta_{\si_j}^{(n)}},
\frac{\Delta_{\si_j}^{(n)}}{\Delta_{\so_i}^{(n)}}\right\}
\end{equation}
that privilege pairs of nodes where either the ``out-strength deficit'' of node $i$ is 
large with respect to the ``in-strength deficit'' of node $j$ or {\em vice-versa}.

\item Link the two selected nodes via a connection weighting

\begin{equation}\label{eq.wMD}
w_{ij}^{(n)}=\min\{\Delta_{\so_i}^{(n)},\Delta_{\si_j}^{(n)}\},
\end{equation}
hence corresponding to the largest volume that this pair of nodes can exchange. This 
step, coupled with the previous one, ensures that each new link is assigned the 
maximum possible weight needed to satisfy either the marginal $\hso_i$ or the 
marginal $\hsi_j$. Moreover, as the authors explicitly notice, this updating 
rule also induces a disassortative topology, in order to reproduce a structural 
feature observed in many real-world networks.

\item Decide if the proposed update must be retained, by evaluating the 
objective function

\begin{equation}
V(\mW^{(n)})=-cL^{(n)}-\sum_{i=1}^N\left[\alpha_i\left(\Delta_{\so_i}^{(n)}
\right)^2+\beta_i\left(\Delta_{\si_i}^{(n)}\right)^2\right],
\end{equation}
with $c$ quantifying the cost of establishing a link. If 
$\Delta_V^{(n)}=V(\mW^{(n)}+w_{ij}^{(n)})>V(\mW^{(n-1)})$ the link 
is retained (note that networks with lower densities have larger values of 
$V$). If instead $\Delta_V^{(n)}<0$, the likelihood of observing the resulting 
configuration is evaluated, \ie, the proposed weight is retained with the Metropolis-Hasting probability 
$P(\mW^{(n)}+w_{ij}^{(n)})/P(\mW^{(n-1)})\propto 
e^{\Delta_V^{(n)}}$.

\item Repeat the aforementioned steps until all marginals have been allocated.
\end{itemize}

As the other probabilistic methods discussed so far, the MD method can be employed to generate a whole ensemble of networks, 
characterized by a value of link density close to the minimum possible one (close because of the variability introduced by the last step) but 
different topological structures. Note however that real networks with such a low density values are rarely observed. 
Thus, the main rationale behind the algorithm is not network reconstruction. 
Rather, the algorithm can be very useful to find an upper bound to systemic risk. 
As such, it can be successfully employed in conjunction with MaxEnt---which, instead, 
provides the lower bound, in order to obtain the interval in which the true value of systemic risk must lie.
Additionally, in order to test for the supposed dependence of systemic risk on the network 
density, the authors also provide a more general rule for updating weights, by replacing eq. (\ref{eq.wMD}) 
with $w_{ij}^{(n)}=\theta\min\{\Delta_{\so_i}^{(n)},\Delta_{\si_j}^{(n)}\}$: the parameter $\theta\in[0,1]$ 
is used to allocate percentages of marginals, thus relaxing the request of creating a network with 
minimum density. By tuning $\theta$, the whole range of link density values can be explored.

\section{Testing the network reconstruction}\label{secfin}

Now that we described a number of algorithms aiming at reconstructing a given network structure, 
we focus on some useful methods to \emph{test} the effectiveness of the achieved reconstruction. 
In particular, three different kinds of indicators will be considered, of \emph{statistical}, \emph{topological} and \emph{dynamical} nature.

\subsection{Statistical indicators}

With this name, we refer to the entries of the so-called {\it confusion matrix}, 
the $4\times4$ table whose elements are the number of {\it true positives}, 
{\it false negatives}, {\it true negatives} and {\it false positives}. The use 
of these indicators is justified upon considering that, from a purely 
topological perspective, a reconstruction method acts as a binary classifier 
``deciding'' if a given pair of nodes should be linked or not, and whose 
performance can be thus evaluated in terms of the indices above \cite{fawcett2006roc}.

To be more explicit, consider the problem of retrieving the position of 
both present and missing links of an observed binary matrix $\hmA$. 
Denoting an individual network obtained by a given reconstruction method $\mA$, for each node pair four different alternatives are possible: 
{\bf a)} $\ha_{ij}=a_{ij}=1$: in this case, an observed link has been correctly predicted 
(we have a {\it true positive}); {\bf b)} $\ha_{ij}=1$ but $a_{ij}=0$: in 
this case, an observed link has been incorrectly predicted as being missing (we 
have a {\it false negative}); {\bf c)} $\ha_{ij}=a_{ij}=0$: in this case, 
a missing link has been correctly predicted (we have a {\it true 
negative}); {\bf d)} $\ha_{ij}=0$ but $a_{ij}=1$: in this case, a missing link 
has been incorrectly predicted as being present (we have a {\it false 
positive}).

The total number of events within one of these four categories can be 
straightforwardly computed as follows:

\begin{equation}\label{etp}
TP=\mathbf{1}(\hmA\circ\mA)\mathbf{1}^{\text{T}},
\end{equation}
\begin{equation}\label{efn}
FN=\mathbf{1}(\hmA\circ(\mathbf{I}-\mA))\mathbf{1}^{\text{T}},
\end{equation}
\begin{equation}\label{etn}
TN=\mathbf{1}((\mathbf{I}-\hmA)\circ(\mathbf{I}-\mA))\mathbf{1}^{
\text{T}},
\end{equation}
\begin{equation}\label{efp}
FP=\mathbf{1}((\mathbf{I}-\hmA)\circ\mA)\mathbf{1}^{\text{T}}
\end{equation}
where the symbol $\circ$ indicates the element-wise product of two matrices, 
$\mathbf{1}=(1,1\dots1)$ is the $N$-th dimensional row-vector whose 
entries are all ones and $\mathbf{I}$ is the $N\times N$ matrix whose generic 
entry reads $I_{ij}=1-\delta_{i,j}$. For instance, 
the total number of true positives reads $TP=\sum_{i}\sum_{j(\neq 
i)}\ha_{ij}a_{ij}$. Note that since these four indices sum up to the total number 
of nodes $N$, only three of them are independent. In particular, the last three ones can 
be compactly re-written in terms of $TP$. We have $FN=\sum_{i}\sum_{j(\neq 
i)}\ha_{ij}(1-a_{ij})=\hL-TP$, $TN=\sum_{i}\sum_{j(\neq i)}(1-\ha_{ij})(1-a_{ij})=N(N-1)-L-\hL+TP$ 
and $FP=\sum_{i}\sum_{j(\neq i)}(1-\ha_{ij})a_{ij}=L-TP$, where $\hL$ and $L$ are the total number of links in the observed 
and reconstructed network, respectively.

The four indices above provide absolute numbers which, by themselves, are of 
limited usefulness. This is the reason why the information provided by $TP$, 
$FN$, $TN$ and $FP$ is often combined to define ``relative'' indices. The first of them 
is the {\it sensitivity} (or {\it true positive rate}) \cite{fawcett2006roc}, defined as

\begin{equation}
TPR=\frac{TP}{TP+FN}=\frac{TP}{\hL}
\end{equation}
and quantifying the percentage of observed links that are correctly recovered. 
Note that for the performance a reconstruction method to be deemed satisfactory, 
the condition of a $TPR$ value close to 1 is \emph{necessary}, but not \emph{sufficient}. 
Indeed, a method that overestimates the number of links achieves a high $TPR$ value by construction 
(for a fully connected reconstructed network, it is $TPR=1$ by definition), 
but also lacks the ability to identify missing connections. The latter is quantified by 
the {\it specificity} (or {\it true negative rate}) \cite{fawcett2006roc}, defined as

\begin{equation}
SPC=\frac{TN}{FP+TN}=\frac{TN}{N(N-1)-\hL}
\end{equation}
\ie, the percentage of observed missing links that are correctly recovered. 
The {\it false positive rate} $FPR=1-SPC$ is the complementary index to $SPC$ \cite{fawcett2006roc}.

Thus, intuitively, any ``good'' reconstruction algorithm should be characterized by a 
large $TPR$ value and a low $FPR$ value (\ie, the better its performance, the 
closer the $TPR$ index to 1 \emph{and} the $FPR$ index to 0). This observation 
leads to the classical ``graphical'' way to visualize the performance of a classifier, 
by representing it as a point of coordinates $(FPR,\:TPR)$ within the unit square of coordinates $(0,0)$, $(0,1)$, $(1,1)$, $(1,0)$. 
Any perfect classifier is thus to be found on the top-left corner of the square, 
whereas, a random classifier (predicting an equal number of present and missing links) lies on its main diagonal.
The performance of a given reconstruction algorithm can then be quantified as the {\em area under the 
curve} (AUC) \cite{fawcett2006roc} identified by the three points of coordinates $(0,0)$, $(FPR,\:TPR)$, $(1,1)$. 
A perfect classifier is then characterized by an AUC of 1, a random classifier by an AUC equal to $1/2$, 
and in general for a non-random classifier it is $1/2<\text{AUC}\leq1$.

An alternative way of evaluating the performance of a reconstruction method is 
plotting its $TPR$ versus a fourth index, its {\it precision} (or {\it 
positive predicted value}) \cite{fawcett2006roc}

\begin{equation}
PPV=\frac{TP}{TP+FP}=\frac{TP}{L}
\end{equation}
which measures the percentage of correctly placed links with respect to the total 
number $L$ of predicted links. In other words, the $PPV$ index quantifies the 
``ability'' of a given classifier to predict \emph{only} true 
positives. Thus, and contrarily to the $TPR$, a large $PPV$ cannot be trivially obtained 
by dense reconstruction methods.

Finally we consider an index measure the overall performance of a classifier in correctly placing both ones and zeros: 
the {\it accuracy} \cite{metz1978roc}, defined as

\begin{equation}\label{aindex}
ACC=\frac{TP+TN}{TP+TN+FP+FN}=\frac{TP+TN}{N(N-1)}.
\end{equation}

\medskip

Whenever a reconstruction method defines an entire ensemble of candidate 
matrices, the above quantities have to be evaluated as averages over such ensemble. 
This can be done using, in eqs. (\ref{etp}-\ref{efp}), the average quantities $\avg{ a_{ij}}=p_{ij}$ 
and $\avg{ \mA}=\mathbf{P}$ instead of those corresponding to a single realization $a_{ij}$ and $\mA$: 

\begin{equation}
\avg{ TP}=\mathbf{1}(\hmA\circ\mathbf{P})\mathbf{1}^{\text{T}},
\end{equation}
\begin{equation}
\avg{ 
FN}=\mathbf{1}(\hmA\circ(\mathbf{I}-\mathbf{P}))\mathbf{1}^{\text{T}
},
\end{equation}
\begin{equation}
\avg{ 
TN}=\mathbf{1}((\mathbf{I}-\hmA)\circ(\mathbf{I}-\mathbf{P}))\mathbf
{1}^{\text{T}},
\end{equation}
\begin{equation}
\avg{ 
FP}=\mathbf{1}((\mathbf{I}-\hmA)\circ\mathbf{P})\mathbf{1}^{\text{T}
}.
\end{equation}

Using ensemble averages, the difference between dense and sparse reconstruction methods 
can be better discussed quantitatively. MaxEnt, which we take as the representative of 
dense reconstruction algorithms, satisfies a relation of the kind 
$\avg{ a_{ij}^{\mbox{\tiny ME}}}=p_{ij}=p\simeq1$ $\forall \neq j$, leading to 

\begin{equation}
\avg{ TP^{\mbox{\tiny ME}}}\simeq \hL \mbox{ and }\avg{ FP^{\mbox{\tiny ME}}}\simeq N(N-1)-\hL.
\end{equation}

As a consequence, $\avg{ PPV^{\mbox{\tiny ME}}}\simeq \hL/N(N-1)$, \ie, 
the power of the method coincides with the network density. In order to fully 
understand the importance of this result, we now consider the Directed Random 
Graph Model (DRGM), defined by the prescription 
$p_{ij}=p=\hL/N(N-1)$ $\forall i\neq j$. We have $\avg{ 
TPR^{\mbox{\tiny DRGM}}}=\avg{ FPR^{\mbox{\tiny DRGM}}}=p$ and, most 
importantly, $\avg{ PPV^{\mbox{\tiny DRGM}}}=p$. In other words, a ``random'' 
classifier is not necessarily an algorithm guessing the (binary) value of each 
entry with probability 1/2: more generally, it is a reconstruction method 
defined by the DRGM recipe, and whose $PPV$ represents a lower bound for any reconstruction algorithm. 
Notice that the MaxEnt method attains such value, 
thus confirming the weakness of its performance---unless very dense 
configurations are considered, since $\avg{ ACC^{\mbox{\tiny ME}}}=p$.

\medskip

Other indicators that have been extensively used to measure the goodness of a 
reconstruction algorithm are the Hamming distance $H$, the Jaccard similarity 
$J$, the cosine similarity $\vartheta$, the Jensen-Shannon divergence $JS$, 
between $\hmA$ and $\mA$ \cite{wang2014similarity}. 
Remarkably, whenever dealing with binary matrices, 
these indices can be rewritten in terms of the four basic quantities $TP$, $FN$, 
$TN$, $FP$:

\begin{equation}\label{hindex}
H=FN+FP,
\end{equation}
\begin{equation}\label{jindex}
J=\frac{TP}{FN+TN+FP},
\end{equation}
\begin{equation}\label{tindex}
\vartheta=\frac{TP}{\hL},
\end{equation}
\begin{equation}
JS=\frac{FN}{2\hL}\ln(2\hL)+\frac{FP}{2L}\ln(2L)-TP\left(\frac{\hL+L}{2\hL L}
\right)\ln\left(\frac{\hL+L}{2\hL L}\right)-\frac{\ln(\hL L)}{2}
\end{equation}
(whenever a whole ensemble of configurations must be considered, 
the expressions above must be averaged accordingly).

\medskip

Moving further, for testing the effectiveness of a given algorithm 
in reconstructing weights, a tempting possibility would be to simply extend 
some of the measures defined in the binary case to the weighted one. 
However, the non-binary nature of the entries makes it difficult to devise a best choice. 
Nevertheless, the most popular metric is the weighted counterpart of the cosine similarity 
\cite{squartini2017network,wang2014similarity}, reading 

\begin{equation}\label{csw}
\vartheta_w=\frac{\mathbf{1}(\hmW\circ\mW)\mathbf{1}^{\text{T}}}{||\hmW||_2\:||\mW||_2}
\end{equation}
with $||\dots||$ indicating the $\text{L}_2$ entry-wise matrix norm.
In other words, these two matrices are treated as vectors of real numbers, 
whose overlap is approximated by a fictitious angle with values ranging from -1 indicating maximum dissimilarity 
to $+1$ meaning exact similarity, and with $0$ indicating absence of correlation. 

Additional indices are the $\text{L}_1$ and the $\text{L}_2$ entry-wise matrix distances 
\cite{gandy2017adjustable}, respectively defined as

\begin{eqnarray}\label{nlw}
||\hmW-\mW||_1&=&\sum_{i=1}^N\sum_{j(\neq 
i)=1}^N|\hw_{ij}-w_{ij}|,\\
||\hmW-\mW||_2&=&\sqrt{\sum_{i=1}^N\sum_{j(\neq 
i)=1}^N(\hw_{ij}-w_{ij})^2}
\end{eqnarray}
and the so-called ``error measure'' \cite{baral2012estimating} reading

\begin{equation}\label{errmm}
\epsilon=\frac{\sum_{i=1}^N\sum_{j=1}^N|\hw_{ij}-w_{ij}|}{\sum_{i=1}^N\sum_{
j=1}^Nw_{ij}}.
\end{equation}

When ensemble methods are considered, eqs (\ref{csw})-(\ref{errmm}) can 
still be used, upon substituting $\mW$ with $\avg{\mW}$. 
Notice, however, that the major drawback of norm-like quantities lies in the fact that they are 
unbounded, which makes it difficult to employ them for comparing different candidate matrices.

\subsection{Topological indicators}

The second set of indicators is represented by quantities of topological nature 
providing a ``coarse-grained'' description of the network under consideration, such as 
degree-degree correlations and the mesoscale community structure. 

\subsubsection{Testing weights reconstruction}

The most straightforward way to compare observed weights with their corresponding estimates is to scatter-plot them. 
However in order to consistently compare only realized connections, 
it is preferable to scatter (realized) observed weights versus \emph{conditional} weights

\begin{equation}
\avg{ w_{ij}|a_{ij}=1}=\frac{\avg{ w_{ij}}}{p_{ij}}
\end{equation}
which to some extent encode the (available) structural information. This prescription is particularly useful 
to compare algorithms generating the same expected weights but predicting different topological structures 
(\eg, MaxEnt and one of the exact density methods) \cite{squartini2017network}.

\subsubsection{Testing higher-order patterns reconstruction}\label{pattern}

Besides reconstructing link weights, a good reconstruction method is also expected to reproduce the 
higher-order trends characterizing the network $\hmG$ under observation. 
To this aim, the observed value of a generic quantity of interest $X(\hmG)$ 
is usually compared with the corresponding prediction obtained by the reconstruction algorithm. 
Importantly, whenever dealing with ensemble methods, the entire set $\ensG$ of configurations must be accounted for, 
whence the need to find statistical measures compactly describing all possible (alternative) outcomes. 
The most basic and useful choices are the ensemble average and standard deviation of $X$ \cite{squartini2011analytical}, namely

\begin{equation}\label{ensave}
\avg{ X}=\sum_{\mG\in\ensG}X(\mG)P(\mG),
\end{equation}
\begin{equation}\label{stdvec}
\sigma_X=\sqrt{\sum_{\mG\in\ensG}(X(\mG)-\avg{ 
X})^2P(\mG)}.
\end{equation}

The evaluation of eqs. (\ref{ensave}) and (\ref{stdvec}) in principle requires the knowledge of the whole ensemble $\ensG$. 
Since listing all the configurations belonging of the ensemble is simply not feasible, 
analytical or numerical techniques can be used to tackle this problem. 

In the first case, a simple remedy is provided by the \emph{delta method} \cite{casella2002statistical}, 
based on the Taylor expansion of the observed value $X(\mG)$ around the expected value of the variables it depends on:

\begin{equation}\label{taylorvec}
X(\mG)=X(\avg{\mG})+\sum_{i=1}^N\sum_{j(\neq 
i)=1}^N(g_{ij}-\avg{ g_{ij}})\left(\frac{\partial X}{\partial 
g_{ij}}\right)_{\mG=\avg{\mG}}+\dots
\end{equation}
Equation (\ref{taylorvec}) is a ``tensorial'' Taylor expansion, since 
each entry of the adjacency matrix $\mG$ is an independent random 
variable. By taking the expected value of both sides of eq. (\ref{taylorvec}), 
one recovers the delta method prescription to calculate the expected value of 
the quantity $X$, \ie,

\begin{equation}\label{meanvec}
\avg{ X}\simeq X(\avg{ \mG}).
\end{equation}
The standard deviation $\sigma_X$ is then estimated by inserting 
eqs. (\ref{taylorvec}-\ref{meanvec}) into eq. (\ref{stdvec}):

\begin{equation}\label{taylorvec2}
\sigma_X\simeq\sqrt{\sum_{i=1}^N\sum_{j(\neq i)=1}^N\sum_{t=1}^N\sum_{s(\neq 
t)=1}^N\text{Cov}[g_{ij},g_{ts}]\left(\frac{\partial X}{\partial 
g_{ij}}\frac{\partial X}{\partial 
g_{ts}}\right)_{\mG=\avg{\mG}}}.
\end{equation}

Remarkably, eqs. (\ref{meanvec}) and (\ref{taylorvec2}) are exact in the case of 
linear constraints represented by degrees and strengths. Other examples of topological 
quantities whose ensemble averages and standard deviations can be computed 
exactly are the so-called \emph{dyadic motifs}, defined by the expressions 

\begin{eqnarray}
N^{\leftrightarrow}&=&\sum_{i=1}^N\sum_{j(\neq i)=1}^Na_{ij}a_{ji},\\
N^{\rightarrow}&=&\sum_{i=1}^N\sum_{j(\neq i)=1}^Na_{ij}(1-a_{ji}),\\
N^{\nleftrightarrow}&=&\sum_{i=1}^N\sum_{j(\neq i)=1}^N(1-a_{ij})(1-a_{ji}).
\end{eqnarray}
Upon considering that distinct dyads are independent, the expected value and standard deviations 
of the expressions above become

\begin{eqnarray}
\avg{ N^{\leftrightarrow}}&=&\sum_{i=1}^N\sum_{j(\neq 
i)=1}^Np_{ij}p_{ji},\\
\avg{ N^{\rightarrow}}&=&\sum_{i=1}^N\sum_{j(\neq 
i)=1}^Np_{ij}(1-p_{ji}),\\
\avg{ N^{\nleftrightarrow}}&=&\sum_{i=1}^N\sum_{j(\neq 
i)=1}^N(1-p_{ij})(1-p_{ji})
\end{eqnarray}
\begin{eqnarray}
\sigma_{N^{\leftrightarrow}}&=&\sum_{i=1}^N\sum_{j(\neq 
i)=1}^N2p_{ij}p_{ji}(1-p_{ij}p_{ji}),\\
\sigma_{N^{\rightarrow}}&=&\sum_{i=1}^N\sum_{j(\neq 
i)=1}^Np_{ij}(1-p_{ji})[1-p_{ij}(1-p_{ji})-p_{ji}(1-p_{ij})],\\
\sigma_{N^{\nleftrightarrow}}&=&\sum_{i=1}^N\sum_{j(\neq 
i)=1}^N2(1-p_{ij})(1-p_{ji})[1-(1-p_{ij})(1-p_{ji})]
\end{eqnarray}

Other quantities, however, can be dealt with far less ease. As last resort, 
it is possible to proceed numerically by explicitly sampling $\ensG$. Once a 
(properly sampled) subset $\mathcal{\tilde{G}}$ has been obtained, the ensemble average 
$\avg{ X}$ can be approximated by the arithmetic mean

\begin{equation}
\avg{ X}\simeq 
\overline{X}=\sum_{\mG\in\mathcal{\tilde{G}}}X(\mG)F(\mG),
\end{equation}
where $P(\mG)$ is replaced by the sampling frequency $F(\mG)=\frac{N_\mG}{|\mathcal{\tilde{G}}|}$, 
and $N_\mG$ is the number of networks in the sample whose adjacency matrix is equal to $\mG$. 
Analogously, the standard deviation $\sigma_X$ becomes

\begin{equation}
\sigma_X\simeq\sqrt{\sum_{\mG\in\mathcal{\tilde{G}}}(X(\mG
)-\overline{X})^2F(\mG)}.
\end{equation}

Finally, once an estimate for $X$ (together with some measure about its uncertainty) has been obtained, 
the comparison between $X(\hmG)$ and $\avg{ X}$ can be carried out 
by checking whether the observed value $X(\hmG)$ lies within the region delimited by the values $\avg{ X}\pm z\sigma_X$, 
with $z$ set to determine a desired level of statistical significance. More compactly, this is expressed as the z-score\footnote{The z-scores assumes 
a Gaussian distribution for the random variable under consideration. If deviations form this hypothesis are expected, a different statistical test should be employed.}

\begin{equation}
z_X=\frac{X(\mG)-\avg{ X}}{\sigma_X}
\end{equation}
which measures the difference between the observed and the expected value in units of standard deviation. 
A z-score whose numerical value is close to zero then indicates that the chosen reconstruction algorithm generates an expected value 
of $X$ that is close enough to the observed one: More generally, 
whenever $|z|\leq z_{th}$ (with $z_{th}$ usually being 1, 2 or 3), the discrepancy 
between the two values cannot be considered significant 
(with a confidence interval of 0.683, 0.954 and 0.997 respectively). 
Whenever dealing with the models defined within the ERG formalism, 
this further implies that the structure of the real network 
$\hmG$, proxied by the quantity $X$, is completely explained by the 
(topological information encoded into the) imposed constraints. By contrast, if 
$|z|> z_{th}$ the observed value $X(\hmG)$ lies outside the 
chosen confidence interval: the structure of the observed network determining $X$ 
is not completely explained by the specific constraints imposed, 
and further model specifications should be used (\ie, additional or more complex constraints) \cite{squartini2011analytical,luu2017structural}.

\subsubsection{What may and what may not be reconstructed}\label{wmb}

Even when a given method does not provide a good reconstruction of an observed 
network, it is nevertheless useful to understand what kind of information 
it can provide. To this end, let us consider the z-score again. Whenever $z_X$ 
is significantly positive, $X$ is said to be over-represented in $\hmG$, 
meaning that the network under analysis shows a positive tendency towards it. 
For instance, chain-like motifs are significantly abundant (\ie, they are found 
more often than expected) in food-webs. Analogously, whenever the z-score is significantly negative, 
$X$ is said to be under-represented. Again in the case of food webs, 
loop-like motifs are significantly missing (\ie, they are found less often than expected).

z-scores (and statistical tests in general) provide information also on ongoing structural changes 
of a given network. A particularly interesting issue concerns the detection of 
early-warning signals of upcoming critical events. As shown in 
\cite{squartini2013early}, this can be done by computing $z_X$ for 
each temporal snapshot of the considered system\footnote{In 
\cite{squartini2013early}, the monitored quantity is precisely the abundance of 
dyadic motifs, $X=N^{\leftrightarrow}, N^{\rightarrow}, N^{\nleftrightarrow}$.} 
and, then, plotting $z_X(t)$ versus $t$. As far as the discrepancy between 
observed and expected values evolves ``smoothly'' from out-of-equilibrium to 
equilibrium states (or {\em vice-versa}), early-warning signals can be possibly detected 
\cite{saracco2016detecting,gualdi2016statistically}, 

An important aspect of a network to be tested for statistical significance is, without doubts, its mesoscale organization into modules or communities. 
An approach similar to what presented in the previous subsection works as follows \cite{bifone2016surprise}. 
Suppose that we know the community organization of the network, characterized by $\Lambda$ total intra-community links and $\Pi$ intra-community pair of nodes. 
The probability that a random network with $N$ nodes and $L$ links has at least these values of $\Lambda$ and $\Pi$ 
derives from an urn model without reinsertion, and thus is given by the inverse cumulative hyper-geometric distribution
\begin{equation}
\Sigma=\sum_{l=\Lambda}^{L}\frac{\binom{\Pi}{l}\binom{N(N-1)-\Pi}{L-l}}{\binom{N(N-1)}{L}}.
\end{equation}
Hence the smaller the value of $\Sigma$, known as \emph{surprise}, the more significant the mesoscale organization of the considered network \cite{bifone2016surprise}.

More refined benchmarks for mesoscale structures are provided by the Stochastic Block Model (SBM) 
\cite{karrer2011stochastic} and its degree-corrected version (dcSBM) \cite{fronczak2013exponential}. 
The effectiveness of these models in reproducing block structures in economic and financial networks has been investigated in \cite{barucca2016disentangling}. 
In particular, these models allow to interpolate between two alternative kinds of partition structures: core-periphery and bipartite. 
In the context of interbank networks, a core-periphery structure indicates the existence of a set of core-banks acting as intermediaries between periphery-banks, 
whereas, a bipartite structure represents an intermediaries-free market, with banks trading (exclusively) according to their preferences for the counterparties \cite{craig2014interbank}.

Network configurations characterized by a structure of $m$ blocks can be represented by an $m\times m$ symmetric matrix---called the \emph{affinity matrix}, 
whose entries represent the density of links within and between modules:

\begin{center}
\begin{minipage}{\textwidth}
	\centering
	\[\mathcal{A}=\left( 
	\begin{array}{cccc}
	\rho_{g_1g_1} & \rho_{g_1g_2} & \dots & \rho_{g_1g_m}\\
	\rho_{g_1g_2} & \rho_{g_2g_2} & \dots & \rho_{g_2g_m}\\
	\vdots & \vdots & \ddots & \vdots\\
	\rho_{g_1g_m} & \rho_{g_2g_m} & \dots & \rho_{g_mg_m}\\
	\end{array}
	\right).\]
\end{minipage}
\end{center}
Using this representation, the SBM and dcSBM respectively assume that connection probability between any two nodes $i$ and $j$ assume the form
\begin{equation}
p_{ij}^{\mbox{\tiny SBM}}=\rho_{g_ig_j},\qquad 
p_{ij}^{\mbox{\tiny dcSBM}}=\rho_{g_ig_j}x_ix_j,
\end{equation}
where $x_i$ is the parameter controlling for the degree of generic node $i$.
Upon varying the model parameters, a whole range of different topologies can be 
generated. In the simple case of the dcSBM with only two blocks $g_1$ and $g_2$, 
the authors of \cite{barucca2016disentangling} impose a ``background'' bipartite-like 
structure with $\rho_{g_1g_2}>\rho_{g_1g_1}=\rho_{g_2g_2}$, while progressively rising 
the degree heterogeneity of nodes belonging to $g_1$ and $g_2$ respectively. 
Upon running a belief-propagation (BP) algorithm 
\cite{decelle2011inference,newman2014core}, the likelihood of the network

\begin{equation}
\mathcal{L}(\mA|\vec{x})=\ln\left[\prod_{i=1}^N\prod_{j(<i)=1}^N 
p_{ij}^{a_{ij}}(1-p_{ij})^{1-a_{ij}}\right]
\end{equation}
highlights the transition from a purely bipartite structure to a purely 
core-periphery structure. The consistency check is done by comparing the 
numerical value of the likelihood function $\mathcal{L}^{\text{BP}}$ with that 
of $\mathcal{L}^{\text{SBM}}$ and $\mathcal{L}^{\text{dcSBM}}$.

This transition is not a surprise if the generative model is known. 
However, some care must be adopted when studying real-world networks. 
Indeed for the empirical interbank networks considered in \cite{barucca2016disentangling}, 
the kind of emerging mesoscale organization ``depends'' on the used benchmark. 
In particular, while the SBM-induced belief-propagation reveals a bipartite structure 
($\rho_{g_1g_2}>\rho_{g_1g_1}>\rho_{g_2g_2}$) on daily data aggregation scales and a core-periphery 
structure ($\rho_{g_1g_1}>\rho_{g_1g_2}>\rho_{g_2g_2}$) on longer time scales, the 
dcSBM-induced belief-propagation always reveals a bipartite structure.

As also noticed elsewhere \cite{yan2013block}, this behavior is likely due 
to the tendency of SBM to detect homogeneous modules (\eg, blocks of nodes 
with large degree and blocks of nodes with low degree \cite{karrer2011stochastic}). 
As a consequence, the dcSBM should be preferred whenever the degree heterogeneity is strong. 
In the case of sparse networks, it is
\footnote{This estimation can be obtained by maximizing the sparse-case likelihood function 
$\mathcal{L}(\hmA|\vec{\lambda})\simeq\sum_i\sum_{j(<i)}(a_{ij}\ln 
p_{ij}-p_{ij})$ with $p_{ij}=\rho_{g_ig_j}\lambda_i\lambda_j$ 
\cite{karrer2011stochastic}.}

\begin{equation}
p_{ij}^{\mbox{\tiny dcSBM}}\simeq\hL_{g_ig_j}\left(\frac{\hat{k}_i}{\hat{K}_{g_i}}
\right)\left(\frac{\hat{k}_j}{\hat{K}_{g_j}}\right),
\end{equation}
with $\hL_{g_ig_j}$ indicating the (observed) total number of links between 
blocks $g_i$ and $g_j$, and $\hat{K}_{g_i}$ indicating the (observed) total 
degree of nodes belonging to block $g_i$.

It is also worth mentioning that a block-wise extension of the Configuration 
Model can be directly defined by an Hamiltonian constraining the block-specific degree sequences \cite{fronczak2013exponential}. 
This leads to link probability coefficients depending on the imposed block-structure. 
That is, the connection probability between node $i$ belonging to block $r$ and node $j$ belonging to block $s$ reads

\begin{equation}\label{ergblock}
p_{ij}^{rs}=\frac{x_i^{rs}x_j^{rs}}{1+x_i^{rs}x_j^{rs}},
\end{equation}
where the Lagrange multipliers are now expressed in a tensorial form, and have to be numerically determined by solving block-specific likelihood conditions.

\subsection{Dynamical indicators}

The third family of indicators include metrics reflecting the outcome of diffusion processes over the network. 
Here in particular we deal with distress propagation across a financial network and the related issue of \emph{systemic 
risk}, \ie, the possibility that a local event triggers a global instability through a cascading effect. 
This issue received a lot of attention especially after the global financial crisis of 2007/2008. 
Since then, it was realized that the complex pattern of interconnections between financial institutions makes the system as a whole 
inherently fragile: those connections constitute the channels through which financial distress can spread which, eventually, 
lead to amplification effects like default cascades 
\cite{battiston2016price,allen2000financial,brunnermeier2009deciphering,
lau2009assessing,gai2010contagion,haldane2011systemic,battiston2016complexity,bardoscia2017pathways}. 

Indeed, while \emph{interconnectedness} implies \emph{diversification} and as 
such helps reducing the individual risk, it also makes the system as a whole 
more vulnerable 
\cite{beale2011individual,corsi2013when,glasserman2015contagion}. As a 
consequence, both researchers and regulators have started to pay attention to 
the structural features of financial systems 
\cite{boss2004network,nier2007network,iori2008network,
krause2012interbank,georg2013effect}, with the aim of properly estimating the 
\emph{systemicness} (or \emph{impact}) and the \emph{vulnerability} of each 
bank. While the former represents the total loss induced on the system by the 
distress of that bank, the latter is the loss experienced by that bank 
when the whole system is under distress 
\cite{lau2009assessing,greenwood2015vulnerable}.

The shortcoming of requiring data on individual exposures to obtain these 
indicators represents the very motivation for the use of effective network 
reconstruction techniques in finance. In particular, reconstruction methods can generate 
scenarios that are compatible with the available information and, as such, can 
be employed to test the resilience of both single institutions and the system as 
a whole. Clearly, the operative definition of any \emph{dynamical (financial)}  indicator depends on 
the underlying model of shock-propagation assumed. While the literature on the 
topic is extensive, here we only outline basic concepts and provide a few 
illustrative examples that, in our opinion, physicists can easily become 
familiar with. For an in-depth analysis of network-based systemic risk models, 
we remand to the recent reviews and books 
\cite{acemoglu2015systemic,bougheas2015complex,hurd2016contagion}.

\subsubsection{Balance sheets and financial networks}\label{balshee}

The financial position of a given bank $i$ is summed up by its balance sheet, 
which reports total assets $a_i$ and liabilities $l_i$ at a given date. Assets are 
resources with a positive economic value (such as loans, derivatives, stocks, 
bonds, options, real estates, etc.), whereas liabilities have a negative 
economic value (obligations, debits, customer deposits, accrued expenses, etc.). 
The net difference between assets and liabilities defines the bank equity

\begin{equation}
e_i=a_i-l_i
\end{equation}
and the bank is said to be solvent as long as its equity is positive. Indeed, 
negative equity means insolvency, as the bank cannot pay back its liabilities 
even by selling all its assets. In the dedicated literature 
\cite{upper2011simulation,gai2010contagion,nier2007network}, insolvency is 
usually considered as a proxy for default, which in turn occurs when the bank 
actually fails to fulfill a legal obligation.

Let us now consider $N$ banks and $Q$ securities (stocks, bonds, options, etc...). 
The detailed composition of the balance sheet of $i$ can be described, 
schematically, as follows. On the assets side we find the loans granted to other 
banks $\{w_{ij}\}_{j=1}^N$, the securities constituting the investment portfolio 
$\{\omega_{i\alpha}\}_{j=1}^Q$ and other (\ie, fixed and intangible) assets $a_i^{o}$:

\begin{equation}
a_i=\sum_{j(\neq i)=1}^N w_{ij}+\sum_{\alpha=1}^Q \omega_{i\alpha}+a_i^{o}.
\end{equation}
On the liabilities side, there are the loans granted from other banks 
$\{w_{ji}\}_{j=1}^N$, as well as debts to outside parties $l_i^{o}$:

\begin{equation}
l_i=\sum_{j(\neq i)=1}^N w_{ji}+l_i^{o}. 
\end{equation}

Hence, financial networks naturally emerge from the interconnections 
between banks balance sheets. In particular, an interbank lending market 
is represented by a monopartite network of loans between banks, whereas an 
equity market is represented by the bipartite network between banks and owned 
securities. Financial shocks propagate across these networks according to three 
main mechanisms that we describe below.

\subsubsection{Counterparty risk and credit shocks}

Bilateral exposures between banks expose them to what is perhaps the most 
intuitive channel of financial contagion: \emph{counterparty risk}. Suppose that 
bank $i$ undergoes significant losses and defaults, failing to meet its 
contractual obligations: this results in actual losses for the creditors of $i$, 
commonly labeled as \emph{credit shocks} \cite{lau2009assessing,krause2012interbank}. 
In particular, bank $j$ undergoes a 
loss equal to $\varphi w_{ju}$, where $\varphi$ indicates the amount of loss 
given default\footnote{For uncollateralized market (mostly studied in the 
literature), $\varphi=1$. If, instead, any central counterparty guarantees for 
interbank loans, $\varphi=0$ and, in principle, banks face no losses - and the 
risk goes to the counterparty.}. Bank $j$ can, in turn, default if this loss 
exceeds its equity $e_j$, originating a new wave of credit shocks. 

Credit shocks have been extensively studied in literature 
(see, \eg, \cite{eisenberg2001systemic,furfine2003interbank,rogers2013failure}) 
Here we briefly describe the DebtRank model \cite{battiston2012debtrank,bardoscia2015debtrank,barucca2016network}, 
whose peculiarity consists in allowing for credit shocks propagation also in 
absence of defaults, provided that balance sheets are deteriorated. Indeed, 
losses suffered by financial institutions from credit shocks are not only due to 
the actual default of counterparties, but also to the mark-to-market revaluation of 
obligations after the deterioration of counterparties credit-worthiness (counterparties 
which are ``closer'' to default are less likely to pay back their debts at maturity). 
In particular, the Debt Rank assumes that relative 
changes of equity translate linearly into relative changes of asset values, 
resulting in an \emph{impact} of bank $i$ on bank $j$ equal to $\varphi 
w_{ji}/e_j$. Individual banks losses are then obtained by iteratively spreading 
the individual banks distress levels weighted by the potential wealth affected.

Formally, the dynamics of the model consists of several rounds of shock 
propagation, hereafter index by $t$. The state of bank $i$ at each $t$ can be 
compactly described by the relative change of equity $h_i(t)=1-e_i(t)/e_i(0)$, 
which ranges between $0$ and $1$. By definition, $h_i(t)=0$ when no equity 
losses occurred for the bank, $h_i(t)=1$ when the bank defaults and $0<h_i(t)<1$ 
for intermediate distress levels. Starting at $t=0$ from 
$h_i(0)=0,\:\forall\:i$, the first-round losses at $t=1$ consist of exogenous 
shocks decreasing the equity of some banks: $0\le h_i(1)\le 1,\:\forall\:i$. 
Later-round losses from subsequent credit shocks are then computed as:

\begin{equation}\label{eq:acca}
h_i(t+1)=\min\left\{1,\; h_i(t)+\sum_{j\in\mathcal{A}(t)}\frac{\varphi 
w_{ij}}{e_i}\,[h_j(t)-h_j(t-1)]\right\}
\end{equation}
where $\mathcal{A}(t)=\{j:h_j(t-1)<1\}$ is the set of banks that have not 
defaulted up to time $t-1$ and, thus, can still spread their financial distress. 
The dynamics stops at convergence (say $\tilde{t}$), \ie, when no more banks can 
propagate their distress and $\mathcal{A}(\tilde{t})=\emptyset$. Individual bank 
indicators are then computed over an appropriate ensemble of initial conditions 
of the dynamics. In particular, by denoting as $h_j(\tilde{t}|i)$ the final relative 
equity change of $j$ when the initial condition is the single default of bank $i$ 
(\ie, $h_i(1)=1$ and $h_j(1)=0 \forall j\neq i$), whose relative systemic importance is $\nu_i=e_i(0)/\sum_je_j(0)$:

\begin{itemize}
 \item the {\em impact} of bank $i$ is the relative equity loss experienced by 
the system from the initial default of $i$
\begin{equation}
I_i=\frac{\sum_{j(\neq i)=1}^N h_j(\tilde{t}|i)\nu_j}{1-\nu_i};
\end{equation}
\item the {\em vulnerability} of $j$ is the relative equity loss for that 
bank averaged over the initial defaults of all other banks
\begin{equation}
V_i=\frac{\sum_{j(\neq i)=1}^N h_i(\tilde{t}|j)}{N-1}.
\end{equation}
\end{itemize}
In both cases, first-round losses caused by exogenous shocks are explicitly 
excluded to account for network effects only.

\subsubsection{Rollover risk and liquidity shocks}

A more involving channel of financial contagion is related to \emph{rollover 
risk}, faced by banks in need to refinance their debt which is about to mature 
with new debt 
\cite{lau2009assessing,cifuentes2005liquidity,kapadia2012quantifying,
anand2012rollover}. In periods of financial distress, diffuse 
worries on future losses and counterparty credit-worthiness can lead banks to 
adopt a micro-prudential \emph{liquidity hoarding} policy by withdrawing liquidity from the market 
\cite{gale2013liquidity,brunnermeier2009market,acharya2010precautionary}. 
In this situation, banks which are short on liquidity may be unable to borrow all the 
needed money from the market and be forced to sell their illiquid assets. 
However, when assets sales are widespread, the market demand cannot cover for 
the supply: the market prices of illiquid assets decreases (a circumstance known 
as \emph{fire sales}), resulting in effective losses for banks labeled as 
\emph{liquidity shocks} 
\cite{lau2009assessing,krause2012interbank}. 
Note that fire sales spillovers may also originate by the leverage targeting 
policy adopted by banks 
\cite{battiston2016leveraging,greenwood2015vulnerable,adrian2010liquidity} (that is, 
banks may respond to exogenous shocks by selling assets in order to maintain the 
desired level of debt over equity) and be exacerbated 
by indirect exposures between banks due to common assets holdings (see below). 
In any event, liquidity shocks do represent an important dimension of systemic 
risk, comparable to credit shocks \cite{glasserman2015contagion} but traveling 
in the opposite direction.

Suppose that bank $i$ suffers significant losses and defaults, thus stopping its 
liquidity provision to the market. Bank $j$, which would have rolled its debt 
over $j$, replenishes a fraction $(1-\psi)$ of the lost funding with its liquid 
assets or from other sources\footnote{In periods of severe distress, exceptional 
monetary policies are usually implemented and central banks become lenders of 
last resort, corresponding to the case $\psi=0$.} and the remaining fraction 
$\psi$ by selling its illiquid assets. The latter, however trade at a discount, 
so that $j$ must sell assets worth $(1+\chi)\psi w_{ij}$ in book value terms, 
corresponding to an overall loss of $\chi\psi w_{ij}$, where the parameter 
$\chi$ sets the change in asset price\footnote{To compute $\chi$, it is 
usually assumed that assets fire sales generate a linear impact on prices 
\cite{nier2007network,greenwood2015vulnerable,feldhutter2012same}, 
so that the relative assets price change is proportional 
the aggregate amount of assets that need to be liquidated.}. Then, bank $j$ also 
defaults if this loss exceeds its equity $e_j$, originating a new wave of 
liquidity shocks.

As for credit shocks, liquidity shocks do propagate also in absence of defaults: 
equity losses experienced by a bank do imply not only a decreasing value of its 
obligations, but also a decreasing ability and willingness to lend money to the 
market. Thus, liquidity shocks can be smoothly incorporated into the DebtRank 
formalism when the network of interbank exposures is \emph{annealed} (\ie, when 
the dynamics of shock propagation is on the same time scale of contracts 
duration) \cite{cimini2016entangling}. By assuming that the ability of banks to 
lend money decreases proportionally to their equities, the \emph{impact} of bank 
$i$ on bank $j$ reads $\psi\chi w_{ij}/e_j$, which sums to the term $\varphi 
w_{ji}/e_j$in eq. (\ref{eq:acca}) to have a dynamical equation incorporating both 
credit and funding shocks. Financial indicators can then be computed as 
illustrated in the previous section.

\subsubsection{Overlapping portfolios and fire-sales spillovers}

Beyond direct exposures, financial contagion can spread among banks through 
indirect exposures to commonly owned securities, namely \emph{portfolio overlap} 
\cite{gualdi2016statistically,greenwood2015vulnerable,cifuentes2005liquidity,
shleifer2011fire,caccioli2014stability,cont2016fire}. 
Indeed, when the occurrence of financial distress triggers fire sales and prices 
start to fall, losses by banks with overlapping holdings become self-reinforcing 
and trigger further simultaneous sell orders, ultimately leading to downward 
spirals for asset prices. 

Here we discuss a simple linear model of fire sales spillovers due to target 
leveraging by banks and driven by portfolio overlaps 
\cite{greenwood2015vulnerable}. Upon defining $\Omega_i=\sum_{\beta=1}^Q\omega_{i\beta}$ 
as the total portfolio size of bank $i$ and $\tilde{\omega}_{i\alpha}=\omega_{i\alpha}/\Omega_i$ 
as the weight of security $\alpha$ within the portfolio of $i$, the model dynamics 
consist of two time steps. At $t=1$, each bank $i$ collects the return of its 
investments:

\begin{equation}
R_i(1)=\sum_{\alpha=1}^Q \tilde\omega_{i\alpha}f_{\alpha}(1)
\end{equation}
where $f_{\alpha}(1)$ denotes the net return of security $\alpha$. In order to 
simulate exogenous shocks, $f_{\alpha}(1)$ is taken as a negative number so that 
$R_i(1)<0$. Since the equity of $i$ has now changed by $\Omega_iR_i(1)$, in order to 
return to the leverage target $b_i=a_i/e_i$ the bank has to reallocate 
$b_i\Omega_iR_i(1)$ assets on its balance sheet. To this end, it is assumed that 
banks reallocate assets proportionately to existing holdings, so that the net 
purchase of bank $i$ on security $\alpha$ is:

\begin{equation}
\phi_{i\alpha}=\tilde\omega_{i\alpha}b_i\Omega_iR_i(1).
\end{equation}

However, asset sales generate price impact (for simplicity, according to a 
linear model), so that the return of security $\alpha$ is now:

\begin{equation}
f_{\alpha}(2)=\sum_{\beta=1}^QL_{\alpha\beta}\sum_{j=1}^N\phi_{j\beta},
\end{equation}
where $L_{\alpha\beta}$ is a generic entry of the matrix of price impact ratios. 
Note that if all securities are perfectly liquid (meaning that all elements of the matrix 
are zero), then price impact vanishes. The \emph{illiquidity} of security 
$\alpha$ is thus defined as $\Lambda_\alpha=\sum_\beta L_{\alpha\beta}$. Finally, the 
return of bank $i$ at $t=2$ becomes:

\begin{equation}
R_i(2)=\sum_{\alpha=1}^Q\tilde\omega_{i\alpha}f_{\alpha}(2)
\end{equation}
and in principle this process can be iterated multiple times.

Using this framework, individual bank indicators can be computed as follows (see 
also \cite{squartini2017network,digiangi2016assessing}):

\begin{itemize}
\item the \emph{impact} (or \emph{systemicness}) of $i$ is the contribution of 
that bank to the system equity wiped out by bank ``de-leveraging'' due to the 
initial shock

\begin{equation}
I_i=\left[\sum_{\alpha=1}^Q\left(\sum_{j=1}^N\omega_{j\alpha}\right)\Lambda_\alpha 
\omega_{i\alpha} \right]\frac{b_iR_i(1)}{\sum_je_j};
\end{equation}

\item the (indirect) \emph{vulnerability} of $i$ is the impact of the initial 
shock on its equity through the de-leveraging of other banks

\begin{equation}
V_i=\frac{1+b_i}{\Omega_i}\sum_{\alpha=1}^Q\Lambda_\alpha\omega_{i\alpha}\sum_{j=1}^N 
\omega_{j\alpha}b_jR_j(1).
\end{equation}
\end{itemize}

Note that beyond balance sheet quantities and individual positions, the above 
expressions depend on securities illiquidity parameters, which are difficult to 
estimate. However, by assuming that fire sales in one security do not directly 
affect prices in other securities, the matrix of price impact ratios becomes 
diagonal and all illiquidity parameters become equal. In this special case, the 
ratios between reconstructed and empirical indicators assume a particularly 
simple form as, for instance \cite{squartini2017network}, 

\begin{equation}
\frac{\avg{ I_i}}{\hat{I}_i}=\frac{\sum_{j=1}^N\sum_{\alpha=1}^Q\avg{ 
\omega_{i\alpha}\omega_{j\alpha}}}{\sum_{j=1}^N\sum_{\alpha=1}^Q 
\hat{\omega}_{i\alpha}\hat{\omega}_{j\alpha}}.
\end{equation}

\section{Model selection criteria}\label{secrit}

Of course, each indicator described in the previous section can be used to 
assess only a specific aspect of the performance of a given reconstruction algorithm. 
Here we introduce more general information-based criteria, 
able to capture the overall performance of a reconstruction method.

\subsection{The Likelihood Ratio Test}

The very basic criterion to compare different reconstruction algorithms 
consists in comparing their likelihood functions. Since the likelihood represents 
the probability that the observed network is reproduced by the chosen model, the 
closer its value to 1, the better the model\footnote{When considering 
log-likelihood functions, instead, the best algorithm is characterized by the 
closest value to zero.}.

However, since increasing the number of model parameters causes the likelihood function 
to increase as well, this basic criterion completely ignores the over-fitting issue, \ie, 
the risk of introducing unnecessary parameters not providing any relevant 
information but weakening the overall predictive power of the model. Indeed, one 
of the desirable features of any model lies in the possibility to generalize/apply it on different systems. 
By tuning too many parameters over a single specific system may induce a model 
that is able to reproduce every detail of the system itself, without capturing 
more general and essential features potentially shared by similar systems. 
For this reason, a more refined criterion is needed, possibly discounting the number of 
parameters entering into the models definition.

The simplest choice is provided by the Likelihood Ratio Test (LRT), designed to 
compare \emph{pairs} of \emph{nested} models. This means that 1) only two models 
at a time can be compared, 2) the space of the parameters of one model must be a subspace of the 
parameters of the other model. This second requirement 
sheds light on the meaning of the test itself, which is intended to verify the 
\emph{need} of enlarging the parameter space, \ie, of adopting a more complex 
model to describe the observations. A concrete example is provided by the pair 
DECM and DWCM, defined by the Hamiltonians of eqs. (\ref{H-DECM}) and (\ref{H-DWCM}), respectively. 
By switching off the DECM Lagrange multipliers controlling for the degrees 
(\ie, setting $\xxo_i=\xxi_i=1$ $\forall i$), the likelihood 
function of the DECM reduces to the likelihood function of the DWCM:

\begin{eqnarray}
p_{ij}^{\mbox{\tiny DECM}}=\frac{\xxo_i\xxi_j\yo_i\yi_j}{1+\xxo_i\xxi_j\yo_i\yi_j-\yo_i\yi_j}&\longrightarrow&p_{ij}^{\mbox{\tiny DWCM}}=\yo_i\yi_j,\\
\avg{ w_{ij}}^{\mbox{\tiny DECM}}=\frac{p_{ij}^{\mbox{\tiny DECM}}}{1-\yo_i\yi_j}&\longrightarrow&\avg{ 
w_{ij}}^{\mbox{\tiny DWCM}}=\frac{\yo_i\yi_j}{1-\yo_i\yi_j}.
\end{eqnarray}

Provided that $r_1$ is the model with the lower number of parameters (\ie, the DWCM) 
and $r_2$ is the model with the larger number of parameters (\ie, the DECM), 
the LRT compares the quantity

\begin{equation}
D=2\mathcal{L}_{r_2}(\hmG|\vec{\hat{\lambda}}_{(r_2)})-2\mathcal{L}_{r_1}(\hat{\mG}|\vec{\hat{\lambda}}_{(r_1)})
\end{equation}
to some properly-defined threshold value $D_{th}$. The latter is determined by 
the Wilks' theorem \cite{wilks1938large}, stating that the probability 
distribution of $D$ is approximately a chi-squared distribution with a number of 
degrees of freedom equal to $|\vec{\lambda}_{(r_2)}|-|\vec{\lambda}_{(r_1)}|$, \ie, 
to the difference between the number of parameters of model $r_2$ and model $r_1$.

\subsection{The Akaike Information Criterion}

In order to allow for the comparison of more than two models, the more refined 
Akaike Information Criterion (AIC) \cite{burnham2002model,burnham2004multimodel,akaike1974new} 
can be used. Among a set of competing models, the best performing $r$ is characterized by the largest value of

\begin{equation}\label{aic}
\mbox{AIC}_r=2M_r-2\mathcal{L}_r(\hmG|\vec{\hat{\lambda}}_{(r)}).
\end{equation}
AIC is thus a model-specific index that is (proportional to) the difference between the 
number of parameters $M_r$ of the model and its maximum log-likelihood. 
Adding the number of parameters to the log-likelihood function 
allows to get rid of the overfitting issue, and AIC represents an 
attempt of finding an optimal trade-off between explanatory power and simplicity.

\medskip

Equation (\ref{aic}) provides the baseline for other similar criteria 
that have been subsequently defined. As an example, whenever the number $n$ of 
empirical observations becomes too small with respect to the number of 
parameters (a rule of thumb being $n/M_r<40$ 
\cite{burnham2002model,burnham2004multimodel}), the modified quantity

\begin{equation}
\mbox{AICc}_r=\mbox{AIC}_r+\frac{2M_r(M_r+1)}{n-M_r-1}
\end{equation}
should be employed. AICc penalizes models with too many parameters even 
more severely than AIC; consistently, whenever $n\gg M_r^2$, AICc converges to AIC and 
eq. (\ref{aic}) is recovered. 

\subsection{The Bayesian Information Criterion}

An alternative criterion to AICc is the \emph{Bayesian Information Criterion} 
(BIC) \cite{burnham2002model,burnham2004multimodel,akaike1974new}. The 
difference between the two lies in the functional form of the term to be added 
to the maximized likelihood. The BIC discounts not only the 
number of parameters but also the number of observations:

\begin{equation}
\mbox{BIC}_r=M_r\ln 
n-2\mathcal{L}_r(\hmG|\vec{\hat{\lambda}}_{(r)}).
\end{equation}

The extra term $\ln n$ is believed to make BIC more restrictive than AIC, as the 
former tends to select models with a lower number of parameters than those 
selected by the latter \cite{burnham2002model,burnham2004multimodel}. However, 
which criterion performs best, and under which conditions, is still a debated issue.

\medskip

As a final comment, we would like to stress the general applicability of 
the aforementioned criteria. In fact, all of them can be extended to quantum-inspired 
entropic measures \cite{dedomenico2016spectral}. Additionally, although all these criteria are 
likelihood-based (\ie, they can be used to compare only models defined by means of a likelihood 
function), they can be also employed to consistently compare probabilistic as well 
deterministic algorithms (it is enough to set the likelihood function 
of these algorithms to zero). In any case, despite their formal differences, 
all the described information-based criteria convey the same message: 
a ``good'' reconstruction algorithm is not only required to accurately fit the observed 
data but also to avoid over-fitting them, thus encompassing a good trade-off between 
accuracy and parsimony.

\subsection{A quick look at multimodel averaging}

Beside individuating the best model within a basket of alternatives, AIC, AICc 
and BIC also allow to quantify the relative improvement brought by each model. 
This is achieved by computing the \emph{Akaike} (or, equivalently, the 
\emph{Bayesian}) \emph{weights}, reading

\begin{equation}
w_r=\frac{e^{-\Delta_r/2}}{\sum_{s=1}^R e^{-\Delta_s/2}},
\end{equation}
where $\Delta_r=\mbox{AIC}_r-\min\{\mbox{AIC}_s\}_{s=1}^R$ (or, 
in the BIC case, $\Delta_r=\mbox{BIC}_r-\min\{\mbox{BIC}_s\}_{s=1}^R$), 
with $R$ being the total number of competing models.

The Akaike (or Bayesian) weight of a certain model is usually interpreted as the 
probability that the corresponding model is the most appropriate one. In 
particular, models with $\Delta\leq 2$ have substantial 
statistical support; models with $4\leq\Delta\leq7$ have less 
support and models with $\Delta>10$ have essentially no support 
(remarkably, confidence intervals can also be also defined) 
\cite{burnham2002model,burnham2004multimodel,akaike1974new}. Finally, in order 
to quantify how better a given model $r_1$ is with respect to a competitor model 
$r_2$, the ratio $w_{r_1}/w_{r_2}$ can be computed.

In table \ref{tab_w} we provide a sample test of this kind for the (undirected) ECM and WCM on several empirical networks. 
We see that, apart from the first two social networks, the ECM is always superior to the WCM, achieving unit probability (within 
machine precision). A closer inspection of the networks, for which the opposite result holds, reveals that these networks are (almost) fully connected. 
In these cases, the degree sequence represents a redundant constraint and therefore a model with less parameters is preferable. 
These results provide additional evidence that degrees convey an information which is not reducible to that of strengths in order to reconstruct a network with non-trivial topology. 

\begin{table}[!h]\footnotesize
\begin{tabular}{lcc}
\hline
\text{\bf Networks} & $w^{\text{AIC}}_{\text{WCM}}$ & $w^{\text{AIC}}_{\text{ECM}}$\\
\hline
\hline
\text{Office social network} & 1 & 0\\
\text{Research group social network} & 1 & 0\\
\text{Fraternity social network} & 0 & 1\\
\text{Maspalomas Lagoon food} & 0 & 1\\ 
\text{Chesapeake Bay food web} & 0 & 1\\
\text{Crystal River (control) food web} & 0 & 1\\
\text{Crystal River food web} & 0 & 1\\
\text{Michigan Lake food web} & 0 & 1\\
\text{Mondego Estuary food web} & 0 & 1\\
\text{Everglades Marshes food web} & 0 & 1\\
\text{Italian Interbank network (in 1999) } & 0 & 1\\
\text{World Trade Web (in 2000)} & 0 & 1\\
\hline
\end{tabular}
\caption{Akaike weights for the WCM and ECM applied to reconstruct the empirical networks listed in the first column \cite{mastrandrea2014enhanced}. 
Except for the first two networks that are basically fully connected, the inclusion of degrees information is non-redundant, and the ECM drastically outperforms the WCM.}\label{tab_w}
\end{table}

Finally, whenever no winning model emerges from these tests, Akaike (or Bayesian) weights can still be used 
to account for several estimates of the same parameter---say, $\{\hat{\mu}\}$. 
A prescription retaining the explanatory power of the models providing these estimates reads

\begin{equation}\label{mma}
\overline{\hat{\mu}}=\sum_rw_{r}^{\text{AIC}}\hat{\mu}_{(r)},
\end{equation}
and consists in averaging the estimates themselves according to the 
``relevance'' of the model they derive from (quantified by, \eg, the Akaike 
weights). Equation (\ref{mma}) illustrates the concept of {\em multimodel 
average}: the estimation to be employed for reconstructing the network is, thus, 
$\overline{\hat{\mu}}$.

\section{Conclusions and perspectives}\label{conc}

Networks are increasingly pervasive in our life, hence network models and methods are becoming 
and will become more and more important in science and society. 
And whatever the development and technological level of this world will become, 
we shall always struggle (probably more and more) to get the information necessary to describing it. 
Already now, for systems as large as the WWW, it is essentially impossible 
to collect anything but partial information. 
We therefore believe that the knowledge of basic instruments needed to deal with 
partial information will be more and more needed in the future, paving the 
way for a novel use of ensemble methods in modern Statistical Physics.

The state of the art so far shows clearly that the performance of a 
reconstruction method crucially depends on several factors. In order to 
detecting the best method for the problem at hand, a great effort has 
been recently devoted to compare different algorithms on a large set of empirical networks. 
Such ``horse-races'' are intended to 
quantify the performance of the various algorithms with respect to the families 
of indicators introduced in section \ref{secfin}, paying particular attention to 
the topological ones. 
Below we briefly summarize the current state of the art presented above and the perspectives for 
future works.

\subsection{Comparing different reconstruction algorithms on real-world 
networks}

Although the rationale of the MaxEnt method is rooted into the empirical 
observation of financial systems (banks tend to maximize their diversification 
since, in case of distress propagation, a complete market is believed to be more 
robust than an incomplete market \cite{allen2000financial}), it is widely 
recognized that MaxEnt performs very poorly in reproducing the topological details 
of a given network. On the other hand, both the MaxEnt and the ``copula'' 
approach are strong performers in reproducing the observed weights, as indicated 
by the (weighted) cosine similarity \cite{anand2017missing}.

However, real financial markets are sparse \cite{gai2011complexity}. Thus, MaxEnt method 
must be complemented by a prescription able to reproduce the topological details 
of a given network structure. Although the density-corrected DWCM constraints 
both the marginals and the link density (thus reproducing both), it may 
fail in reproducing higher-order topological quantities \cite{mazzarisi2017methods}. 
A better performance is achieved by those algorithms estimating the topological 
details independently from the weights (\eg, in a two-step fashion). For example, 
the fitness-induced ERG model described by eq. (\ref{L}) not only reproduces (to 
a large extent) the in-degree and out-degree sequences of empirical systems 
\cite{cimini2015estimating}, but is also characterized by a functional form 
of link probabilities guaranteeing that 
the observed disassortative trends are correctly replicated 
\cite{mazzarisi2017methods}. Additionally, this method has been shown to 
satisfactorily reproduce the structural details of the same networks 
\cite{squartini2017network}, thus outperforming other methods taking as input 
the same kind of information \cite{mazzarisi2017methods}.

\begin{table}[p]\footnotesize
\begin{tabularx}{\textwidth}{lcllXcc}
\toprule
{\bf Name} & {\bf ME} & {\bf Type} & {\bf Category} & {\bf Brief description} & 
{\bf Sec.} & {\bf Ref.} \\
\midrule\addlinespace
MaxEnt & \cmark & Dense & Deterministic & Maximizes Shannon entropy on network 
entries by constraining marginals & \ref{maxent} & 
\cite{wells2004financial,upper2011simulation} \\
\addlinespace
IPF & \cmark & Tunable & Deterministic & Minimizes the KL divergence from MaxEnt 
& \ref{rasalgo} & \cite{bacharach1965estimating} \\
\addlinespace
MECAPM & \cmark & Dense & Probabilistic & Constrains matrix entries to match, on 
average, MaxEnt values & \ref{digiangi} & \cite{digiangi2016assessing} \\
\addlinespace
Drehmann \& Tarashev & \cmark	& Tunable	& Probabilistic & Randomly 
perturbs the MaxEnt reconstruction & \ref{drehmann} & 
\cite{drehmann2013measuring} \\
\addlinespace
Mastromatteo et al. & \cmark & Tunable	& Probabilistic & Explores the space of 
network structures with the message-passing algorithm & \ref{mastro} & 
\cite{mastromatteo2012reconstruction} \\
\addlinespace
 Moussa \& Cont & \cmark & Tunable & Probabilistic & Implements IPF on 
non-trivial topologies & \ref{mussa_cont} & \cite{moussa2011contagion} \\
\addlinespace
 Fitness-induced ERG 	 & \cmark	& Exact & Probabilistic & Uses the 
fitness ansatz to inform an exponential random graph model	 & \ref{FiERG} & 
\cite{cimini2015systemic,cimini2015estimating} \\
\addlinespace
Copula approach & \xmark & Dense & Deterministic & Generates a network via a 
copula function of the marginals & \ref{copulaapp} & \cite{baral2012estimating} 
\\
\addlinespace
Gandy \& Veraart & \xmark	& Tunable & Probabilistic & Implements an 
adjustable Bayesian reconstruction & \ref{veraart} & \cite{gandy2016bayesian} 
\\
\addlinespace
Montagna \& Lux & \xmark	& Tunable & Probabilistic & Assumes {\em ad-hoc} 
connection probabilities depending on marginals & \ref{montagna} & 
\cite{montagna2017contagion} \\
\addlinespace
Ha\l{}aj \& Kok & \xmark & & Probabilistic & Uses external information to define a 
(geographical) probability map & \ref{halaj} & \cite{halaj2013assessing} 
\\
\addlinespace
Minimum-Density & \xmark & Sparse & Probabilistic & Minimizes the network 
density while satisfying the marginals & \ref{min_dens} 
& \cite{anand2014filling} \\
\addlinespace\bottomrule
\end{tabularx}
\caption{Overview of the reconstruction methods reviewed in the present 
work.}\label{tab_meth}
\end{table}

A more detailed analysis, focusing on statistical indicators, has been 
carried out in \cite{anand2017missing} where the indices defined by eqs. 
(\ref{aindex}), (\ref{hindex}), (\ref{jindex}) and (\ref{tindex}) 
have been used to rank different reconstruction algorithms. What emerges is that 
measures ``emphasizing'' the link structure between financial institutions favor 
methods that produce sparser networks, whereas measures that ``emphasize'' the 
magnitude of bilateral exposures favor methods that allocate exposures as evenly 
as possible. More specifically, both the fitness-induced ERG model and the 
Minimum-Density algorithm are strong performers according to the accuracy index, 
the Hamming distance and the Jaccard distance. The former, however, is ``the 
clear winner among the ensemble methods [$\dots$] across all measures of interest'' 
\cite{anand2017missing}.

An additional round of comparisons has been carried out in \cite{gandy2017adjustable}, where the performance of the Bayesian approach(es) has been compared with the performance of the fitness-induced ERG model. To this aim, the authors have employed the ``empirical'' approach discussed in subsection \ref{veraart}, where parameters are tuned to match the actual network density. Both the ``empirical'' Bayesian approach and the fitness-induced ERG model are strong performers under the accuracy, the sensitivity and the specificity indices. This confirms that, as long as the binary network topology is concerned, the fitness-induced ERG model seems to represent the best algorithm available so far. For what concerns weights reconstruction, the fitness-induced ERG model achieves the best score under the $\text{L}_1$ and $\text{L}_2$ norms, but is outperformed by the Bayesian approach under the PTS index. The latter quantifies the probability of finding, within a numerically generated sample of networks, reconstructed weights whose magnitude lies within the 10\% of the observed value. As noticed 
in \cite{gandy2017adjustable}, the reason of the success of the Bayesian 
approach lies in the fact that it allows one to generate configurations 
characterized by a whole range of different weights, whereas the fitness-induced 
ERG model assigns weights via a simple Bernoulli distribution, whence the 
smaller probability of finding values that satisfy the requirements above.
It should, however, be noticed that the same successful result is achieved by 
using the Weighted Random Graph Model (which performs even better than the 
Bayesian method), pointing out that any algorithm which is sufficiently generic 
seems to be able to achieve a high score under the PTS index. In other words, the requirement to perform 
satisfactorily under this index does not seem to pertain to any algorithm 
specifically designed for reconstruction. Overall, the versatility of 
the Bayesian approach (\ie, its capability of producing a large number of 
different topological structures and weights distributions) make it a 
good method for designing possible scenarios over which running stress tests, 
rather than for reconstructing a particular network configuration.

In \cite{ramadiah2017reconstructing}, the authors adjust several methods 
described in the previous sections to deal with the reconstruction of bipartite 
bank-firm credit networks. The main difference between the 
original and the modified versions of such algorithms lies in the weights 
allocation step, realized through the IPF algorithm. Generally speaking, what 
the authors observe is that the best performance depends on the specific 
indicator and the level of aggregation. However, apart from the trivial result 
that the MaxEnt and the Minimum-Density methods achieve, respectively, the 
highest sensitivity and the highest specificity, the authors find that the 
considered variants of ERG models (\ie, the methods inspired by \cite{saracco2015} and \cite{battiston2016leveraging}) ``consistently perform best'' \cite{ramadiah2017reconstructing} and ``are able to reconstruct adjacency matrices and weighted networks relatively well, and they are capable to preserve the statistical properties of the actual network at all (data) aggregation levels'' \cite{ramadiah2017reconstructing}. Interestingly, the performance of reconstruction methods in reproducing \emph{dynamical (financial)} indicators is also tested. As the authors notice, even null models preserving degrees fail to accurately reproduce the actual level of systemic risk (defined as the probability of default of a bank \cite{ramadiah2017reconstructing}). However, the model inspired by \cite{saracco2015} (followed by the model inspired by \cite{battiston2016leveraging} and MaxEnt) ``has the closest behavior to the actual network overall, while Minimum-Density shows an inconsistent performance across different aggregation levels'' \cite{ramadiah2017reconstructing}.

Generally speaking the fitness-induced ERG model performs well in replicating the binary topology because it provides a realistic estimate of the degrees in the network. Some studies have shown that methods that take as input local topological properties can outperform models that take as input more information of non-topological nature (\eg, geographical distances in the case of the WTW) \cite{picciolo2012role}.

\subsection{Is link density really needed?}

All aforementioned ``horse-races'' point out the superior performance of the 
algorithms that can be calibrated to reproduce the observed link density over 
the ones which cannot: the network link density, in other words, constitutes a 
piece of information that \emph{must} be taken into account, in order to achieve 
an accurate reconstruction.

However, as stressed by the authors in \cite{gandy2017adjustable}, the network 
density cannot be deduced from the marginals alone. As a consequence, it must be 
known from the beginning. If this is not the case (as it often happens to be), 
dealing with the issue of estimating link density is not always straightforward. 
Although the knowledge of just a network subgraph, in fact, often ensures that 
the actual link density is accurately estimated, such entries must be selected 
carefully. As recommended in \cite{blagus2015empirical}, the random selection 
scheme should be employed, in order to avoid possible biases, which may 
arise when nodes are sampled according to some criterion. As the authors in 
\cite{squartini2017network} show, if (sets of) nodes were selected according 
to their total strength, $\hat{s}_i^{\mbox{\tiny tot}}=\hso_i+\hsi_i$, the resulting 
density estimate would be strongly dependent on the specific subset value 
$\sum_{i\in I}\hat{s}_i^{\mbox{\tiny tot}}$. Nodes characterized by large total strengths, in fact, tend 
to cluster into densely-connected groups whereas nodes characterized by small 
total strengths tend to cluster into loosely-connected groups. The analysis in 
\cite{squartini2017network} thus suggests that a sampling-based reconstruction 
procedure should rest upon a ``balanced'' sampling of the nodes set, biased 
neither towards ``core'' nodes nor towards ``peripheral'' nodes.

In order to further show how relevant the role played by topological information 
can be in providing accurate estimates of structural quantities, let us compare 
the statistical fluctuations of weights estimates, output by the (bipartite 
versions of the) MECAPM and degree-corrected gravity model (the latter named ``Enhanced'' CAPM) 
\cite{squartini2017stock}. Provided that

\begin{equation}
(\sigma^2_{w_{i\alpha}})^{\text{\tiny MECAPM}}=w_{i\alpha}^{\text{\tiny 
ME}}(1+w_{i\alpha}^{\text{\tiny ME}})
\end{equation}
and
\begin{equation}
(\sigma^2_{w_{i\alpha}})^{\text{\tiny dcGM}}_{w_{i\alpha}}=(w_{i\alpha}^{\text{\tiny 
ME}})^2\left[\frac{1}{p_{i\alpha}}-1\right]
\end{equation}
one finds that 

\begin{equation}
\frac{(\sigma^2_{w_{i\alpha}})^{\text{\tiny dcGM}}}{(\sigma^2_{w_{i\alpha}})^{\text{\tiny MECAPM}}}\simeq\sqrt{\frac{1}{p_{i\alpha}}-1},
\end{equation}
a ratio that is (strictly) smaller than 1 whenever $p_{i\alpha}>1/2$. In other 
words, provided that the MaxEnt method satisfactorily estimates the observed 
weights, adding topological information helps reducing the error affecting these 
estimates by ``shrinking'' the ensemble over which higher-order properties must 
be estimated. As a consequence, the error accompanying the latter ones is 
reduced as well.

The example above also helps clarifying the role played by purely weighted 
information in the whole reconstruction process. Loosely speaking, we may say 
that the information encoded into nodes strengths is not \emph{per se} of 
``lower quality'' with respect to the information encoded into link density (or 
into nodes degrees). What emerges is rather that it should not be used to 
\emph{directly} reconstruct a given network, but first to estimate nodes 
degrees, and only after be enforced as a complementary constraint.

\subsection{Policy-making implications: the case of the OTC market}

As mentioned in the Introduction, the more information is available about the 
interconnections shaping a given economic or financial network, the more 
effective a regulatory intervention can be whenever a systemic event is detected 
in the system. As a consequence, discovering which kind of information plays a major role in 
reconstructing a given economic or financial network is not only interesting 
from an academic point of view, but of paramount importance also for regulators 
and policy-makers, and has profound societal implications. 

A currently debated topic is thus what kind of information should be disclosed 
by financial institutions. As an example, consider the case of 
the OTC market. In order to improve the transparency of this market, new 
reporting rules have been introduced in the aftermath of the crisis \cite{ESMA} 
establishing that, beside marginals, a certain number of entries of the 
adjacency matrix have to be made accessible (\ie, the ones exceeding a given 
threshold). This constitutes a challenge that can be approached by employing 
(some of) the algorithms reviewed in the present work. 
For example, constraining single entries of the adjacency matrix can be easily 
implemented both within the ERG formalism and by slightly modifying the IPF 
algorithm. While, in the second case, it is enough to subtract the known entries 
from the marginals and redistribute what remains on the unconstrained entries, 
in the first case the additional constraints can be dealt with by keeping the 
corresponding Lagrange multipliers in the tensor form (the latter would be 
estimated by solving the equation $\hw_{ij}=\avg{w_{ij}}$ for each 
known entry).

In any case, and whatever algorithm is chosen, the adaptation of existing reconstruction 
methods to newly-disclosed kinds of information is not only important to 
determine the effectiveness of the new reporting rules, but can also lead to the design better 
data-sharing agreements.

\newpage

\appendix

\section*{Appendix A. A combinatorial derivation of Shannon entropy}
\addcontentsline{toc}{section}{Appendix A. A combinatorial derivation of Shannon 
entropy}
\setcounter{section}{1}

Since Shannon entropy is the quantity underlying many of the approaches reviewed 
in the present work, in what follows we present a very intuitive derivation of 
it.

Let us start by considering the set of sequences of binary symbols (0, 1) 
whose length is $n$ (\ie, the sequences like 01001001\dots). We can ask 
how many bits are needed in order to transmit any message of this 
particular kind. Upon considering that there are $|\mathcal{M}|=2^n$ messages 
satisfying the aforementioned properties, the answer is simply $n=\log_2 
|\mathcal{M}|$, that is the length of the message itself.

Let us now consider a less trivial situation, by allowing our messages to be 
composed by a number of symbols larger than two (\eg, $R_1,R_2\dots R_N$) and to 
be emitted by a source sequentially, according to the probability coefficients 
$P_1,P_2\dots P_N$ (with the latter satisfying the normalization condition 
$\sum_iP_i=1$). In this case, the set of admittable messages has cardinality $|\mathcal{M}'|=N^n$, 
and its generic element is composed by $n_i$ symbols of the $R_i$ kind. Since each symbol is emitted 
independently from the others, when $n\rightarrow\infty$ we can apply the law of 
large numbers and say that the number of times $n_i$ in which the outcome $R_i$ 
is observed satisfies the limit $n_i/n\rightarrow P_i$. As a consequence, when 
$n$ becomes sufficiently large, the sequences characterized by an occurrence of 
symbol $R_i$ which is exactly $P_i$ become overwhelmingly more likely to occur 
than the others (\ie, a negligible amount of information is lost upon discarding 
sequences \emph{not} satisfying $n_i/n\simeq P_i$). We can thus restrict 
ourselves to consider the set of messages of length $\sum_in_i=n$ characterized 
by probability coefficients reading

\begin{equation}
P(R_{i_1}R_{i_2}\dots R_{i_n})=P_1^{n_1}P_2^{n_2}\dots P_N^{n_N}.
\end{equation}

The messages above constitute a set of 

\begin{equation}
|\mathcal{M}'|=\frac{n!}{n_1!n_2!\dots n_N!}
\end{equation}
equiprobable elements: as for the binary case considered at the beginning of 
this Appendix, the number of bits needed to transmit one of these messages can 
be estimated as $I'=\log_2 |\mathcal{M}'|$, that is

\begin{equation}
I'\simeq -n\sum_i P_i\log_2 P_i
\end{equation}
(upon using the Stirling approximation $\ln(x!)\simeq x\ln x-x$). As a 
consequence, one can define the quantity

\begin{equation}\label{app1}
S=-\sum_i P_i\log_2 P_i\simeq\frac{I'}{n}
\end{equation}
as the Shannon entropy. In other words, Shannon entropy is proportional to the 
logarithm of the probability of a typical (long) sequence, divided by the number 
of symbols composing the sequence itself. Otherwise stated, $S$ quantifies the 
(average) number of bits needed to transmit a typical sequence (\ie, one of those 
belonging to $\mathcal{M}'$).

The reasoning leading to eq. (\ref{app1}) can be restated more precisely by 
invoking the law of large numbers to gain insight into the asymptotic behavior 
of the i.i.d. variables describing our symbols $R_1,R_2\dots R_N$:

\begin{equation}
\frac{1}{n}\log_2\frac{1}{P(R_{i_1}R_{i_2}\dots 
R_{i_n})}=-\frac{1}{n}\log_2\prod_iP_i(R_i)=-\frac{1}{n}\sum_i\log_2 
P_i(R_i)\rightarrow S.
\end{equation}
As a consequence, the set of all possible messages is split in two: with high 
probability, the observed messages will belong to ``a typical'' subset, whose 
members are described by a coefficient approaching

\begin{equation}
P(R_{i_1}R_{i_2}\dots R_{i_n})\simeq 2^{-n S}
\end{equation}
and inducing an uniform distribution over this set. This result is known as 
\emph{asymptotic equipartition property} (AEP).

Althought the early derivation of Shannon entropy rested upon the concept of 
\emph{bit}, thus forcing the base of the logarithm to be 2, from a purely 
numerical viewpoint it is more convenient to make use of the natural 
logarithm (adopted in the present review). This means measuring information in \emph{nats} 
(1 nat equals $\log_2e$ bits).

\section*{Appendix B. Sketching a principled derivation of Shannon entropy}
\addcontentsline{toc}{section}{Appendix B. Sketching a principled derivation of 
Shannon entropy}
\setcounter{section}{2}
\setcounter{equation}{0}

A possible derivation of Shannon entropy from the Shannon-Khinchin axioms is 
shown below (alternative proofs can be found in 
\cite{shannon1948mathematical1,shannon1948mathematical2,jaynes1957information}). 
The fourth axiom implies that whenever a combination of two independent 
subsystems is considered, Shannon entropy reads $S(W_AW_B)=S(W_A)+S(W_B)$. Upon 
deriving it twice, the first time with respect to $W_A$ and the second time with 
respect to $W_B$, the expression $S'(W_AW_B)+W_AW_BS''(W_AW_B)=0$ is obtained. 
Upon posing $W_AW_B\equiv W$, the expression above can be rearranged as

\begin{equation}
(WS'(W))'=0,
\end{equation}
the latter derivative being taken with respect to $W$. The derived function is, 
thus, a constant. Solving this differential 
equation leads to the celebrated logarithmic functional form $S(W)=k\ln W$. 

Let us now consider the case of dependent subsets. Upon posing $W_{A+B}=W$, it 
is enough to consider that the requirement $S(W_{A+B})=S(W_{A})+S(W_{B|A})$ 
induces the definition of conditional entropy \cite{kin1938math}, as a 
(weighted) average of the number of configurations of subsystem $B$ (say $V_1$, 
$V_2$\dots), the weights being provided by the fraction of configurations of 
subsystem $A$ inducing them (\ie, $w_1=\frac{V_1}{V}$, $w_2=\frac{V_2}{V}$\dots):

\begin{equation}
S(W_{A+B})=S(W_A)+w_1S(V_1)+w_2S(V_2).
\end{equation}
Upon noticing that $S(W_{A+B})=k\ln V$, we obtain the expression

\begin{equation}
S(W_A)=-k(w_1\ln w_1)-k(w_2\ln w_2)\dots
\end{equation}

\section*{Appendix C. Two relevant properties of Shannon entropy}
\addcontentsline{toc}{section}{Appendix C. Two relevant properties of Shannon 
entropy}
\setcounter{section}{3}
\setcounter{equation}{0}

Some of the models reviewed in this work are defined within the ERG formalism. 
These are exponential distributions satisfying the so-called Gibbs property, \ie, 
that of being maximally non-committal with respect to the missing information 
\cite{jaynes1957information}. In fact, upon substituting 

\begin{equation}
P(\mG|\vec{\lambda})=\frac{e^{-\sum_m\lambda_mC_m(\mG)}}{
Z(\vec{\lambda})}
\end{equation}
within $S=-\sum_{\mG\in\ensG}P(\mG)\ln P(\mG)$, we 
obtain 

\begin{equation}
S_P=\sum_m\lambda_m\avg{C_m}+\ln Z(\vec{\lambda}).
\end{equation}

Let us now calculate the quantity $S_P-S_Q$, that is the difference between the 
Shannon entropy of an exponential distribution and the entropy of a generic 
distribution $Q(\mG)$ which satisfies the same constraints as  
$P(\mG)$ (\ie, $\sum_{\mG\in\ensG}Q(\mG)=1$ and 
$\sum_{\mG\in\ensG}Q(\mG)C_m(\mG)=\avg{C_m}$ $\forall m$):

\begin{eqnarray}
S_{P}-S_{Q}&=&\sum_m\lambda_m\avg{C_m}+\ln 
Z(\vec{\lambda})+\sum_{\mG\in\ensG}Q(\mG)\ln 
Q(\mG)=\nonumber\\
&=&\sum_{\mG\in\ensG}Q(\mG)\left[\sum_m\lambda_mC_m
+\ln Z(\vec{\lambda})+\ln Q(\mG)\right]=\nonumber\\
&=&\sum_{\mG\in\ensG}Q(\mG)\left[-\ln P(\mG)+\ln 
Q(\mG)\right]=\nonumber\\
&=&\sum_{\mG\in\ensG}Q(\mG)\left[-\ln\left(\frac{P(\mathbf{G
})}{Q(\mG)}\right)\right]\geq\sum_{\mG\in\ensG}Q(\mG
)\left[1-\frac{P(\mG)}{Q(\mG)}\right]=\nonumber\\
&=&1-1=0.
\end{eqnarray}
The result above shows that the entropy of an exponential distribution is larger 
than the entropy of any other distribution satisfying the same constraints (the 
equality is valid if and only if the probability coefficients of the two distributions coincide).

\medskip

Unconstrained Shannon entropy instead attains its maximum in correspondence of the uniform 
distribution. In fact, 

\begin{eqnarray}
S_{U}-S_{P}&=&-\sum_{\mG\in\ensG}\frac{1}{|\ensG|}\ln 
\frac{1}{|\ensG|}+\sum_{\mG\in\ensG}P(\mG)\ln 
P(\mG)=\nonumber\\
&=&-\sum_{\mG\in\ensG}P(\mG)\ln\frac{1}{|\ensG|}-\sum_
{\mG\in\ensG}P(\mG)\ln\left(\frac{1}{P(\mG)}
\right)=\nonumber\\
&=&\sum_{\mG\in\ensG}P(\mG)\left[-\ln\left(\frac{1}{
|\ensG|P(\mG)}\right)\right]\geq\sum_{\mG\in\ensG}
P(\mG)\left[1-\frac{1}{|\ensG|P(\mG)}\right]=\nonumber\\
&=&1-1=0.
\end{eqnarray}

This can be also verified by calculating the stationary point and the Hessian 
matrix of the functional

\begin{equation}
\mathscr{L}[P]=S-\lambda_0\left[\sum_{\mG\in\ensG}P(\mG
)-1\right],
\end{equation}
whose generic entries read, respectively

\begin{equation}
\frac{\partial \mathscr{L}[P]}{\partial 
P(\mG)}=\frac{1}{|\ensG|}\quad ;\quad\frac{\partial^2\mathscr{L}[
P]}{\partial P(\mG)\partial 
P(\mG')}=-\frac{\delta_{\mG\mG'}}{P(\mG)\ln2},
\end{equation}
and evaluating the second derivative in correspondence of the uniform distribution $P(\mG)=\frac{1}{|\ensG|}$. 
Since the Hessian matrix is diagonal and its entries are strictly negative (for those 
distributions satisfying $P(\mG)>0,\:\forall\:\mG$), such a matrix 
is negative-definite: the stationary point of the Shannon entropy is a maximum.

\section*{Appendix D. A notable, continuous case}
\addcontentsline{toc}{section}{Appendix D. A notable, continuous case}
\setcounter{section}{4}
\setcounter{equation}{0}

So far, we have considered distributions obtained by constraining only first moments. 
Let us now discuss the case in which the second moment is constrained as well. 
In order to do so, let us imagine we have a one-dimensional 
real variable $x\in \mathbb{R}$ described by the probability density function 
$p(x)$. Upon rewriting our constraints as

\begin{equation}
1=\int_{-\infty}^{+\infty}p(x)\:dx,
\label{shan3} 
\end{equation}
\begin{equation}
\mu=\int_{-\infty}^{+\infty}x\:p(x)\:dx
\label{shan4} 
\end{equation}
and
\begin{equation}
\mu_2=\int_{-\infty}^{+\infty}x^2\:p(x)\:dx
\label{shan5} 
\end{equation}
we may ask what is the least-biased pdf that contains information on the mean 
value and the variance of $x$. Let us apply the entropy-maximization 
prescription to the continuous version of Shannon entropy, by defining the 
functional 

\begin{eqnarray}
\mathscr{L}[p]&=&-\int_{-\infty}^{+\infty}p(x)\ln 
p(x)\:dx-\lambda_0\left[\int_{-\infty}^{+\infty}p(x)\:dx-1\right]- 
\lambda_1\left[\int_{-\infty}^{+\infty}x p(x)\:dx-\mu\right]\nonumber\\
&&- \lambda_2\left[\int_{-\infty}^{+\infty}x^2 p(x)\:dx-\mu_2\right]
\end{eqnarray}
with $\lambda_0$, $\lambda_1$, $\lambda_2$ being the Lagrange multipliers 
corresponding to the three conditions (\ref{shan3}), (\ref{shan4}) and 
(\ref{shan5}). Maximizing the functional above means looking for the function 
$p(x)$ which makes the functional derivative $\frac{\delta\mathscr{L}[p]}{\delta 
p(x)}$ vanish, \ie,

\begin{equation}
\frac{\delta\mathscr{L}[p]}{\delta p(x)}=-\ln 
p(x)-1-\lambda_0-\lambda_1x-\lambda_2x^2=0.
\end{equation} 
The solution is 

\begin{equation}
p(x)=e^{\left[-1-\lambda_0-\lambda_1x-\lambda_2x^2\right]}
\label{shan8} 
\end{equation}
which is of course a Gaussian probability density function. The Lagrange 
multipliers can be found using the normalization condition 
$e^{1+\lambda_0}=\int_{-\infty}^{+\infty}e^{-\lambda_2x^2-\lambda_1x}=
\sqrt{\frac{\pi}{\lambda_2}}e^{\frac{\lambda_1^2}{4\lambda_2}}$ 
together with eqs. (\ref{shan4}) and (\ref{shan5}). 
Upon identifying $\sigma^2=\mu_2-\mu^2$ one gets 

\begin{equation}
p(x)=\frac{e^{-\frac{(x-\mu)^2}{2\sigma^2}}}{\sqrt{2\pi\sigma^2}}.
\end{equation}

Finally, in order to verify that the Gaussian distribution is actually a maximum 
of the constrained entropy, it is sufficient to verify (in analogy with the 
discrete case presented above) that the second functional derivative is 
negative-definite as well. This is indeed true because $\frac{\delta^2 
\mathscr{L}[p]}{\delta p(x)\delta p(x')}=-\frac{\delta(x-x')}{p(x)}$.

\newpage

\section*{Acknowledgements}

TS, GC and GC acknowledge support from the EU projects CoeGSS (grant num. 
676547) and SoBigData (grant num. 654024) and the FET project DOLFINS (grant 
num. 640772). DG acknowledges support from the Econophysics foundation 
(Stichting Econophysics, Leiden, the Netherlands). AG acknowledges support from 
the Italian PNR project CRISIS-Lab.

\section*{References}

\bibliography{Bibliography}

\begin{thebibliography}{100}
\expandafter\ifx\csname url\endcsname\relax
  \def\url#1{\texttt{#1}}\fi
\expandafter\ifx\csname urlprefix\endcsname\relax\def\urlprefix{URL }\fi
\expandafter\ifx\csname href\endcsname\relax
  \def\href#1#2{#2} \def\path#1{#1}\fi

\bibitem{albert2002statistical}
R.~Albert, A.-L. Barab{\'{a}}si, {\em Statistical mechanics of complex
  networks}, Reviews of Modern Physics {\bf 74}~(1) (2002) 47--97.
\newblock \href {http://dx.doi.org/10.1103/RevModPhys.74.47}
  {\path{doi:10.1103/RevModPhys.74.47}}.

\bibitem{boccaletti2006complex}
S.~Boccaletti, V.~Latora, Y.~Moreno, M.~Chavez, D.~Hwang, {\em Complex
  networks: Structure and dynamics}, Physics Reports {\bf 424}~(4-5) (2006)
  175--308.
\newblock \href {http://dx.doi.org/10.1016/j.physrep.2005.10.009}
  {\path{doi:10.1016/j.physrep.2005.10.009}}.

\bibitem{caldarelli2007scale}
G.~Caldarelli, {\em Scale-Free Networks: Complex Webs in Nature and
  Technology}, Oxford University Press, 2007.

\bibitem{carlson2014koenisberg}
S.~C. Carlson, {\em K{\"{o}}nigsberg Bridge Problem}, Encyclopedia Britannica.

\bibitem{biggs1998graph}
N.~L. Biggs, E.~K. Lloyd, R.~J. Wilson, {\em Graph Theory 1736-1936}, Clarendon
  Press, Oxford, 1998.

\bibitem{moreno1934sociometry}
J.~L. Moreno, H.~H. Jennings, {\em Who Shall Survive?: A New Approach to the
  Problem of Human Interrelations}, Nervous and mental disease monograph
  series, Nervous and mental disease publishing co., 1934.

\bibitem{moreno1941sociometry}
J.~L. Moreno, {\em Foundations of sociometry: An introduction}, Sociometry {\bf
  4}~(1) (1941) 15--35.
\newblock \href {http://dx.doi.org/10.2307/2785363}
  {\path{doi:10.2307/2785363}}.

\bibitem{lazer2009social}
D.~Lazer, A.~Pentland, L.~A. Adamic, S.~Aral, A.-L. Barab{\'{a}}si, D.~Brewer,
  N.~Christakis, N.~Contractor, J.~Fowler, M.~Gutmann, T.~Jebara, G.~King,
  M.~Macy, D.~Roy, M.~{Van Alstyne}, {\em Computational social science},
  Science {\bf 323}~(5915) (2009) 721--723.
\newblock \href {http://dx.doi.org/10.1126/science.1167742}
  {\path{doi:10.1126/science.1167742}}.

\bibitem{erdos1959random}
P.~Erd\"os, A.~R{\'{e}}nyi, {\em On random graphs}, Publicationes Mathematicae
  Debrecen {\bf 6} (1959) 290--297.

\bibitem{erdos1960evolution}
P.~Erd\"os, A.~R{\'{e}}nyi, {\em On the evolution of random graphs},
  Publications of the Mathematical Institute of the Hungarian Academy of
  Sciences {\bf 5} (1960) 17--61.

\bibitem{bollobas1979graph}
B.~Bollob{\'{a}}s, {\em Graph Theory, An Introductory Course}, 1st Edition,
  Springer, 1979.

\bibitem{bollobas1985random}
B.~Bollob{\'{a}}s, {\em Random Graphs}, Academic Press, London, 1985.

\bibitem{faloutsos1999power}
M.~Faloutsos, P.~Faloutsos, C.~Faloutsos, {\em On power-law relationships of
  the Internet topology}, SIGCOMM Proceeding (1999) 251--262\href
  {http://dx.doi.org/10.1.1.37.234} {\path{doi:10.1.1.37.234}}.

\bibitem{caldarelli2000fractal}
G.~Caldarelli, R.~Marchetti, L.~Pietronero, {\em The fractal properties of
  internet}, Europhysics Letters {\bf 52}~(4) (2000) 386--391.
\newblock \href {http://dx.doi.org/10.1209/epl/i2000-00450-8}
  {\path{doi:10.1209/epl/i2000-00450-8}}.

\bibitem{huffaker2002distance}
B.~Huffaker, M.~Fomenkov, D.~Plummer, D.~Moore, Claffy, K.~Claffy, {\em
  Distance metrics in the Internet}, IEEE International Telecommunications
  Symposium, Sept. 2002 (2002) 200--202.

\bibitem{page1999pagerank}
L.~Page, S.~Brin, R.~Motwami, T.~Winograd, R.~Motwani, {\em The PageRank
  citation ranking: Bringing order to the web}, Unpublished work, Stanford
  InfoLab (1999).

\bibitem{kleinberg1999authoritative}
J.~M. Kleinberg, {\em Authoritative sources in a hyperlinked environment},
  Journal of the ACM {\bf 46}~(5) (1999) 604--632.
\newblock \href {http://dx.doi.org/10.1145/324133.324140}
  {\path{doi:10.1145/324133.324140}}.

\bibitem{guimera2003self}
R.~Guimer\`a, L.~Danon, A.~D\'{\i}az-Guilera, F.~Giralt, A.~Arenas, {\em
  Self-similar community structure in a network of human interactions},
  Physical Review E {\bf 68} (2003) 065103.
\newblock \href {http://dx.doi.org/10.1103/PhysRevE.68.065103}
  {\path{doi:10.1103/PhysRevE.68.065103}}.

\bibitem{newman2002email}
M.~E.~J. Newman, S.~Forrest, J.~Balthrop, {\em Email networks and the spread of
  computer viruses}, Physical Review E {\bf 66} (2002) 035101.
\newblock \href {http://dx.doi.org/10.1103/PhysRevE.66.035101}
  {\path{doi:10.1103/PhysRevE.66.035101}}.

\bibitem{caldarelli2004preferential}
G.~Caldarelli, F.~Coccetti, P.~{De Los Rios}, {\em Preferential exchange:
  Strengthening connections in complex networks}, Physical Review E {\bf
  70}~(2) (2004) 27102.
\newblock \href {http://dx.doi.org/10.1103/PhysRevE.70.027102}
  {\path{doi:10.1103/PhysRevE.70.027102}}.

\bibitem{onnela2007structure}
J.-P. Onnela, J.~Saram{\"a}ki, J.~Hyv{\"o}nen, G.~Szab{\'o}, D.~Lazer,
  K.~Kaski, J.~Kert{\'e}sz, A.-L. Barab{\'a}si, {\em Structure and tie
  strengths in mobile communication networks}, Proceedings of the National
  Academy of Sciences {\bf 104}~(18) (2007) 7332--7336.
\newblock \href {http://dx.doi.org/10.1073/pnas.0610245104}
  {\path{doi:10.1073/pnas.0610245104}}.

\bibitem{eagle2009inferring}
N.~Eagle, A.~S. Pentland, D.~Lazer, {\em Inferring friendship network structure
  by using mobile phone data}, Proceedings of the National Academy of Sciences
  {\bf 106}~(36) (2009) 15274--15278.
\newblock \href {http://dx.doi.org/10.1073/pnas.0900282106}
  {\path{doi:10.1073/pnas.0900282106}}.

\bibitem{miritello2011dynamical}
G.~Miritello, E.~Moro, R.~Lara, {\em Dynamical strength of social ties in
  information spreading}, Physical Review E {\bf 83} (2011) 045102.
\newblock \href {http://dx.doi.org/10.1103/PhysRevE.83.045102}
  {\path{doi:10.1103/PhysRevE.83.045102}}.

\bibitem{mislove2007measurements}
A.~Mislove, M.~Marcon, K.~P. Gummadi, P.~Druschel, B.~Bhattacharjee, {\em
  Measurement and analysis of online social networks}, in: Proceedings of the
  7th ACM SIGCOMM Conference on Internet Measurement, IMC '07, ACM, New York,
  NY, USA, 2007, pp. 29--42.
\newblock \href {http://dx.doi.org/10.1145/1298306.1298311}
  {\path{doi:10.1145/1298306.1298311}}.

\bibitem{borgatti2009network}
S.~P. Borgatti, A.~Mehra, D.~J. Brass, G.~Labianca, {\em Network analysis in
  the social sciences}, Science {\bf 323}~(5916) (2009) 892--895.
\newblock \href {http://dx.doi.org/10.1126/science.1165821}
  {\path{doi:10.1126/science.1165821}}.

\bibitem{kwak2010what}
H.~Kwak, C.~Lee, H.~Park, S.~Moon, {\em What is Twitter, a social network or a
  news media?}, in: Proceedings of the 19th International Conference on World
  Wide Web, WWW '10, ACM, New York, NY, USA, 2010, pp. 591--600.
\newblock \href {http://dx.doi.org/10.1145/1772690.1772751}
  {\path{doi:10.1145/1772690.1772751}}.

\bibitem{delvicario2016spreading}
M.~{Del Vicario}, A.~Bessi, F.~Zollo, F.~Petroni, A.~Scala, G.~Caldarelli,
  H.~E. Stanley, W.~Quattrociocchi, {\em The spreading of misinformation
  online}, Proceedings of the National Academy of Sciences {\bf 113} (2016)
  554--559.
\newblock \href {http://dx.doi.org/10.1073/pnas.1517441113}
  {\path{doi:10.1073/pnas.1517441113}}.

\bibitem{mason2007network}
O.~Mason, M.~Verwoerd, {\em Graph theory and networks in Biology}, IET Systems
  Biology {\bf 1} (2007) 89--119.
\newblock \href {http://dx.doi.org/10.1049/iet-syb:20060038}
  {\path{doi:10.1049/iet-syb:20060038}}.

\bibitem{STRING}
D.~Szklarczyk, A.~Franceschini, S.~Wyder, K.~Forslund, D.~Heller,
  J.~Huerta-Cepas, M.~Simonovic, A.~Roth, A.~Santos, K.~P. Tsafou, M.~Kuhn,
  P.~Bork, L.~J. Jensen, C.~von Mering, {\em STRING v10: protein-protein
  interaction networks, integrated over the tree of life}, Nucleic Acids
  Research {\bf 43}~(D1) (2015) D447--D452.
\newblock \href {http://dx.doi.org/10.1093/nar/gku1003}
  {\path{doi:10.1093/nar/gku1003}}.

\bibitem{koh2012analyzing}
G.~C. K.~W. Koh, P.~Porras, B.~Aranda, H.~Hermjakob, S.~E. Orchard, {\em
  Analyzing protein-protein interaction networks}, Journal of Proteome Research
  {\bf 11}~(4) (2012) 2014--2031.
\newblock \href {http://dx.doi.org/10.1021/pr201211w}
  {\path{doi:10.1021/pr201211w}}.

\bibitem{williams2000simple}
R.~J. Williams, N.~D. Martinez, {\em Simple rules yield complex food webs},
  Nature {\bf 404}~(6774) (2000) 180--183.
\newblock \href {http://dx.doi.org/10.1038/35004572}
  {\path{doi:10.1038/35004572}}.

\bibitem{dunne2002food}
J.~A. Dunne, R.~J. Williams, N.~D. Martinez, {\em Food-web structure and
  network theory: The role of connectance and size}, Proceedings of the
  National Academy of Sciences {\bf 99}~(20) (2002) 12917--12922.
\newblock \href {http://dx.doi.org/10.1073/pnas.192407699}
  {\path{doi:10.1073/pnas.192407699}}.

\bibitem{proulx2005network}
S.~R. Proulx, D.~E. Promislow, P.~C. Phillips, {\bf Network thinking in ecology
  and evolution}, Trends in Ecology \& Evolution {\bf 20}~(6) (2005) 345--353.
\newblock \href {http://dx.doi.org/10.1016/j.tree.2005.04.004}
  {\path{doi:10.1016/j.tree.2005.04.004}}.

\bibitem{garlaschelli2005structure}
D.~Garlaschelli, M.~I. Loffredo, {\em Structure and evolution of the world
  trade network}, Physica A: Statistical Mechanics and its Applications {\bf
  355}~(1) (2005) 138--144.
\newblock \href {http://dx.doi.org/10.1016/j.physa.2005.02.075}
  {\path{doi:10.1016/j.physa.2005.02.075}}.

\bibitem{hidalgo2007product}
C.~A. Hidalgo, B.~Klinger, A.-L. Barab{\'{a}}si, R.~Hausmann, A.-L.
  Barab{\`{a}}si, R.~Hausmann, {\em The product space conditions the
  development of nations}, Science {\bf 317} (2007) 482--487.
\newblock \href {http://dx.doi.org/10.1126/science.1144581}
  {\path{doi:10.1126/science.1144581}}.

\bibitem{schweitzer2009economic}
F.~Schweitzer, G.~Fagiolo, D.~Sornette, F.~Vega-Redondo, A.~Vespignani, D.~R.
  White, {\em Economic networks: The new challenges}, Science {\bf 325}~(5939)
  (2009) 422--425.
\newblock \href {http://dx.doi.org/10.1126/science.1173644}
  {\path{doi:10.1126/science.1173644}}.

\bibitem{tacchella2012ec}
A.~Tacchella, M.~Cristelli, G.~Caldarelli, A.~Gabrielli, L.~Pietronero, {\em A
  new metrics for countries' fitness and products' complexity}, Scientific
  Reports {\bf 2} (2012) 723.
\newblock \href {http://dx.doi.org/10.1038/srep00723}
  {\path{doi:10.1038/srep00723}}.

\bibitem{cimini2014science}
G.~Cimini, A.~Gabrielli, F.~Sylos~Labini, {\em The scientific competitiveness
  of nations}, PLoS ONE {\bf 9}~(12) (2014) e113470.
\newblock \href {http://dx.doi.org/10.1371/journal.pone.0113470}
  {\path{doi:10.1371/journal.pone.0113470}}.

\bibitem{pugliese2017unfolding}
E.~Pugliese, G.~Cimini, A.~Patelli, A.~Zaccaria, L.~Pietronero, A.~Gabrielli,
  {\em Unfolding the innovation system for the development of countries:
  co-evolution of Science, Technology and Production},
  \url{https://arxiv.org/abs/1707.05146} (2017).

\bibitem{iori2008network}
G.~Iori, G.~{De Masi}, O.~V. Precup, G.~Gabbi, G.~Caldarelli, {\em A network
  analysis of the Italian overnight money market}, Journal of Economic Dynamics
  and Control {\bf 32}~(1) (2008) 259--278.
\newblock \href {http://dx.doi.org/10.1016/j.jedc.2007.01.032}
  {\path{doi:10.1016/j.jedc.2007.01.032}}.

\bibitem{glattfelder2009backbone}
J.~Glattfelder, S.~Battiston, {\em Backbone of complex networks of
  corporations: The flow of control}, Physical Review E 80~({\bf 3}) (2009)
  036104.
\newblock \href {http://dx.doi.org/10.1103/PhysRevE.80.036104}
  {\path{doi:10.1103/PhysRevE.80.036104}}.

\bibitem{bordino2012web}
I.~Bordino, S.~Battiston, G.~Caldarelli, M.~Cristelli, A.~Ukkonen, I.~Weber,
  {\em Web search queries can predict stock market volumes.}, PLoS ONE {\bf
  7}~(7) (2012) e40014.
\newblock \href {http://dx.doi.org/10.1371/journal.pone.0040014}
  {\path{doi:10.1371/journal.pone.0040014}}.

\bibitem{barabasi2009scale}
A.-L. Barab{\'{a}}si, {\em Scale-free networks: A decade and beyond}, Science
  {\bf 325}~(5939) (2009) 412--413.
\newblock \href {http://dx.doi.org/10.1126/science.1173299}
  {\path{doi:10.1126/science.1173299}}.

\bibitem{barabasi1999emergence}
A.-L. Barab{\'{a}}si, R.~Albert, {\em Emergence of scaling in random networks},
  Science {\bf 286}~(5439) (1999) 509--512.
\newblock \href {http://dx.doi.org/10.1126/science.286.5439.509}
  {\path{doi:10.1126/science.286.5439.509}}.

\bibitem{watts1998collective}
D.~J. Watts, S.~H. Strogatz, {\em Collective dynamics of 'small-world'
  networks.}, Nature {\bf 393}~(6684) (1998) 440--442.
\newblock \href {http://dx.doi.org/10.1038/30918} {\path{doi:10.1038/30918}}.

\bibitem{szabo2004clustering}
G.~Szab{\'{o}}, M.~Alava, J.~Kert{\'{e}}sz, {\em Clustering in Complex
  Networks}, Vol.~{\bf 650}, Springer, 2004, pp. 139--162.

\bibitem{bonacich1987power}
P.~Bonacich, {\em Power and centrality : A family of measures}, American
  Journal of Sociology {\bf 92}~(5) (1987) 1170--1182.
\newblock \href {http://dx.doi.org/10.1086/228631} {\path{doi:10.1086/228631}}.

\bibitem{newman2002assortative}
M.~E.~J. Newman, {\em Assortative mixing in networks}, Physical Review Letters
  {\bf 2}~(4) (2002) 1--5.
\newblock \href {http://dx.doi.org/10.1103/PhysRevLett.89.208701}
  {\path{doi:10.1103/PhysRevLett.89.208701}}.

\bibitem{almeida2008consistent}
M.~Almeida-Neto, P.~Guimar{\~a}es, P.~R. Guimar{\~a}es, R.~D. Loyola,
  W.~Ulrich, {\em A consistent metric for nestedness analysis in ecological
  systems: reconciling concept and measurement}, Oikos {\bf 117}~(8)
  1227--1239.
\newblock \href {http://dx.doi.org/10.1111/j.0030-1299.2008.16644.x}
  {\path{doi:10.1111/j.0030-1299.2008.16644.x}}.

\bibitem{bascompte2003nestedness}
J.~Bascompte, P.~Jordano, C.~J. Meli{\'a}n, J.~M. Olesen, {\em The nested
  assembly of plant-animal mutualistic networks}, Proceedings of the National
  Academy of Sciences {\bf 100}~(16) (2003) 9383--9387.
\newblock \href {http://dx.doi.org/10.1073/pnas.1633576100}
  {\path{doi:10.1073/pnas.1633576100}}.

\bibitem{johnson2013factors}
S.~Jonhson, V.~Dom\'inguez-Garc\'ia, M.~A. Mu{\~n}oz, {\em Factors determining
  nestedness in complex networks}, PLoS ONE {\bf 8}~(9) (2013) e74025.
\newblock \href {http://dx.doi.org/10.1371/journal.pone.0074025}
  {\path{doi:10.1371/journal.pone.0074025}}.

\bibitem{caldarelli2002scale}
G.~Caldarelli, A.~Capocci, P.~{De Los Rios}, M.~Mu{\~{n}}oz, {\em Scale-free
  networks from varying vertex intrinsic fitness}, Physical Review Letters {\bf
  89}~(25) (2002) 258702.
\newblock \href {http://dx.doi.org/10.1103/physrevlett.89.258702}
  {\path{doi:10.1103/physrevlett.89.258702}}.

\bibitem{leskovec2008microscopic}
J.~Leskovec, L.~Backstrom, R.~Kumar, A.~Tomkins, Microscopic evolution of
  social networks, in: Proceedings of the 14th ACM SIGKDD International
  Conference on Knowledge Discovery and Data Mining, KDD '08, ACM, New York,
  NY, USA, 2008, pp. 462--470.
\newblock \href {http://dx.doi.org/10.1145/1401890.1401948}
  {\path{doi:10.1145/1401890.1401948}}.

\bibitem{medo2011temporal}
M.~Medo, G.~Cimini, S.~Gualdi, {\em Temporal effects in the growth of
  networks}, Physical Review Letters {\bf 107} (2011) 238701.
\newblock \href {http://dx.doi.org/10.1103/PhysRevLett.107.238701}
  {\path{doi:10.1103/PhysRevLett.107.238701}}.

\bibitem{park2004statistical}
J.~Park, M.~E.~J. Newman, {\em The statistical mechanics of networks}, Physical
  Review E {\bf 70} (2004) 66117.
\newblock \href {http://dx.doi.org/10.1103/PhysRevE.70.066117}
  {\path{doi:10.1103/PhysRevE.70.066117}}.

\bibitem{bianconi2008entropy}
G.~Bianconi, {\em The entropy of randomized network ensembles}, Europhysics
  Letters {\bf 81}~(2) (2008) 28005.
\newblock \href {http://dx.doi.org/10.1209/0295-5075/81/28005}
  {\path{doi:10.1209/0295-5075/81/28005}}.

\bibitem{garlaschelli2008maximum}
D.~Garlaschelli, M.~I. Loffredo, {\em Maximum likelihood: Extracting unbiased
  information from complex networks}, Physical Review E {\bf 78}~(1) (2008)
  015101.
\newblock \href {http://dx.doi.org/10.1103/PhysRevE.78.015101}
  {\path{doi:10.1103/PhysRevE.78.015101}}.

\bibitem{squartini2011analytical}
T.~Squartini, D.~Garlaschelli, {\em Analytical maximum-likelihood method to
  detect patterns in real networks}, New Journal of Physics {\bf 13} (2011)
  083001.
\newblock \href {http://dx.doi.org/10.1088/1367-2630/13/8/083001}
  {\path{doi:10.1088/1367-2630/13/8/083001}}.

\bibitem{fronczak2014exponential}
A.~Fronczak, {\em Exponential Random Graph Models}, Springer-Verlag New York,
  2014, pp. 500--517.

\bibitem{battiston2016price}
S.~Battiston, G.~Caldarelli, R.~May, T.~Roukny, J.~Stiglitz, {\em The price of
  complexity in financial networks}, Proceedings of the National Academy of
  Sciences {\bf 110} (2016) 10031--10036.
\newblock \href {http://dx.doi.org/10.1073/pnas.1521573113}
  {\path{doi:10.1073/pnas.1521573113}}.

\bibitem{anand2017missing}
K.~Anand, I.~van Lelyveld, A.~Banai, S.~Friedrich, R.~Garratt, G.~H. laj,
  J.~Fique, I.~Hansen, S.~M. Jaramillo, H.~Lee, J.~L. Molina-Borboa, S.~Nobili,
  S.~Rajan, D.~Salakhova, T.~C. Silva, L.~Silvestri, S.~R.~S. de~Souza, {\em
  The missing links: A global study on uncovering financial network structures
  from partial data}, Journal of Financial Stability {\bf 35} (2018) 107--119.
\newblock \href {http://dx.doi.org/10.1016/j.jfs.2017.05.012}
  {\path{doi:10.1016/j.jfs.2017.05.012}}.

\bibitem{guimera2009missing}
R.~Guimer{\`a}, M.~Sales-Pardo, {\em Missing and spurious interactions and the
  reconstruction of complex networks}, Proceedings of the National Academy of
  Sciences {\bf 106}~(52) (2009) 22073--22078.
\newblock \href {http://dx.doi.org/10.1073/pnas.0908366106}
  {\path{doi:10.1073/pnas.0908366106}}.

\bibitem{lu2011predicting}
L.~L\"u, T.~Zhou, {\em Link prediction in complex networks: A survey}, Physica
  A: Statistical Mechanics and its Applications {\bf 390} (2011) 1150--1170.
\newblock \href {http://dx.doi.org/10.1016/j.physa.2010.11.027}
  {\path{doi:10.1016/j.physa.2010.11.027}}.

\bibitem{song2005self}
C.~Song, S.~Havlin, H.~A. Makse, {\em Self-similarity of complex networks.},
  Nature {\bf 433}~(7024) (2005) 392--395.
\newblock \href {http://dx.doi.org/10.1038/nature03248}
  {\path{doi:10.1038/nature03248}}.

\bibitem{cover2006elements}
T.~M. Cover, J.~A. Thomas, {\em Elements of Information Theory},
  Wiley-Interscience, 2006.

\bibitem{hanel2014multiplicity}
R.~Hanel, S.~Thurner, M.~Gell-Mann, {\em How multiplicity determines entropy
  and the derivation of the maximum entropy principle for complex systems},
  Proceedings of the National Academy of Sciences {\bf 111}~(19) (2014)
  6905--6910.
\newblock \href {http://dx.doi.org/10.1073/pnas.1406071111}
  {\path{doi:10.1073/pnas.1406071111}}.

\bibitem{presse2013principles}
S.~Press{\'{e}}, K.~Ghosh, J.~Lee, K.~A. Dill, {\em Principles of maximum
  entropy and maximum caliber in statistical physics}, Review of Modern Physics
  {\bf 85}~(3) (2013) 1115--1141.
\newblock \href {http://dx.doi.org/10.1103/RevModPhys.85.1115}
  {\path{doi:10.1103/RevModPhys.85.1115}}.

\bibitem{dewar2003information}
R.~Dewar, {\em Information theory explanation of the fluctuation theorem,
  maximum entropy production and self-organized criticality in non-equilibrium
  stationary states}, Journal of Physics A: Mathematical and General {\bf
  36}~(3) (2003) 631.
\newblock \href {http://dx.doi.org/10.1088/0305-4470/36/3/303}
  {\path{doi:10.1088/0305-4470/36/3/303}}.

\bibitem{wissner2013causal}
A.~D. Wissner-Gross, C.~E. Freer, {\em Causal entropic forces}, Physical Review
  Letters {\bf 110}~(16) (2013) 168702.
\newblock \href {http://dx.doi.org/10.1103/PhysRevLett.110.168702}
  {\path{doi:10.1103/PhysRevLett.110.168702}}.

\bibitem{anand2009entropy}
K.~Anand, G.~Bianconi, {\em Entropy measures for complex networks: toward an
  information theory of complex topologies}, Physical Review E {\bf 80} (2009)
  045102(R).
\newblock \href {http://dx.doi.org/10.1103/PhysRevE.80.045102}
  {\path{doi:10.1103/PhysRevE.80.045102}}.

\bibitem{shannon1948mathematical1}
C.~E. Shannon, {\em A mathematical theory of communication}, Bell System
  Technical Journal {\bf 27} (1948) 379--423.
\newblock \href {http://dx.doi.org/10.1002/j.1538-7305.1948.tb01338.x}
  {\path{doi:10.1002/j.1538-7305.1948.tb01338.x}}.

\bibitem{shannon1948mathematical2}
C.~E. Shannon, {\em A mathematical theory of communication}, Bell System
  Technical Journal {\bf 27} (1948) 623--656.
\newblock \href {http://dx.doi.org/110.1002/j.1538-7305.1948.tb00917.x}
  {\path{doi:110.1002/j.1538-7305.1948.tb00917.x}}.

\bibitem{jaynes1957information}
E.~T. Jaynes, {\em Information theory and statistical mechanics}, Physical
  Review {\bf 106}~(4) (1957) 620--630.
\newblock \href {http://dx.doi.org/10.1103/PhysRev.106.620}
  {\path{doi:10.1103/PhysRev.106.620}}.

\bibitem{lesne2014shannon}
A.~Lesne, {\em Shannon entropy: A rigorous notion at the crossroads between
  probability, information theory, dynamical systems and statistical physics},
  Mathematical Structures in Computer Science {\bf 24}~(3).
\newblock \href {http://dx.doi.org/10.1017/S0960129512000783}
  {\path{doi:10.1017/S0960129512000783}}.

\bibitem{squartini2015information}
T.~Squartini, E.~Ser-Giacomi, D.~Garlaschelli, G.~Judge, {\em Information
  recovery in behavioral networks}, PLoS ONE {\bf 10}~(5) (2015) e0125077.
\newblock \href {http://dx.doi.org/10.1371/journal.pone.0125077}
  {\path{doi:10.1371/journal.pone.0125077}}.

\bibitem{wells2004financial}
S.~Wells, {\em Financial interlinkages in the United Kingdom's interbank market
  and the risk of contagion}, Working Paper 230, Bank of England (2004).

\bibitem{upper2011simulation}
C.~Upper, {\em Simulation methods to assess the danger of contagion in
  interbank markets}, Journal of Financial Stability {\bf 7}~(3) (2011)
  111--125.
\newblock \href {http://dx.doi.org/10.1016/j.jfs.2010.12.001}
  {\path{doi:10.1016/j.jfs.2010.12.001}}.

\bibitem{mistrulli2011assessing}
P.~E. Mistrulli, {\em Assessing financial contagion in the interbank market:
  Maximum entropy versus observed interbank lending patterns}, Journal of
  Banking {\&} Finance {\bf 35}~(5) (2011) 1114--1127.
\newblock \href {http://dx.doi.org/10.1016/j.jbankfin.2010.09.018}
  {\path{doi:10.1016/j.jbankfin.2010.09.018}}.

\bibitem{mastromatteo2012reconstruction}
I.~Mastromatteo, E.~Zarinelli, M.~Marsili, {\em Reconstruction of financial
  networks for robust estimation of systemic risk}, Journal of Statistical
  Mechanics: Theory and Experiment {\bf 2012}~(03) (2012) P03011.
\newblock \href {http://dx.doi.org/10.1088/1742-5468/2012/03/P03011}
  {\path{doi:10.1088/1742-5468/2012/03/P03011}}.

\bibitem{squartini2017network}
T.~Squartini, G.~Cimini, A.~Gabrielli, D.~Garlaschelli, {\em Network
  reconstruction via density sampling}, Applied Network Science {\bf 2}~(1)
  (2017) 3.
\newblock \href {http://dx.doi.org/10.1007/s41109-017-0021-8}
  {\path{doi:10.1007/s41109-017-0021-8}}.

\bibitem{mazzarisi2017methods}
P.~Mazzarisi, F.~Lillo, {\em Methods for reconstructing interbank networks from
  limited information: A comparison}, Springer International Publishing, 2017,
  pp. 201--215.
\newblock \href {http://dx.doi.org/10.1007/978-3-319-47705-3_15}
  {\path{doi:10.1007/978-3-319-47705-3_15}}.

\bibitem{squartini2013jan}
T.~Squartini, D.~Garlaschelli, {\em Jan Tinbergen's legacy for economic
  networks: From the gravity model to quantum statistics}, 2013, pp. 161--186.
\newblock \href {http://dx.doi.org/10.1007/978-3-319-00023-7}
  {\path{doi:10.1007/978-3-319-00023-7}}.

\bibitem{sharpe1964capm}
W.~F. Sharpe, {\em Capital asset prices: A theory of market equilibrium under
  conditions of risk}, The Journal of Finance {\bf 19}~(3) (1964) 425--442.
\newblock \href {http://dx.doi.org/10.2307/2977928}
  {\path{doi:10.2307/2977928}}.

\bibitem{digiangi2016assessing}
D.~{Di Gangi}, F.~Lillo, D.~Pirino, {\em Assessing systemic risk due to fire
  sales spillover through maximum entropy network reconstruction},
  \url{http://arxiv.org/abs/1509.00607} (2016).

\bibitem{kullback1951information}
S.~Kullback, R.~A. Leibler, {\em On information and sufficiency}, The Annals of
  Mathematical Statistics {\bf 22}~(1) (1951) 79--86.
\newblock \href {http://dx.doi.org/10.1214/aoms/1177729694}
  {\path{doi:10.1214/aoms/1177729694}}.

\bibitem{bacharach1965estimating}
M.~Bacharach, {\em Estimating nonnegative matrices from marginal data},
  International Economic Review {\bf 6}~(3) (1965) 294--310.
\newblock \href {http://dx.doi.org/10.2307/2525582}
  {\path{doi:10.2307/2525582}}.

\bibitem{bishop2007discrete}
Y.~M. Bishop, S.~E. Fienberg, P.~W. Holland, {\em Discrete Multivariate
  Analysis: Theory and Practice}, Springer-Verlag New York, 2007.
\newblock \href {http://dx.doi.org/10.1007/978-0-387-72806-3}
  {\path{doi:10.1007/978-0-387-72806-3}}.

\bibitem{drehmann2013measuring}
M.~Drehmann, N.~Tarashev, {\em Measuring the systemic importance of
  interconnected banks}, Journal of Financial Intermediation {\bf 22}~(4)
  (2013) 586--607.
\newblock \href {http://dx.doi.org/10.1016/j.jfi.2013.08.001}
  {\path{doi:10.1016/j.jfi.2013.08.001}}.

\bibitem{ESMA}
{\em ESMA Trade Reporting},
  \url{https://www.esma.europa.eu/policy-rules/post-trading/trade-reporting}.

\bibitem{moussa2011contagion}
A.~Moussa, {\em Contagion and systemic risk in financial networks}, Ph.D.
  thesis, Columbia University PhD Thesis (2011).

\bibitem{bollobas2003directed}
B.~Bollob{\'{a}}s, C.~Borgs, J.~Chayes, O.~Riordan, {\em Directed scale-free
  graphs}, in: Proceedings of the Fourteenth Annual ACM-SIAM Symposium on
  Discrete Algorithms, Society for Industrial and Applied Mathematics, 2003,
  pp. 132--139.

\bibitem{cimini2015estimating}
G.~Cimini, T.~Squartini, A.~Gabrielli, D.~Garlaschelli, {\em Estimating
  topological properties of weighted networks from limited information},
  Physical Review E {\bf 92}~(4) (2015) 040802.
\newblock \href {http://dx.doi.org/10.1103/PhysRevE.92.040802}
  {\path{doi:10.1103/PhysRevE.92.040802}}.

\bibitem{gandy2017adjustable}
A.~Gandy, L.~A.~M. Veraart, {\em Adjustable network reconstruction with
  applications to CDS exposures}, \url{https://ssrn.com/abstract=2895754}
  (2017).

\bibitem{mastrandrea2014enhanced}
R.~Mastrandrea, T.~Squartini, G.~Fagiolo, D.~Garlaschelli, {\em Enhanced
  reconstruction of weighted networks from strengths and degrees}, New Journal
  of Physics {\bf 16}~(4) (2014) 043022.
\newblock \href {http://dx.doi.org/10.1088/1367-2630/16/4/043022}
  {\path{doi:10.1088/1367-2630/16/4/043022}}.

\bibitem{almog2015double}
A.~Almog, T.~Squartini, D.~Garlaschelli, {\em The double role of GDP in shaping
  the structure of the International Trade Network},
  \url{http://arxiv.org/abs/1512.02454} (2015).

\bibitem{garlaschelli2004fitness}
D.~Garlaschelli, M.~Loffredo, {\em Fitness-dependent topological properties of
  the World Trade Web}, Physical Review Letters {\bf 93}~(18) (2004) 1--4.
\newblock \href {http://dx.doi.org/10.1103/PhysRevLett.93.188701}
  {\path{doi:10.1103/PhysRevLett.93.188701}}.

\bibitem{demasi2006fitness}
G.~{De Masi}, G.~Iori, G.~Caldarelli, G.~D. Masi, {\em Fitness model for the
  Italian interbank money market}, Physical Review E {\bf 74}~(6) (2006)
  066112.
\newblock \href {http://dx.doi.org/10.1103/PhysRevE.74.066112}
  {\path{doi:10.1103/PhysRevE.74.066112}}.

\bibitem{cimini2015systemic}
G.~Cimini, T.~Squartini, D.~Garlaschelli, A.~Gabrielli, {\em Systemic risk
  analysis on reconstructed economic and financial networks.}, Scientific
  Reports {\bf 5} (2015) 15758.
\newblock \href {http://dx.doi.org/10.1038/srep15758}
  {\path{doi:10.1038/srep15758}}.

\bibitem{garlaschelli2005scale}
D.~Garlaschelli, S.~Battiston, M.~Castri, V.~D. Servedio, G.~Caldarelli, {\em
  The scale-free topology of market investments}, Physica A: Statistical and
  Theoretical Physics {\bf 350}~(2-4) (2005) 491--499.
\newblock \href {http://dx.doi.org/doi: 10.1016/j.physa.2004.11.040}
  {\path{doi:doi: 10.1016/j.physa.2004.11.040}}.

\bibitem{squartini2017stock}
T.~Squartini, A.~Almog, G.~Caldarelli, I.~van Lelyveld, D.~Garlaschelli,
  G.~Cimini, {\em Enhanced capital-asset pricing model for the reconstruction
  of bipartite financial networks}, Physical Review E {\bf 96} (2017) 032315.
\newblock \href {http://dx.doi.org/10.1103/PhysRevE.96.032315}
  {\path{doi:10.1103/PhysRevE.96.032315}}.

\bibitem{barrat2004architecture}
A.~Barrat, M.~Barthelemy, R.~Pastor-Satorras, A.~Vespignani, {\em The
  architecture of complex weighted networks}, Proceedings of the National
  Academy of Sciences {\bf 101}~(11) (2004) 3747--3752.
\newblock \href {http://dx.doi.org/10.1073/pnas.0400087101}
  {\path{doi:10.1073/pnas.0400087101}}.

\bibitem{musmeci2013bootstrapping}
N.~Musmeci, S.~Battiston, G.~Caldarelli, M.~Puliga, A.~Gabrielli, {\em
  Bootstrapping topology and systemic risk of complex network using the fitness
  model}, Journal of Statistical Physics {\bf 151} (2013) 720--734.
\newblock \href {http://dx.doi.org/10.1007/s10955-013-0720-1}
  {\path{doi:10.1007/s10955-013-0720-1}}.

\bibitem{blagus2015empirical}
N.~Blagus, L.~{\v{S}}ubelj, M.~Bajec, {\em Empirical comparison of network
  sampling techniques}, \url{https://arxiv.org/abs/1506.02449} (2015).

\bibitem{battiston2016leveraging}
S.~Battiston, G.~Caldarelli, M.~D'Errico, S.~Gurciullo, {\em Leveraging the
  network: a stress-test framework based on DebtRank}, Statistics and Risk
  Modeling {\bf 33} (2016) 117--138.
\newblock \href {http://dx.doi.org/10.1515/strm-2015-0005}
  {\path{doi:10.1515/strm-2015-0005}}.

\bibitem{ruzzenenti2012spatial}
F.~Ruzzenenti, F.~Picciolo, R.~Basosi, D.~Garlaschelli, {\em Spatial effects in
  real networks: Measures, null models, and applications}, Physical Review E
  {\bf 86}~(6) (2012) 66110.
\newblock \href {http://dx.doi.org/10.1103/PhysRevE.86.066110}
  {\path{doi:10.1103/PhysRevE.86.066110}}.

\bibitem{duenas2013modeling}
M.~Duenas, G.~Fagiolo, {\em Modeling the International-Trade Network: A gravity
  approach}, Journal of Economic Interaction and Coordination {\bf 8}~(1)
  (2013) 155--178.
\newblock \href {http://dx.doi.org/10.1007/s11403-013-0108-y}
  {\path{doi:10.1007/s11403-013-0108-y}}.

\bibitem{anand2011shannon}
K.~Anand, G.~Bianconi, S.~Severini, {\em Shannon and von Neumann entropy of
  random networks with heterogeneous expected degree}, Physical Review E {\bf
  83} (2011) 036109.
\newblock \href {http://dx.doi.org/10.1103/PhysRevE.83.036109}
  {\path{doi:10.1103/PhysRevE.83.036109}}.

\bibitem{dedomenico2016spectral}
M.~{De Domenico}, J.~Biamonte, {\em Spectral entropies as information-theoretic
  tools for complex network comparison}, Physical Review X {\bf 6}~(4) (2016)
  41062.
\newblock \href {http://dx.doi.org/10.1103/PhysRevX.6.041062}
  {\path{doi:10.1103/PhysRevX.6.041062}}.

\bibitem{judge2011information}
G.~G. Judge, R.~C. Mittelhammer, {\em An Information Theoretic Approach to
  Econometrics}, Cambridge University Press, 2011.

\bibitem{cressie1984multinomial}
N.~Cressie, T.~Read, {\em Multinomial goodness-of-fit tests}, Journal of the
  Royal Statistical Society. Series B (Methodological) {\bf 46}~(3) (1984)
  440--464.

\bibitem{cho2015information}
W.~K.~T. Cho, G.~Judge, {\em An information theoretic approach to network
  tomography}, Applied Economics Letters {\bf 22}~(1) (2015) 1--6.
\newblock \href {http://dx.doi.org/10.1080/13504851.2013.866199}
  {\path{doi:10.1080/13504851.2013.866199}}.

\bibitem{renyi1961entropy}
A.~R\'enyi, {\em On Measures of Entropy and Information}, in: Proceedings of
  the Fourth Berkeley Symposium on Mathematical Statistics and Probability,
  University of California Press, Berkeley, 1961, pp. 547--561.

\bibitem{tsallis1988possible}
C.~Tsallis, {\em Possible generalization of Boltzmann-Gibbs statistics},
  Journal of Statistical Physics {\bf 52}~(1) (1988) 479--487.
\newblock \href {http://dx.doi.org/10.1007/BF01016429}
  {\path{doi:10.1007/BF01016429}}.

\bibitem{piantadosi2012copulas}
J.~Piantadosi, P.~Howlett, J.~Borwein, {\em Copulas with maximum entropy},
  Optimization Letters {\bf 6}~(1) (2012) 99--125.
\newblock \href {http://dx.doi.org/10.1007/s11590-010-0254-2}
  {\path{doi:10.1007/s11590-010-0254-2}}.

\bibitem{baral2012estimating}
P.~Baral, J.~P. Fique, {\em Estimation of bilateral exposures - A Copula
  approach}, Mimeo, Indiana University, University of Porto (2012).

\bibitem{gandy2016bayesian}
A.~Gandy, L.~A.~M. Veraart, {\em A Bayesian methodology for systemic risk
  assessment in financial networks}, Management Science (articles in advances
  (2016) 4428--4446.
\newblock \href {http://dx.doi.org/10.1287/mnsc.2016.2546}
  {\path{doi:10.1287/mnsc.2016.2546}}.

\bibitem{schrijver2002history}
A.~Schrijver, {\em On the history of the transportation and maximum flow
  problems}, Mathematical Programming {\bf 91}~(3) (2002) 437.
\newblock \href {http://dx.doi.org/10.1007/s101070100259}
  {\path{doi:10.1007/s101070100259}}.

\bibitem{montagna2017contagion}
M.~Montagna, T.~Lux, {\em Contagion risk in the interbank market: A
  probabilistic approach to cope with incomplete structural information},
  Quantitative Finance {\bf 17}~(1) (2017) 101--120.
\newblock \href {http://dx.doi.org/10.1080/14697688.2016.1178855}
  {\path{doi:10.1080/14697688.2016.1178855}}.

\bibitem{nier2007network}
E.~Nier, J.~Yang, T.~Yorulmazer, A.~Alentorn, {\em Network models and financial
  stability}, Journal of Economic Dynamics and Control {\bf 31}~(6) (2007)
  2033--2060.
\newblock \href {http://dx.doi.org/doi:10.1016/j.jedc.2007.01.014}
  {\path{doi:doi:10.1016/j.jedc.2007.01.014}}.

\bibitem{halaj2013assessing}
G.~Halaj, C.~Kok, {\em Assessing interbank contagion using simulated networks},
  Computational Management Science {\bf 10}~(2) (2013) 157--186.
\newblock \href {http://dx.doi.org/10.1007/s10287-013-0168-4}
  {\path{doi:10.1007/s10287-013-0168-4}}.

\bibitem{anand2014filling}
K.~Anand, B.~Craig, G.~von Peter, {\em Filling in the blanks: Network structure
  and interbank contagion}, Quantitative Finance {\bf 15}~(4) (2015) 625--636.
\newblock \href {http://dx.doi.org/10.1080/14697688.2014.968195}
  {\path{doi:10.1080/14697688.2014.968195}}.

\bibitem{fawcett2006roc}
T.~Fawcett, {\em An introduction to ROC analysis}, Pattern Recognition Letters
  {\bf 27} (2006) 861--874.
\newblock \href {http://dx.doi.org/10.1016/j.patrec.2005.10.010}
  {\path{doi:10.1016/j.patrec.2005.10.010}}.

\bibitem{metz1978roc}
C.~E. Metz, {\em Basic principles of ROC analysis}, Seminars in Nuclear
  Medicine {\bf 8}~(4) (1978) 283--298.
\newblock \href {http://dx.doi.org/10.1016/S0001-2998(78)80014-2}
  {\path{doi:10.1016/S0001-2998(78)80014-2}}.

\bibitem{wang2014similarity}
J.~Wang, H.~T. Shen, J.~Song, J.~Ji, {\em Hashing for similarity search: A
  survey}, \url{https://arxiv.org/abs/1408.2927} (2014).

\bibitem{casella2002statistical}
G.~Casella, R.~L. Berger, {\em Statistical Inference}, Vol.~2, Duxbury Pacific
  Grove, CA, 2002.

\bibitem{luu2017structural}
D.~T. Luu, T.~Lux, B.~Yanovski, {\em Structural correlations in the Italian
  overnight money market: An analysis based on network configuration models},
  Entropy {\bf 19}~(6) (2017) 259.
\newblock \href {http://dx.doi.org/10.3390/e19060259}
  {\path{doi:10.3390/e19060259}}.

\bibitem{squartini2013early}
T.~Squartini, I.~van Lelyveld, D.~Garlaschelli, {\em Early-warning signals of
  topological collapse in interbank networks.}, Scientific reports {\bf 3}
  (2013) 3357.
\newblock \href {http://dx.doi.org/10.1038/srep03357}
  {\path{doi:10.1038/srep03357}}.

\bibitem{saracco2016detecting}
F.~Saracco, R.~{Di Clemente}, A.~Gabrielli, T.~Squartini, {\em Detecting early
  signs of the 2007--2008 crisis in the world trade}, Scientific Reports {\bf
  6} (2016) 30286.
\newblock \href {http://dx.doi.org/10.1038/srep30286}
  {\path{doi:10.1038/srep30286}}.

\bibitem{gualdi2016statistically}
S.~Gualdi, G.~Cimini, K.~Primicerio, R.~D. Clemente, D.~Challet, {\em
  Statistically validated network of portfolio overlaps and systemic risk},
  Scientific Reports {\bf 6} (2016) 39467.
\newblock \href {http://dx.doi.org/10.1038/srep39467}
  {\path{doi:10.1038/srep39467}}.

\bibitem{bifone2016surprise}
C.~Nicolini, A.~Bifone, {\em Modular structure of brain functional networks:
  breaking the resolution limit by Surprise}, Scientific Reports {\bf 6} (2016)
  19250.
\newblock \href {http://dx.doi.org/10.1038/srep19250}
  {\path{doi:10.1038/srep19250}}.

\bibitem{karrer2011stochastic}
B.~Karrer, M.~E.~J. Newman, {\em Stochastic blockmodels and community structure
  in networks}, Physical Review E {\bf 83}~(1) (2011) 16107.
\newblock \href {http://dx.doi.org/10.1103/PhysRevE.83.016107}
  {\path{doi:10.1103/PhysRevE.83.016107}}.

\bibitem{fronczak2013exponential}
P.~Fronczak, A.~Fronczak, M.~Bujok, {\em Exponential random graph models for
  networks with community structure}, Physical Review E {\bf 88}~(3) (2013)
  32810.
\newblock \href {http://dx.doi.org/10.1103/PhysRevE.88.032810}
  {\path{doi:10.1103/PhysRevE.88.032810}}.

\bibitem{barucca2016disentangling}
P.~Barucca, F.~Lillo, {\em Disentangling bipartite and core-periphery structure
  in financial networks}, Chaos, Solitons {\&} Fractals {\bf 88} (2016)
  244--253.
\newblock \href {http://dx.doi.org/10.1016/j.chaos.2016.02.004}
  {\path{doi:10.1016/j.chaos.2016.02.004}}.

\bibitem{craig2014interbank}
B.~Craig, G.~von Peter, {\em Interbank tiering and money center banks}, Journal
  of Financial Intermediation {\bf 23}~(3) (2014) 322--347.
\newblock \href {http://dx.doi.org/http://dx.doi.org/10.1016/j.jfi.2014.02.003}
  {\path{doi:http://dx.doi.org/10.1016/j.jfi.2014.02.003}}.

\bibitem{decelle2011inference}
A.~Decelle, F.~Krzakala, C.~Moore, L.~Zdeborov{\'{a}}, {\em Inference and phase
  transitions in the detection of modules in sparse networks}, Physical Review
  Letters {\bf 107}~(6) (2011) 65701.
\newblock \href {http://dx.doi.org/10.1103/PhysRevLett.107.065701}
  {\path{doi:10.1103/PhysRevLett.107.065701}}.

\bibitem{newman2014core}
X.~Zhang, T.~Martin, M.~Newman, {\em Identification of core-periphery structure
  in networks}, Physical Review E {\bf 91} (2014) 0321803.
\newblock \href {http://dx.doi.org/10.1103/PhysRevE.91.032803}
  {\path{doi:10.1103/PhysRevE.91.032803}}.

\bibitem{yan2013block}
X.~Yan, C.~R. Shalizi, J.~E. Jnesen, F.~Krzakala, C.~Moore, L.~Zdeborova,
  P.~Zhang, Y.~Zhu, {\em Model selection for degree-corrected block models},
  Journal of Statistical Mechanics 2014 (2014) P05007.
\newblock \href {http://dx.doi.org/10.1088/1742-5468/2014/05/P05007}
  {\path{doi:10.1088/1742-5468/2014/05/P05007}}.

\bibitem{allen2000financial}
F.~Allen, D.~Gale, {\em Financial contagion}, Journal of Political Economy {\bf
  108}~(1) (2000) 1--33.
\newblock \href {http://dx.doi.org/10.1086/262109} {\path{doi:10.1086/262109}}.

\bibitem{brunnermeier2009deciphering}
M.~K. Brunnermeier, {\em Deciphering the liquidity and credit crunch
  2007-2008}, Journal of Economic Perspectives {\bf 23}~(1) (2009) 77--100.
\newblock \href {http://dx.doi.org/10.1257/jep.23.1.77}
  {\path{doi:10.1257/jep.23.1.77}}.

\bibitem{lau2009assessing}
J.~A. Chan-Lau, M.~Espinosa, K.~Giesecke, J.~A. Sol{\'{e}}, {\em Assessing the
  systemic implications of financial linkages}, Global financial stability
  report (chapter 2), IMF Monetary and Capital Markets Development (2009).

\bibitem{gai2010contagion}
P.~Gai, S.~Kapadia, {\em Contagion in financial networks}, Proceedings of the
  Royal Society A: Mathematical, Physical and Engineering Sciences {\bf
  466}~(2120) (2010) 2401--2423.
\newblock \href {http://dx.doi.org/10.1098/rspa.2009.0410}
  {\path{doi:10.1098/rspa.2009.0410}}.

\bibitem{haldane2011systemic}
A.~G. Haldane, R.~M. May, {\em Systemic risk in banking ecosystems.}, Nature
  {\bf 469}~(7330) (2011) 351--355.
\newblock \href {http://dx.doi.org/10.1038/nature09659}
  {\path{doi:10.1038/nature09659}}.

\bibitem{battiston2016complexity}
S.~Battiston, J.~D. Farmer, A.~Flache, D.~Garlaschelli, A.~G. Haldane,
  H.~Heesterbeek, C.~Hommes, C.~Jaeger, R.~May, M.~Scheffer, {\em Complexity
  theory and financial regulation}, Science {\bf 351}~(6275) (2016) 818--819.
\newblock \href {http://dx.doi.org/10.1126/science.aad0299}
  {\path{doi:10.1126/science.aad0299}}.

\bibitem{bardoscia2017pathways}
M.~Bardoscia, S.~Battiston, F.~Caccioli, G.~Caldarelli, {\em Pathways towards
  instability in financial networks}, Nature Communications {\bf 8} (2017)
  14416.
\newblock \href {http://dx.doi.org/10.1038/ncomms14416}
  {\path{doi:10.1038/ncomms14416}}.

\bibitem{beale2011individual}
N.~Beale, D.~G. Rand, H.~Battey, K.~Croxson, R.~M. May, M.~A. Nowak, {\em
  Individual versus systemic risk and the Regulator's Dilemma}, Proceedings of
  the National Academy of Sciences {\bf 108}~(31) (2011) 12647--52.
\newblock \href {http://dx.doi.org/10.1073/pnas.1105882108}
  {\path{doi:10.1073/pnas.1105882108}}.

\bibitem{corsi2013when}
F.~Corsi, S.~Marmi, F.~Lillo, {\em When micro prudence increases macro risk:
  The destabilizing effects of financial innovation, leverage, and
  diversification}, Operations Research {\bf 64}~(5) (2016) 1073--1088.
\newblock \href {http://dx.doi.org/10.1287/opre.2015.1464}
  {\path{doi:10.1287/opre.2015.1464}}.

\bibitem{glasserman2015contagion}
P.~Glasserman, H.~P. Young, {\em How likely is contagion in financial
  networks?}, Journal of Banking and Finance {\bf 50} (2015) 383--399.
\newblock \href {http://dx.doi.org/10.1016/j.jbankfin.2014.02.006}
  {\path{doi:10.1016/j.jbankfin.2014.02.006}}.

\bibitem{boss2004network}
M.~Boss, H.~Elsinger, M.~Summer, S.~Thurner, {\em Network topology of the
  interbank market}, Quantitative Finance {\bf 4}~(6) (2004) 677--684.
\newblock \href {http://dx.doi.org/10.1080/14697680400020325}
  {\path{doi:10.1080/14697680400020325}}.

\bibitem{krause2012interbank}
A.~Krause, S.~Giansante, {\em Interbank lending and the spread of bank
  failures: A network model of systemic risk}, Journal of Economic Behavior
  {\&} Organization {\bf 83}~(3) (2012) 583--608.

\bibitem{georg2013effect}
C.-P. Georg, {\em The effect of the interbank network structure on contagion
  and common shocks}, Journal of Banking {\&} Finance 37~(7) (2013) 2216--2228.
\newblock \href {http://dx.doi.org/10.1016/j.jbankfin.2013.02.032}
  {\path{doi:10.1016/j.jbankfin.2013.02.032}}.

\bibitem{greenwood2015vulnerable}
R.~Greenwood, A.~Landier, D.~Thesmar, {\em Vulnerable banks}, Journal of
  Financial Economics {\bf 115}~(3) (2015) 471--485.
\newblock \href {http://dx.doi.org/10.1016/j.jfineco.2014.11.006}
  {\path{doi:10.1016/j.jfineco.2014.11.006}}.

\bibitem{acemoglu2015systemic}
D.~Acemoglu, A.~Ozdaglar, A.~Tahbaz-Salehi, {\em Systemic risk and stability in
  financial networks}, American Economic Review {\bf 105}~(2) (2015) 564.
\newblock \href {http://dx.doi.org/10.1257/aer.20130456}
  {\path{doi:10.1257/aer.20130456}}.

\bibitem{bougheas2015complex}
S.~Bougheas, A.~Kirman, {\em Complex Financial Networks and Systemic Risk: A
  Review}, Springer International Publishing, 2015, pp. 115--139.
\newblock \href {http://dx.doi.org/10.1007/978-3-319-12805-4_6}
  {\path{doi:10.1007/978-3-319-12805-4_6}}.

\bibitem{hurd2016contagion}
T.~R. Hurd, {\em Contagion! Systemic Risk in Financial Networks},
  SpringerBriefs in Quantitative Finance, Springer International Publishing,
  2016.
\newblock \href {http://dx.doi.org/10.1007/978-3-319-33930-6}
  {\path{doi:10.1007/978-3-319-33930-6}}.

\bibitem{eisenberg2001systemic}
L.~Eisenberg, T.~H. Noe, {\em Systemic risk in financial systems}, Management
  Science {\bf 47}~(2) (2001) 236--249.
\newblock \href {http://dx.doi.org/10.1287/mnsc.47.2.236.9835}
  {\path{doi:10.1287/mnsc.47.2.236.9835}}.

\bibitem{furfine2003interbank}
C.~H. Furfine, {\em Interbank exposures: Quantifying the risk of contagion},
  Journal of Money, Credit and Banking {\bf 35}~(1) (2003) 111--128.
\newblock \href {http://dx.doi.org/10.1353/mcb.2003.0004}
  {\path{doi:10.1353/mcb.2003.0004}}.

\bibitem{rogers2013failure}
L.~C.~G. Rogers, L.~A.~M. Veraart, {\em Failure and rescue in an interbank
  network}, Management Science {\bf 59}~(4) (2013) 882--898.
\newblock \href {http://dx.doi.org/10.1287/mnsc.1120.1569}
  {\path{doi:10.1287/mnsc.1120.1569}}.

\bibitem{battiston2012debtrank}
S.~Battiston, M.~Puliga, R.~Kaushik, P.~Tasca, G.~Caldarelli, {\em DebtRank:
  Too central to fail? Financial networks, the FED and systemic risk.},
  Scientific Reports {\bf 2} (2012) 541.
\newblock \href {http://dx.doi.org/10.1038/srep00541}
  {\path{doi:10.1038/srep00541}}.

\bibitem{bardoscia2015debtrank}
M.~Bardoscia, S.~Battiston, F.~Caccioli, G.~Caldarelli, {\em DebtRank: A
  microscopic foundation for shock propagation.}, PLoS ONE {\bf 10}~(6) (2015)
  e0130406.
\newblock \href {http://dx.doi.org/10.1371/journal.pone.0130406}
  {\path{doi:10.1371/journal.pone.0130406}}.

\bibitem{barucca2016network}
P.~Barucca, M.~Bardoscia, F.~Caccioli, M.~D'Errico, G.~Visentin, G.~Caldarelli,
  S.~Battiston, {\em Network valuation in financial systems}.

\bibitem{cifuentes2005liquidity}
R.~Cifuentes, G.~Ferrucci, H.~S. Shin, {\em Liquidity risk and contagion},
  Journal of the European Economic Association {\bf 3}~(2-3) (2005) 556--566.
\newblock \href {http://dx.doi.org/10.1162/jeea.2005.3.2-3.556}
  {\path{doi:10.1162/jeea.2005.3.2-3.556}}.

\bibitem{kapadia2012quantifying}
S.~Kapadia, M.~Drehmann, J.~Elliott, G.~Sterne, {\em Liquidity risk, cash flow
  constraints, and systemic feedbacks}, University of Chicago Press, 2012, pp.
  29--61.
\newblock \href {http://dx.doi.org/10.7208/chicago/9780226921969.003.0003}
  {\path{doi:10.7208/chicago/9780226921969.003.0003}}.

\bibitem{anand2012rollover}
K.~Anand, P.~Gai, M.~Marsili, {\em Rollover risk, network structure and
  systemic financial crises}, Journal of Economic Dynamics and Control {\bf
  36}~(8) (2012) 1088--1100.
\newblock \href {http://dx.doi.org/10.1016/j.jedc.2012.03.005}
  {\path{doi:10.1016/j.jedc.2012.03.005}}.

\bibitem{gale2013liquidity}
D.~Gale, T.~Yorulmazer, {\em Liquidity hoarding}, Theoretical Economics {\bf
  8}~(2) (2013) 291--324.
\newblock \href {http://dx.doi.org/10.3982/te1064} {\path{doi:10.3982/te1064}}.

\bibitem{brunnermeier2009market}
M.~K. Brunnermeier, L.~H. Pedersen, {\em Market liquidity and funding
  liquidity}, Review of Financial Studies {\bf 22}~(6) (2009) 2201--2238.
\newblock \href {http://dx.doi.org/10.1093/rfs/hhn098}
  {\path{doi:10.1093/rfs/hhn098}}.

\bibitem{acharya2010precautionary}
V.~V. Acharya, O.~Merrouche, {\em Precautionary hoarding of liquidity and
  interbank markets: Evidence from the subprime crisis}, Review of Finance {\bf
  17}~(1) (2013) 107--160.
\newblock \href {http://dx.doi.org/10.1093/rof/rfs022}
  {\path{doi:10.1093/rof/rfs022}}.

\bibitem{adrian2010liquidity}
T.~Adrian, H.~S. Shin, {\em Liquidity and leverage}, Journal of Financial
  Intermediation 19~({\bf 3}) (2010) 418--437.
\newblock \href {http://dx.doi.org/10.1016/j.jfi.2008.12.002}
  {\path{doi:10.1016/j.jfi.2008.12.002}}.

\bibitem{feldhutter2012same}
P.~Feldh{\"{u}}tter, {\em The same bond at different prices: Identifying search
  frictions and selling pressures}, Review of Financial Studies {\bf 25}~(4)
  (2012) 1155--1206.
\newblock \href {http://dx.doi.org/10.1093/rfs/hhr093}
  {\path{doi:10.1093/rfs/hhr093}}.

\bibitem{cimini2016entangling}
G.~Cimini, M.~Serri, {\em Entangling credit and funding shocks in interbank
  markets}, PLoS ONE {\bf 11}~(8) (2016) e0161642.
\newblock \href {http://dx.doi.org/10.1371/journal.pone.0161642}
  {\path{doi:10.1371/journal.pone.0161642}}.

\bibitem{shleifer2011fire}
A.~Shleifer, R.~Vishny, {\em Fire sales in finance and macroeconomics}, Journal
  of Economic Perspectives {\bf 25}~(1) (2011) 29--48.
\newblock \href {http://dx.doi.org/10.1257/jep.25.1.29}
  {\path{doi:10.1257/jep.25.1.29}}.

\bibitem{caccioli2014stability}
F.~Caccioli, M.~Shrestha, C.~Moore, J.~D. Farmer, {\em Stability analysis of
  financial contagion due to overlapping portfolios}, Journal of Banking {\&}
  Finance {\bf 46} (2014) 233--245.
\newblock \href {http://dx.doi.org/10.1016/j.jbankfin.2014.05.021}
  {\path{doi:10.1016/j.jbankfin.2014.05.021}}.

\bibitem{cont2016fire}
R.~Cont, L.~Wagalath, {\em Fire sales forensics: Measuring endogenous risk},
  Mathematical Finance {\bf 26}~(4) (2016) 835--866.
\newblock \href {http://dx.doi.org/10.1111/mafi.12071}
  {\path{doi:10.1111/mafi.12071}}.

\bibitem{wilks1938large}
S.~S. Wilks, {\em The large-sample distribution of the likelihood ratio for
  testing composite hypotheses}, The Annals of Mathematical Statistics {\bf
  9}~(1) (1938) 60--62.
\newblock \href {http://dx.doi.org/10.1214/aoms/1177732360}
  {\path{doi:10.1214/aoms/1177732360}}.

\bibitem{burnham2002model}
K.~P. Burnham, D.~R. Anderson, {\em Model Selection and Multimodel Inference: A
  Practical Information-Theoretic Approach}, Springer-Verlag New York, 2002.
\newblock \href {http://dx.doi.org/10.1007/b97636} {\path{doi:10.1007/b97636}}.

\bibitem{burnham2004multimodel}
K.~P. Burnham, D.~R. Anderson, {\em Multimodel inference}, Sociological Methods
  {\&} Research {\bf 33}~(2) (2004) 261--304.
\newblock \href {http://dx.doi.org/10.1177/0049124104268644}
  {\path{doi:10.1177/0049124104268644}}.

\bibitem{akaike1974new}
H.~Akaike, {\em A new look at the statistical model identification}, IEEE
  Transactions on Automatic Control {\bf 19}~(6) (1974) 716--723.
\newblock \href {http://dx.doi.org/10.1109/TAC.1974.1100705}
  {\path{doi:10.1109/TAC.1974.1100705}}.

\bibitem{gai2011complexity}
P.~Gai, A.~Haldane, S.~Kapadia, {\em Complexity, concentration and contagion},
  Journal of Monetary Economics {\bf 58}~(5) (2011) 453--470.
\newblock \href {http://dx.doi.org/10.1016/j.jmoneco.2011.05.005}
  {\path{doi:10.1016/j.jmoneco.2011.05.005}}.

\bibitem{ramadiah2017reconstructing}
A.~Ramadiah, F.~Caccioli, D.~Fricke, {\em Reconstructing and stress testing
  credit networks}, \url{https://dx.doi.org/10.2139/ssrn.3084543} (2017).

\bibitem{saracco2015}
F.~Saracco, R.~Di~Clemente, A.~Gabrielli, T.~Squartini, {\em Randomizing
  bipartite networks: The case of the World Trade Web}, Scientific Reports {\bf
  5} (2015) 10595.
\newblock \href {http://dx.doi.org/doi:10.1038/srep10595}
  {\path{doi:doi:10.1038/srep10595}}.

\bibitem{picciolo2012role}
F.~Picciolo, T.~Squartini, F.~Ruzzenenti, R.~Basosi, D.~Garlaschelli, {\em The
  role of distances in the World Trade Web}, in: 2012 Eighth International
  Conference on Signal Image Technology and Internet Based Systems, 2012, pp.
  784--792.
\newblock \href {http://dx.doi.org/10.1109/SITIS.2012.118}
  {\path{doi:10.1109/SITIS.2012.118}}.

\bibitem{kin1938math}
A.~I. Khintchine, {\em Mathematical Foundations of Information Theory}, Dover
  Publications (New York), 1957.

\end{thebibliography}

\end{document}